\documentclass[twocolumn]{aastex63}
\pdfoutput = 1

\usepackage{hhline}
\usepackage{multirow}
\usepackage{amssymb}
\usepackage{amsmath}
\usepackage{natbib}

\newcommand{\lya}{\textrm{Ly}\ensuremath{\alpha}}
\def\oii{[O~{\sc ii]}}
\def\hi{H~{\sc i}}
\newcommand{\msun}{\textrm{M}\ensuremath{_{\odot}}}
\newcommand{\fesc}{\ensuremath{f_{esc}^{\rm Ly\alpha}}}
\newcommand{\bw}{\ensuremath{B_{W}}}
\newcommand{\ia}{\ensuremath{IA445}}

\shorttitle{\lya~Escape from LAEs}
\shortauthors{Pucha et al.}

\begin{document}
%\linenumbers
\title{Lyman-Alpha Escape from Low-Mass, Compact, High-Redshift Galaxies}

\correspondingauthor{Ragadeepika Pucha}
\email{rpucha@arizona.edu}

\author[0000-0002-4940-3009]{Ragadeepika Pucha}
\affil{Steward Observatory, The University of Arizona, 933 North Cherry Avenue, Tucson, AZ 85721, USA}

\author[0000-0001-9687-4973]{Naveen A. Reddy}
\affil{Department of Physics \& Astronomy, University of California, 900 University Avenue, Riverside, CA 92521, USA}

\author[0000-0002-4928-4003]{Arjun Dey}
\author[0000-0002-0000-2394]{St\'ephanie Juneau}
\affiliation{NSF's NOIRLab, 950 N. Cherry Ave., Tucson, AZ 85719, USA}

\author[0000-0003-3004-9596]{Kyoung-Soo Lee}
\affil{Department of Physics \& Astronomy, Purdue University, 525 Northwestern Avenue, West Lafayette, IN 47907, USA}

\author[0000-0001-8302-0565]{Moire K. M. Prescott}
\affiliation{Department of Astronomy, New Mexico State University, P.O.Box 30001, MSC 4500, Las Cruces, NM, 88003, USA}

\author[0000-0003-4702-7561]{Irene Shivaei}
\affil{Steward Observatory, The University of Arizona, 933 North Cherry Avenue, Tucson, AZ 85721, USA}

\author[0000-0001-9991-8222]{Sungryong Hong}
\affiliation{Korea Astronomy and Space Science Institute, 776 Daedeokdae-ro, Yuseong-gu, Daejeon 34055, Korea}

\begin{abstract}

We investigate the effects of stellar populations and sizes on \lya~escape in 27 spectroscopically confirmed and 35 photometric Lyman-Alpha Emitters (LAEs) at z $\approx$ 2.65 in seven fields of the Bo\"otes region of the NOAO Deep Wide-Field Survey. We use deep {\it HST}/WFC3 imaging to supplement ground-based observations and infer key galaxy properties. Compared to typical star-forming galaxies (SFGs) at similar redshifts, the  LAEs are less massive ($M_{\star} \approx 10^{7} - 10^{9}~\msun$), younger (ages $\lesssim$ 1 Gyr), smaller ($r_{e} <$ 1 kpc), less dust-attenuated (E(B$-$V) $\le$ 0.26 mag), but have comparable star-formation-rates (SFRs $\approx 1 - 100~\msun {\rm yr^{-1}}$). Some of the LAEs in the sample may be very young galaxies having low nebular metallicities (${\rm Z_{neb} \lesssim 0.2 Z_{\odot}}$) and/or high ionization parameters ($\log{(\rm U)} \gtrsim -2.4$). Motivated by previous studies, we examine the effects of the concentration of star formation and gravitational potential on \lya~escape, by computing star-formation-rate surface density, $\Sigma_{\rm SFR}$ and specific star-formation-rate surface density, $\Sigma_{\rm sSFR}$. For a given $\Sigma_{\rm SFR}$, the \lya~escape fraction is higher for LAEs with lower stellar masses. LAEs have higher $\Sigma_{\rm sSFR}$ on average compared to SFGs. Our results suggest that compact star formation in a low gravitational potential yields conditions amenable to the escape of \lya~photons. These results have important implications for the physics of \lya~radiative transfer and for the type of galaxies that may contribute significantly to cosmic reionization.
\end{abstract}

%\keywords{}


\section{Introduction} \label{sec:intro}

The Lyman Alpha (\lya) emission line of Hydrogen is one of the strongest emission lines in the Universe, and is invaluable for selecting high-redshift sources \citep{Partridge+1967}. Galaxies selected via this emission line (Lyman Alpha Emitters, or LAEs) are typically fainter in their continuum emission compared to galaxies selected using traditional Lyman Break techniques \citep[Lyman Break Galaxies or LBGs; ][]{Steidel+1993, Giavalisco2002}. LAEs thus probe the low-luminosity end of the galaxy luminosity function. They are also excellent tracers of large-scale structures in the Universe \citep[see][and references therein]{Ouchi+2020}.

Several deep narrowband imaging surveys have targeted LAEs in the past couple of decades. These include the Large Area Lyman Alpha (LALA) survey \citep[e.g.,][]{Rhoads+2000}, the Subaru deep survey \citep[e.g.,][]{Ouchi+2003}, the Hobby-Eberly Telescope Dark Energy Experiment (HETDEX) Pilot Survey \citep[e.g.,][]{Adams+2011}, and the Multi Unit Spectroscopic Explorer (MUSE) Hubble Ultra Deep Field survey \citep[e.g.,][]{Bacon+2017}. Due to their faint continuum, these galaxies are often studied via stacking analysis \citep[][]{Gawiser+2006, Finkelstein+2007, Gawiser+2007, Nilsson+2007, Lai+2008, Finkelstein+2009, Ono+2010, Guaita+2011, Acquaviva+2012, Vargas+2014, Hao+2018, Kusakabe+2018}, which can mask the intrinsic scatter in the distribution of LAE properties. Due to the availability of deep observations, it is now possible to study individual LAEs and their properties \citep[][]{Pirzkal+2007, Nilsson+2011, Hagen+2014, McLinden+2014, Vargas+2014, Sandberg+2015, Hathi+2016, Shimakawa+2017, Hao+2018}. These studies suggest that LAEs are typically low-mass, young, star-forming galaxies with low dust content. LAEs are also considered as important probes for studying low mass galaxies that are primary building blocks of typical present day $L^{\star}$ galaxies \citep{Gawiser+2007, Hao+2018, Khostovan+2019, HerreroAlonso+2021}. Furthermore, morphological studies of LAEs have shown that they are compact (effective radius, $r_{e}~\approx~1~{\rm kpc}$) and the sizes do not evolve with redshift \citep{Bond+2009, Gronwall+2011, Bond+2012, Malhotra+2012, Paulino-Afonso+2018, Shibuya+2019}. 

Despite this progress, the resonant nature of \lya~emission poses many challenges in interpreting the observations \citep{Dijkstra2017}. In fact, \lya~is often used to study the spatial structure of interstellar medium (ISM) and circumgalactic medium (CGM) in and around galaxies \citep[][]{Steidel+2011, Momose+2014, Wisotzki+2018, Herenz+2020, Leclercq+2020, Sanderson+2021}. These photons can be easily scattered by neutral hydrogen (\hi; column density $\gtrsim 10^{18}~{\rm cm^{-2}}$), until they either ``escape'' the galaxy or get absorbed by dust. The fraction of \lya~photons that escape a galaxy, or the \lya~escape fraction (\fesc) is thus a complicated function of the spatial distribution of \hi~gas (i.e., gas covering fraction), gas kinematics, and dust  \citep[][]{Kornei+2010, Hayes+2011, Wofford+2013, Rivera-Thorsen+2015, Trainor+2015, Reddy+2016, Jaskot+2019}. Understanding the physical mechanisms that alter the internal distribution of gas is therefore important to understand the physics of \lya~escape.

Comparing LAEs to continuum-selected star-forming galaxies (SFGs) can lend insight into the key factors regulating the escape of \lya~photons from galaxies. \lya~measurements of SFGs have shown that some have observable \lya~emission, while others do not. \citep[][]{Pentericci+2007, Hathi+2016, DeBarros+2017, Du+2018, Haro+2018, Haro+2020, Weiss+2021}. Several studies have measured \fesc~for LAEs and typical SFGs at different redshifts. LAEs typically have \fesc~$\approx$ 20 $-$ 30 \% \citep{Blanc+2011, Nakajima+2012, Song+2014, Trainor+2015, Matthee+2021}, while the average \fesc~in typical SFGs at z $\approx$ 2$-$3 is less than 10\% \citep{Hayes+2010, Kornei+2010, Matthee+2016, Sobral+2017, Weiss+2021, Reddy+2022}. This difference in the typical \fesc~between LAEs and SFGs indicates a possible correlation between \lya~escape and galaxy properties.

Theoretical and observational studies suggest that star formation in a compact region aids the escape of \lya~(and ionizing) photons \citep[see ][]{Gnedin+2008, Razoumov+2010, Heckman+2011, Borthakur+2014, Izotov+2016, Ma+2016, Reddy+2016, Sharma+2016, Kimm+2019, Marchi+2019, Ma+2020, Naidu+2020, Kakiichi+2021, Reddy+2022}. The radiative, thermal, and mechanical feedback associated with the compactness of star formation can result in strong gas outflows. These outflows, in turn, can lead to ``holes'' or low-column-density channels in the ISM, thereby creating pathways for \lya~photons to escape. This physical mechanism is supported by several studies that suggest a correlation between star-formation-rate surface density ($\Sigma_{\rm SFR}$) and \lya~escape \citep{Heckman+2011, Ma+2016, Sharma+2016, Verhamme+2017, Marchi+2019, Cen2020, Naidu+2020}. \citet{Marchi+2019} provided further evidence of this scenario by studying the role of kinematics and neutral hydrogen column density on \lya~emission using LAEs in the VANDELS survey. They found that the amount of scattering of \lya~photons is smaller for galaxies with higher interstellar gas outflow velocities, proposing that \lya~escape is larger in galaxies with strong feedback. However, they observed no correlation between \lya~escape and star-formation-rate (SFR), suggesting that other factors may be responsible for modulating \lya~escape. More recently, the effect of gravitational potential was investigated by considering the specific star-formation-rate surface density, $\Sigma_{\rm sSFR}$, which is $\Sigma_{\rm SFR}$ normalized by stellar mass \citep[][]{Kim+2020, Reddy+2022}. \citet{Reddy+2022} performed a spectroscopic survey of star-forming galaxies and found that \lya~escape is more efficient in low-mass galaxies coupled with high $\Sigma_{\rm SFR}$. 

Apart from these internal factors, the large-scale environment around the galaxy (within $\sim$10 kpc of the galaxy) may play a role in regulating the gas covering fraction. Interactions with close neighbors are known to lead to starbursts in galaxies \citep{Luo+2014, Knapen+2015, Stierwalt+2015, Moreno+2021}, which may in turn lead to conditions that favor the formation of low-column-density channels in the ISM through which \lya~photons may escape. While some studies find that LAEs in protoclusters have higher observed \lya~luminosities compared to field LAEs \citep{Dey+2016, Shi+2019, Shi+2020, Huang+2021}, other studies find similar or suppressed \lya~emission in protocluster galaxies \citep{Lemaux+2018, Malavasi+2021}. The differences can be partly attributed to selection effects and to variations in the \lya~escape measurements. However, studies that find an enhanced \lya~in dense regions also observe an enhanced star formation in these galaxies.

In this paper, we investigate the role of different galaxy properties, such as stellar masses, SFRs, sizes, $\Sigma_{\rm SFR}$, $\Sigma_{\rm sSFR}$, and local environment on the escape of \lya~using a sample of 62 LAEs at z $\approx$ 2.65 in the Bo\"otes region of the NOAO Wide-Field Survey \citep{Jannuzi+1999}. Twenty-seven of these LAEs have confirmed spectroscopic redshifts, which enables more robust inferences of the stellar populations. The ground-based imaging of the \lya~emission line comes from the Subaru telescope, whose large aperture and sensitive optics provide us with access to some of the faintest LAEs (down to $\approx$ 0.1 $L^{\star}_{\rm Ly\alpha}$) at this redshift. We further use imaging from the {\it Hubble Space Telescope (HST)} in seven subfields, which contributes multi-wavelength data for the analysis and a higher spatial resolution to study the sizes of galaxies. Three of the {\it HST} fields are in comparatively dense regions, providing an opportunity to study LAEs in different environments. To understand how \lya~escape depends on different properties, we compare the LAEs to a sample of 136 SFGs at 2.6 $\le$ z $\le$ 3.8, that have deep rest-frame far-UV spectra from the MOSFIRE Deep Evolution Field (MOSDEF) survey \citep{MOSDEF_Survey, Topping+2020b, Reddy+2022}. 

The outline of this paper is as follows.  Section~\ref{sec:data} describes the data used in this paper. The selection of LAEs and their photometry is outlined in Section~\ref{sec:analysis}. The derivation of different galaxy properties as well as the proxies used to estimate \lya~escape are detailed in Section~\ref{sec:lae_properties}. We describe the dependence of \lya~escape on different properties in Section~\ref{sec:props_dependence} and present the conclusions in Section~\ref{sec:conclusions}. Throughout the paper, we assume a flat universe cosmology with $\Omega$ = 0.287 and $H_{0}$ = 69.3 ${\rm km~s^{-1}~Mpc^{-1}}$ \citep{WMAP9}. All wavelengths are presented in air, and all magnitudes are given in the AB system \citep{AB_Mag_System}.

\section{Observations and Data Reduction} \label{sec:data}

\begin{figure}[t!]
	\centering
 	\includegraphics[width = \columnwidth]{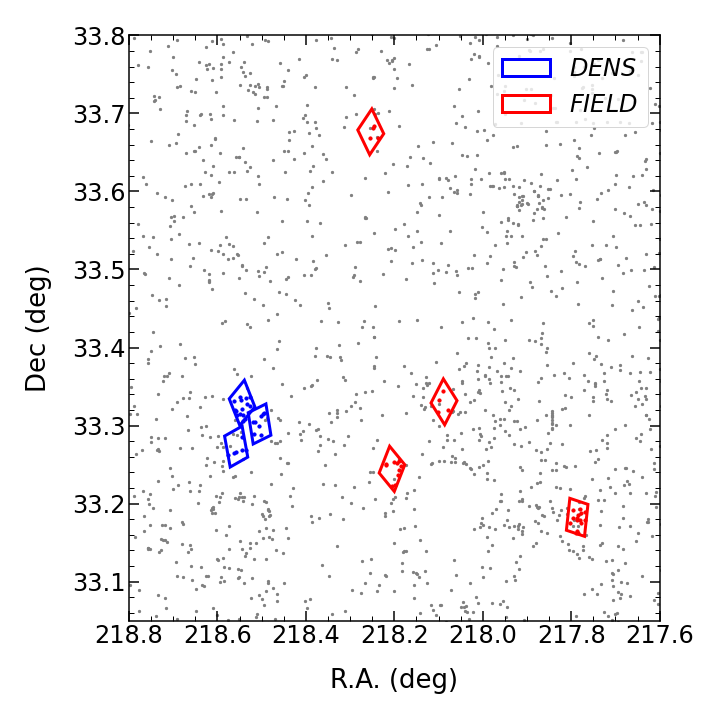}
    \caption{Overlap of the seven {\it HST} fields on the distribution of LAEs (shown in gray) from \citet{Prescott+2008}. The $DENS$ and $FIELD$ fields-of-view are shown in blue and red, respectively. The LAEs studied in this paper (see Section~\ref{subsec:laes_selection}) are marked as blue and red points, depending on their {\it HST} fields.}
    \label{fig:Fields}
\end{figure}

\citet{Prescott+2008} searched for LAEs in a $\approx$ 1 $\deg^2$ region (Figure~\ref{fig:Fields}) around a z $\approx$ 2.656 \lya~blob \citep[LAB; discovered by][]{Dey+2005} in the Bo{\"o}tes field of the NOAO Deep Wide-Field Survey \citep[NDWFS;][]{Jannuzi+1999}. They used the intermediate-band \ia~filter ($\lambda_{c}$ = 4458 \AA, $\Delta\lambda_{\ia}$ = 201 \AA) on the SuprimeCam/Subaru telescope \citep{SuprimeCam}, which probes the \lya~emission line at redshifts of 2.55 $\le$ z $\le$ 2.75 (see Figure~\ref{fig:Filters}). Combining these deep observations with the optical broad-band imaging from NDWFS, they uncovered $\approx$ 2200 candidate LAEs. Follow-up spectroscopy was performed on a subset of these candidates using MMT/Hectospec, which is a 300-fiber multi-object spectrograph with a 1 $\deg$ field-of-view \citep[]{Fabricant+2005}. \citet{Hong+2014} developed an automated algorithm to detect emission lines and to measure redshifts of candidates. This led to 876 confirmed redshifts, out of which 711 are in the aforementioned redshift range. This is the single largest spectroscopic sample of LAEs in such a narrow redshift range to date. Details of the photometric and spectroscopic data are described by \citet{Prescott+2008} and \citet{Hong+2014}, respectively. 

\citet{Prescott+2008} found that the number density of LAEs within $\approx$ 10\arcmin~of the LAB is almost three times higher than that of the LAEs in the field. Based on this local overdensity of LAE candidates, three high-density regions ({\it ``DENS''}) within 3\arcmin~of the LAB and four low-density regions ({\it ``FIELD''}) that are further away ($>$ 15\arcmin) were selected for studying LAEs in different environments. We obtained near-infrared images of these seven fields using the Wide-Field Camera 3 (WFC3) \citep{WFC3} on the {\it Hubble Space Telescope} ({\it HST}). The overlap of these fields on the entire $\approx$ 1 $\deg^{2}$ field-of-view is shown in Figure~\ref{fig:Fields}. In this paper, we focus on the study of LAEs from these {\it HST} fields. The following subsection describes the {\it HST} observations and data reduction. In addition to the \ia~images, we include existing ground-based images in the following filters: \bw ($\lambda_{c}$ = 4135 \AA, $\Delta\lambda_{\bw}$ = 1278 \AA); $R$ ($\lambda_{c}$ = 6514 \AA, $\Delta\lambda_{R}$ = 1512 \AA), and $I$ ($\lambda_{c}$ = 8205 \AA, $\Delta\lambda_{I}$ = 1915 \AA). The 5$\sigma$ magnitude limits in 3\arcsec~diameter apertures in \bw, \ia, $R$, and $I$ bands are 26.4, 26.4, 25.6, and 25.1 mags, respectively.

\subsection{HST Imaging} \label{subsec:hstbased}
\begin{figure}[t!]
	\centering
 	\includegraphics[width = \columnwidth]{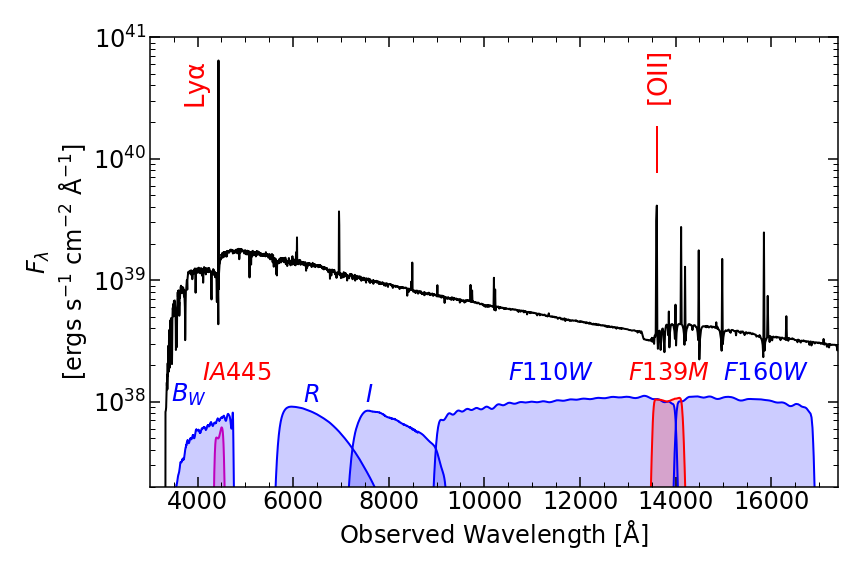}
    \caption{Spectral Energy Distribution (SED) of a model galaxy with age = 100 Myr, stellar metallicity, ${\rm Z}_{\star}~=~{\rm 0.2~Z}_{\odot}$, and a constant star formation history. The model is generated using {\it BAGPIPES} \citep{Bagpipes} and is redshifted to z = 2.65. Wavelengths corresponding to \lya~and \oii~emission lines at this redshift are marked. The broad band filters \bw, $R$, $I$, $F110W$, and $F160W$ that are used in this analysis are shown from left to right in blue below. Similarly, the intermediate band filters, \ia~and $F139M$, which probe the \lya~and \oii~lines, respectively, are shown in red.}
    \label{fig:Filters}
\end{figure}

We obtained {\it HST} imaging of seven (three {\it DENS} and four {\it FIELD}) fields in the Bo{\"o}tes region during {\it HST} Cycle 20 (April$-$November 2013), as part of {\it HST}-GO-13000 (PI S. Hong). Each field was observed in two broad-band filters, $F110W$ ($\lambda_{c}$ = 11515 \AA, $\Delta\lambda_{F110W}$ = 4996 \AA) and $F160W$ ($\lambda_{c}$ = 15434 \AA, $\Delta\lambda_{F160W}$ = 2875 \AA), and a medium-band filter, $F139M$ ($\lambda_{c}$ = 13840 \AA, $\Delta\lambda_{F160W}$ = 652 \AA). The $F139M$ filter covers the redshifted \oii$\lambda\lambda$3726,3729 at 2.5 $\lesssim$ z $\lesssim$ 2.7 (Figure~\ref{fig:Filters}). The field-of-view of each of these fields is 123\arcsec $\times$ 137\arcsec. Observations for each such pointing consisted of 4 orbits ($F110W$[1], $F160W$[1], $F139M$[2]), and each orbit was observed with seven dithers. As a result, we obtained seven individual dithered images in $F110W$ and $F160W$, and fourteen dithered images in $F139M$ for every field.  

We use AstroDrizzle \citep{AstroDrizzle} to align and drizzle all the individual images per filter per field onto a single frame. The output weight maps are effective exposure maps and reflect the relative weight of the individual pixels. These weight maps are used to create rms maps for the space-based images, which are needed to compute photometric uncertainties. We therefore astrometrically align both science and weight images to the \ia~images. For this purpose, we first use Source Extractor \citep{SExtractor} to identify stars on both sets of data (using DETECT\_THRESH = 1.2 and CLASS\_STAR $\ge$ 0.5). These stars are then used to align all the {\it HST} science and weight images to the ground-based images, using the IRAF packages, \emph{ccmap} and \emph{ccsetwcs} \citep{Tody1993}. We also generate flag maps for the drizzled images from the weight maps, as a measure of data quality of pixels. In order to perform multi-band photometry, we register all the ground-based images, including the weight, rms and flag maps to the individual $F110W$ science images in each field, using the IRAF task, \emph{wregister}. The pixels where no {\it HST} data are available are set to zero for all the ground-based images.  In summary, we have seven fields observed in seven different bands ($B_{W}$, $IA445$, $R$, $I$, $F110W$, $F139M$ and $F160W$). Each band has an associated science, exposure, rms, and a flag image. The 5$\sigma$ magnitude limits in 1\arcsec~diameter apertures in $F110W$, $F139M$, and $F160W$ bands are 26.9, 25.6, and 26.4 mags, respectively.

In Figure~\ref{fig:Filters}, we show the spectral energy distribution (SED) of a young model galaxy redshifted to z = 2.65, along with the response curves of all the filters used here. The seven filters together cover the rest-frame ultraviolet and optical region. Specifically, the broad-band filters $B_{W}$, $R$, and $I$ probe the rest-frame UV-continuum of the galaxy; the $F110W$ and $F160W$ filters probe the Balmer break region of the SED; and the medium band filters, $IA445$ and $F139M$, are positioned over the \lya~and \oii~emission lines, respectively. 

\subsection{Comparison Sample of High-Redshift Star-Forming Galaxies} \label{subsec:mosdef}
To evaluate the properties of LAEs relative to typical galaxies at comparable redshifts, we select a sample of high-redshift galaxies from the MOSFIRE Deep Evolution Field (MOSDEF) survey \citep{MOSDEF_Survey}. The survey consists of rest-frame optical spectroscopy of $\approx$ 1500 $H$-band selected galaxies using the MOSFIRE spectrograph \citep{MOSFIRE} on the Keck I telescope. Star-forming galaxies were selected in three redshift ranges: z = 1.37 $-$ 1.70, 2.09 $-$ 2.61, and 2.95 $-$ 3.80, in the CANDELS fields \citep{Grogin+2011, Koekemoer+2011}. The $H$-band limiting magnitudes of 24.0, 24.5, and 25.0 mags, respectively, for these redshift intervals ensure a roughly consistent stellar mass limit of $\sim~10^{9}$~\msun~across all redshifts. Our comparison sample is a subset of these sources and consists of 136 typical star-forming galaxies (SFGs) that have complementary rest-frame far-UV spectra from the Low-Resolution Imaging Spectrometer \citep[LRIS;][]{LRIS} on the Keck telescope. Details regarding this sample can be found in \citet{Topping+2020b} and \citet{Reddy+2022}.

\section{LAE Selection and Photometry} \label{sec:analysis}

\subsection{Source Detection and Photometry} \label{subsec:detection}
We use the source detection algorithm, Source Extractor, in dual-image mode for detecting sources and measuring their photometry \citep{SExtractor}. For this purpose, we combine the \ia~and \bw~images in each field to construct a detection image. Individual images in every band are then taken as measurement images to calculate source fluxes. The parameters include DETECT\_THRESH = 5.0, ANALYSIS\_THRESH = 5.0, and DETECT\_MINAREA = 10 pixels. We perform aperture photometry (FLUX\_APER) in twelve different apertures with diameters ranging from 0.25\arcsec~to 5\arcsec. However, the photometric uncertainties produced by Source Extractor are lower than the rms scatter in the images, suggesting that they are underestimated (FLUXERR\_APER). To overcome this problem, we use the detected source positions to perform photometry using the python {\it photutils} package \citep{Astropy_Photutils}. We compute the corresponding photometric uncertainties using the {\it calc\_total\_error} function that combines the background error with the Poisson noise of the sources, which is then compared to the rms scatter within each image. For every aperture, we consider the rms scatter as a lower limit for the flux error in that band. Based on the curve of growth and visual inspection of sources and the apertures, we use 3\arcsec~diameter apertures for ground-based images. For each source in {\it HST} images, we use either a 0.75\arcsec~or~ 1\arcsec~diameter aperture, selecting the one that yields a higher signal-to-noise ratio.

\subsection{Selection of LAE Candidates} \label{subsec:laes_selection}

\begin{figure}[t!]
	\centering
 	\includegraphics[width = 1.0\columnwidth]{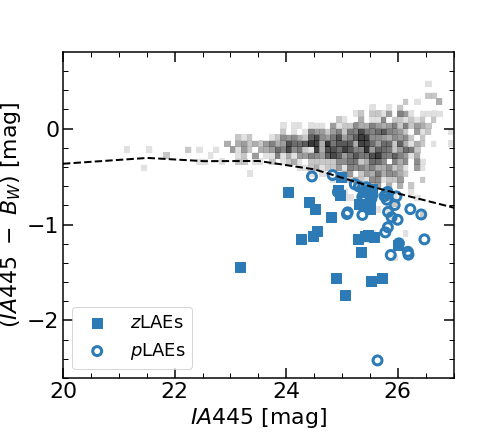}
    \caption{Color Magnitude Diagram ((\ia~$-$~\bw) vs \ia) of all detected sources from the seven {\it HST} fields is shown by the two-dimensional grayscale histogram. The black dashed line denotes the criterion used for the selection of LAE candidates. LAEs with confirmed spectroscopic redshifts between 2.55 $\le$ z $\le$ 2.75 are shown as filled blue squares ($z$LAEs), while the photometric candidates are shown as open blue circles ($p$LAEs).}
    \label{fig:CMD}
\end{figure}

The \lya~emission line falls under the intermediate-band filter, \ia~at the redshifts of interest (Figure~\ref{fig:Filters}). As a result, candidate LAEs will have excess emission in this band, compared to the broad-band, \bw. To identify LAE candidates, we compare the \ia~magnitude to the \bw~magnitude. Even though this selection process of LAEs has been previously performed across the entire 1 $\deg^{2}$ region \citep{Prescott+2008}, we repeat the process on the seven {\it HST} fields.

We begin with all the sources that have \ia~and \bw~magnitudes between 20 mag $\le$ (\bw~and \ia)~$\le$ 30 mag, and also have the Source Extractor output, FLAGS $\le$ 3. This ensures that our sample is clean from objects that are near the image boundaries and/or have saturated pixels. We calculate the rms scatter in (\ia~$-$~\bw) for different bins of \ia~and using these values, we obtain the rms scatter, $\sigma_{(IA445~-~B_{W})}$, as a function of \ia~magnitude. If there is no \lya~emission, the difference between \ia~and \bw~depends on the UV slope of the galaxy SED, and we consider the median value of the difference of all the sources, $\overline{m}_{(IA445~-~B_{W})}$. We select LAE candidates based on the following conditions:

\begin{center}
    \begin{equation*}
        22 \le IA445 \le 28 
    \end{equation*}
    \begin{equation*}
        S/N(\ia) \ge 5
    \end{equation*}
    \begin{equation*}
        S/N(\ia~\ensuremath{-}~\bw) \ge 5
    \end{equation*}
    \begin{equation*}
        (\ia~\ensuremath{-}~\bw) \le \overline{m}_{(IA445~-~B_{W})} \ensuremath{-} 1.5\sigma_{(IA445~-~B_{W})},
    \end{equation*}
\end{center}

\noindent similar to the criterion used in \citet{Hao+2018}. In Figure~\ref{fig:CMD}, we show the color-magnitude diagram between (\ia~$-$~\bw) and \ia, that was used for the selection of LAE candidates. 

\begin{figure}[b!]
    \centering
    \includegraphics[width = 1.0\columnwidth]{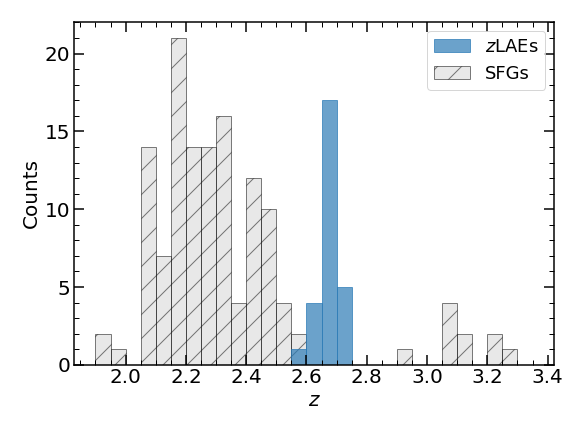}
    \caption{Redshift distribution of $z$LAEs and the comparison SFGs studied in this paper. The LAEs with confirmed spectroscopic redshifts ($z$LAEs) are shown in filled blue, while the SFGs are shown in hatched gray. The median redshift of the $z$LAEs is $\approx$ 2.66, whereas we assume a z = 2.65 for the photometric LAE candidates ($p$LAEs).}
    \label{fig:z_dist}
\end{figure}

The selected LAEs may be contaminated by low-redshift \oii~emitters, where the \oii~emission line falls in the $IA445$ filter ($0.16~\le~z~\le0.24$). Most of these are removed from the sample by applying a cut of (\bw~$-~R$) $\le$ 0.7 \citep[see Figure~2 in][]{Prescott+2008}. After removing these contaminants from our selection, we obtain a total of 74 possible candidates. We remove 10 sources based on visual inspection in both ground-based and space-based images. Two of them are bright and spatially extended in the rest-frame optical regime, and are likely low-redshift contaminants. There is a possibility that they might be true and unusually luminous high-redshift candidates, but we make a conservative decision to remove them from the sample. The remaining eight sources have artifacts in the images, like contamination from nearby sources, or they are located near WFC3/IR artefacts (``blobs'') in the {\it HST} images \citep{WFC3_IR_blobs2010, WFC3_IR_blobs2018}. After removing these objects, we are left with 64 LAE candidates overlapping the {\it HST} fields.

Twenty-nine of these candidates were observed spectroscopically \citep{Hong+2014} and are available in our LAE redshift catalog (Section~\ref{sec:data}). Out of these, 26 sources have confirmed redshifts between 2.55 $\le$ z $\le$ 2.75. From visual inspection of the spectra of the other three sources, we find faint \lya~emission in one of them (DENS3\_118) that is missed by the automated redshift finding algorithm. We manually find the redshift of this source by fitting a Gaussian to the emission line. The remaining two spectra do not show any clear emission line in the wavelength region of interest, and their \ia~photometry may be affected by skyline residuals. We remove these two sources from our catalog.

The final sample consists of 27 sources with confirmed spectroscopic redshifts, and we refer to these galaxies as $z$LAEs throughout this paper. The median redshift of these $z$LAEs is $\approx$ 2.66. We assume a z = 2.65 for the rest of the 35 photometric LAE candidates ($p$LAEs) as this is the median redshift at which the \ia~band covers the \lya~emission line. From Figure~\ref{fig:CMD}, we find that the $p$LAEs have smaller (\ia~-~\bw) excesses and appear to be fainter in the \ia~than the $z$LAEs. In Figure~\ref{fig:z_dist}, we show the distribution of redshifts of $z$LAEs and SFGs (Section~\ref{subsec:mosdef}). 

\subsection{Photometric Measurements for SED Fitting} \label{subsec:final_phot}
From a visual inspection of the LAEs, we find spatial offsets between \lya~emission and the stellar continuum, as has been observed in other LAE samples \citep{Jiang+2013, Hoag+2019, Lemaux+2021}. Since the near-IR emission observed from {\it HST} is dominated by stellar continuum light, whereas the intermediate-band emission is dominated by \lya~emitted along the line of sight, the spatial distributions of the two are strongly affected by the relative distributions of gas, dust, and ionizing sources. In order to measure accurate photometry, we refit the LAE centers using the {\it HST} images. We estimate the source centroids by fitting a 2D Gaussian profile to the light distribution of each source in a 5\arcsec~$\times$ 5\arcsec~cutout. This new center does not affect the photometry of ground-based images because of the larger 3\arcsec aperture used, compared with the 0.75\arcsec~(or 1\arcsec) diameter used for the {\it HST} photometry. 

We repeat the photometry of all the LAEs (27 $z$LAEs and 35 $p$LAEs) using these new centers. Due to the different sensitivity limits of the space and ground-based data, the vastly different PSF sizes, and the varying morphology as a function of wavelength, we choose to measure aperture photometry and select aperture sizes that optimize the S/N. The fluxes and corresponding uncertainties are computed using the python {\it photutils} package as mentioned in Section~\ref{subsec:detection}.
The coordinates, redshifts, and photometry of the final sample of 62 LAEs are presented in the Appendix.

\section{Properties of LAEs} \label{sec:lae_properties}

\subsection{Emission Line Measurements} \label{subsec:em_lines}
In this subsection, we describe the method used to measure equivalent widths and fluxes of both \lya~and \oii~emission lines for all the LAEs. We use photometry from \bw, \ia, $R$, $I$, and $F110W$ for \lya~emission line measurements. We convert magnitudes from these bands to flux densities ($f_{\lambda}$). We then fit a power law to these flux densities to calculate the continuum flux density at the central wavelength of the \ia~filter. We exclude \ia~flux density in this fit, given that it is contaminated by the \lya~emission. We repeat this process 10,000 times by varying each of the photometric measurements within its 1$\sigma$ uncertainties. Since $B_{W}$ does not probe the continuum alone, we use an error bar of $\pm$2$\sigma$ for this band. The outputs of the repeated fits closely follow a normal distribution. Thus, the final continuum flux density probed by the \ia~filter ($f_{\lambda c; IA445}$) and its error ($\delta f_{\lambda c; IA445}$) are taken as the mean and standard deviation of the flux densities calculated from the repeated fits. 

We calculate the \lya~flux and equivalent width using these continuum flux density measurements. Given the flux density ($f_{\lambda; IA445}$) and its error ($\delta f_{\lambda; IA445}$) measured from the \ia~magnitude, the \lya~flux ($F_{\rm Ly\alpha}$) and its error ($\delta F_{\rm Ly\alpha}$) are: 

\begin{center}
    \begin{equation}
        F_{\rm Ly\alpha} = (f_{\lambda; IA445} -  f_{\lambda c; IA445}) \times \Delta \lambda_{IA445} 
    \end{equation} 
    \begin{equation}
        \delta F_{\rm Ly\alpha} = \sqrt{(\delta f_{\lambda; IA445}^{2}) + (\delta f_{\lambda c; IA445}^{2})} \times \Delta \lambda_{IA445}
    \end{equation}
\end{center}

\noindent where $\Delta \lambda_{IA445}$ is the FWHM of the \ia~filter.  Furthermore, we calculate the rest-frame Ly$\alpha$ equivalent width ($W_{\rm Ly\alpha}$) and its error ($\delta W_{\rm Ly\alpha}$) as:

\begin{center}
    \begin{equation}
        W_{\rm Ly\alpha} = \frac{(f_{\lambda; IA445} -  f_{\lambda c; IA445})}{f_{\lambda c; IA445}} \times \frac{\Delta \lambda_{IA445}}{(1+z)} 
    \end{equation}
    
    \begin{equation}
        \begin{split}
            \delta W_{\rm Ly\alpha} =  {\frac{f_{\lambda; IA445}}{f_{\lambda c; IA445}}} \times \sqrt{(\frac{\delta f_{\lambda; IA445}}{f_{\lambda;IA445}})^2 + (\frac{\delta f_{\lambda c; IA445}}{f_{\lambda c;IA445}})^2} \\
            \times \frac{\Delta \lambda_{IA445}}{(1+z)}
        \end{split}     
    \end{equation}
\end{center}    

\noindent We are not accounting for the correction due to the Lyman forest here, due to which we slightly underestimate the \lya~measurements. This estimated flux is used to correct \bw~for emission, as described in Section~\ref{subsubsec:bw_corr}. 

Similarly, we calculate the \oii~flux ($F_{\rm [O~{\textsc{ii}}]}$) and equivalent width ($W_{\rm [O~{\textsc{ii}}]}$) by considering the photometry from $F110W$, $F139M$, and $F160W$ bands. By varying the flux densities from {\it $F110W$} and {\it $F160W$} bands within their uncertainties, we fit a power law to calculate the continuum flux density at the central wavelength of the $F139M$ filter. The \oii~flux and equivalent width, along with their errors, are:

\begin{center}
    \begin{equation}
        F_{\rm [O~{\textsc{ii}}]} = (f_{\lambda; F139M} -  f_{\lambda c; F139M}) \times \Delta \lambda_{F139M} 
    \end{equation} 
    
    \begin{equation}
        \delta F_{\rm [O~{\textsc{ii}}]} = \sqrt{(\delta f_{\lambda; F139M}^{2}) + (\delta f_{\lambda c; F139M}^{2})} \times \Delta \lambda_{F139M}
    \end{equation}
    
    \begin{equation}
        W_{\rm [O~{\textsc{ii}}]} = \frac{(f_{\lambda; F139M} -  f_{\lambda c; F139M})}{f_{\lambda c; F139M}} \times \frac{\Delta \lambda_{F139M}}{(1+z)} 
    \end{equation} 
    
    \begin{equation}
        \begin{split}
            \delta W_{\rm [O~{\textsc{ii}}]} = \frac{f_{\lambda; F139M}}{f_{\lambda c; F139M}} \times \sqrt{(\frac{\delta f_{\lambda; F139M}}{f_{\lambda; F139M}})^2 + (\frac{\delta f_{\lambda c; F139M}}{f_{\lambda c; F139M}})^2} \\ 
            \times \frac{\Delta \lambda_{F139M}}{(1+z)}
        \end{split}  
    \end{equation}
\end{center}

\noindent where $f_{\lambda; F139M}$ and $\delta f_{\lambda; F139M}$ are the flux density and its error measured from the $F139M$ filter; $f_{\lambda c; F139M}$ and $\delta f_{\lambda c; F139M}$ are the continuum flux density and its error computed from the rest-frame optical continuum fitting; and $\Delta \lambda_{F139M}$ is the FWHM of the $F139M$ filter. For the LAEs where there is no clear \oii~emission from $F139M$; i.e., when the flux densities measured by $F139M$ are less than the continuum expected at this filter, we place 3$\sigma$ upper limits on both the flux and equivalent width measurements.  

The \lya~equivalent width and flux measurements for the SFGs are computed using the rest-frame far-UV LRIS spectra \citep[][]{Reddy+2022}, while their \oii~flux measurements are obtained from the rest-frame optical spectra from MOSFIRE \citep[][]{MOSDEF_Survey, Reddy+2018b} of the galaxies. The LRIS and MOSFIRE spectra are corrected for slit loss, so the fluxes and equivalent widths derived should be similar to those obtained from photometry and can be used for the purposes of our comparison. 

\subsubsection{Correcting $B_{W}$ for Ly$\alpha$ Flux} \label{subsubsec:bw_corr}

In order to fit the stellar continuum using the measured photometry, we first need to correct the broad-band filter for contamination from nebular emission lines. For the z $\approx$ 2.65 LAEs, the \bw~filter band pass samples the \lya~emission line, the UV continuum emission, and the Lyman forest absorption. The \bw~flux is corrected for \lya~emission as described below.

Using the flux density and its error from the \bw~filter ($f_{\lambda; \bw}$ and $\delta f_{\lambda; \bw}$), we calculate the flux from the filter, $F_{\bw}$, and its error, $\delta F_{\bw}$:

\begin{center}
    \begin{equation}
        F_{\bw} = f_{\lambda; \bw} \times \Delta \lambda_{\bw}
    \end{equation}
    \begin{equation}
        \delta F_{\bw} = \delta f_{\lambda; \bw} \times \Delta \lambda_{\bw}
    \end{equation}
\end{center}

\noindent where $\Delta \lambda_{\bw}$ is the FWHM of the \bw~filter. The corrected flux density and its error in the \bw~filter ($f_{\lambda corr; \bw}$ and $\delta f_{\lambda corr; \bw}$) are then:

\begin{center}
    \begin{equation}
        f_{\lambda corr; \bw} = \frac{(F_{\bw} - F_{\rm Ly\alpha})}{\Delta \lambda_{\bw}}
    \end{equation}
    
    \begin{equation}
        \delta f_{\lambda corr; \bw} = \frac{\sqrt{(\delta F_{\bw}^2) + (\delta F_{\rm Ly\alpha}^2)}}{\Delta \lambda_{\bw}}
    \end{equation}
\end{center}

This corrected flux density and its error are converted back to the magnitude system, which are then used as proper continuum measurements of the LAEs at the central wavelength of the \bw~filter. While these estimates for the corrections to the \bw~photometry are simplistic, more sophisticated approaches will not significantly change the results given the photometric uncertainties of these faint LAEs.

\begin{figure}
    \begin{minipage}[c][0.65\width]{1.0\columnwidth}
        \centering
        \includegraphics[width=1.0\columnwidth]{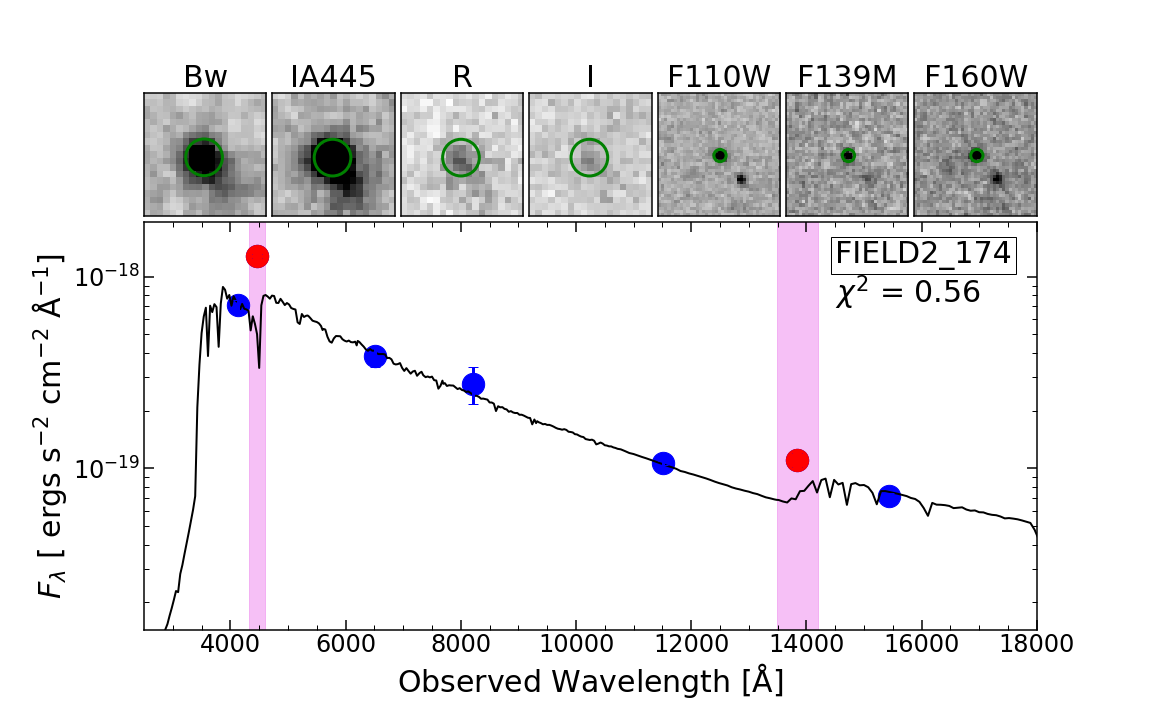}
        %\caption{Choose any caption if you want to add }}
    \end{minipage}
    \hfill
    \begin{minipage}[c][0.65\width]{1.0\columnwidth}
        \centering
        \includegraphics[width=1.0\columnwidth]{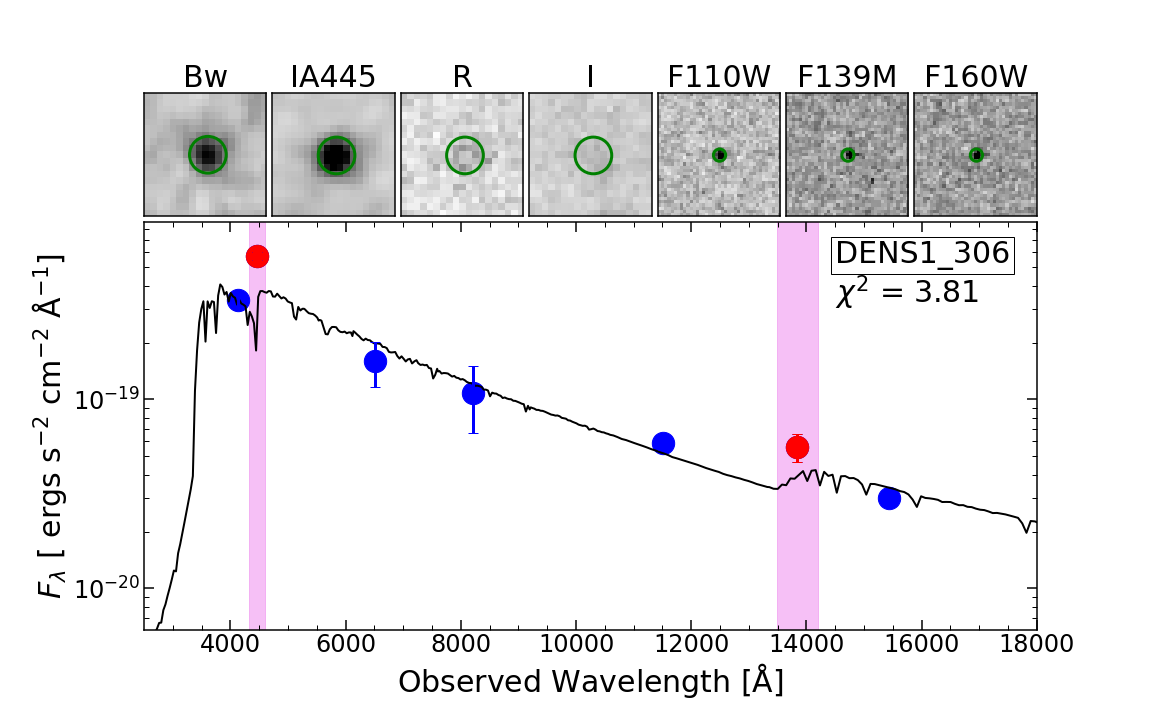}
        %\caption{Choose any caption if you want to add }}
    \end{minipage}
    \hfill
    \begin{minipage}[c][0.65\width]{1.0\columnwidth}
        \centering
        \includegraphics[width=1.0\columnwidth]{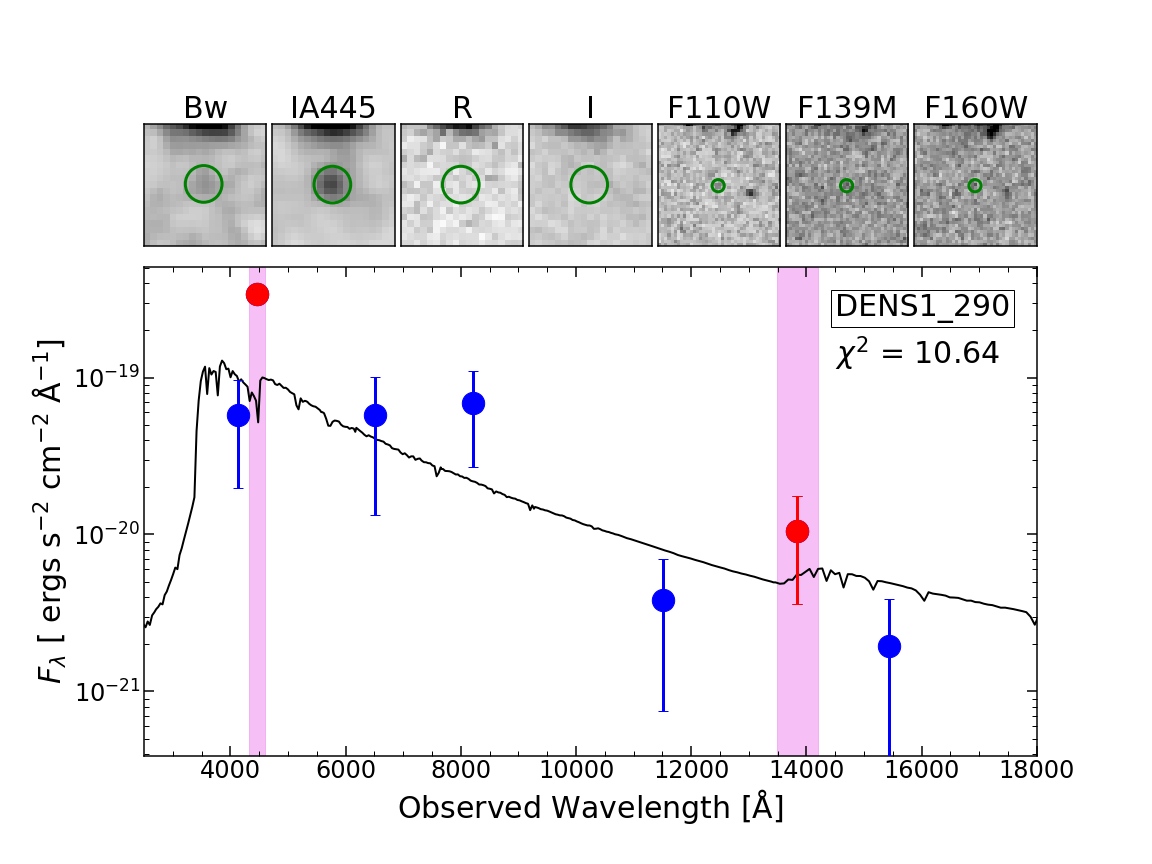}
        %\caption{Choose any caption if you want to add }}
    \end{minipage}
    \hfill
    % \begin{subfigure}
    %     \centering
    %     \includegraphics[width=1.0\columnwidth]{Good_Fit.png}
    %     %\caption{Choose any caption if you want to add }}
    % \end{subfigure}
    % \begin{subfigure}
    %     \centering
    %     \includegraphics[width=1.0\columnwidth]{Medium_Fit.png}
    %     %\caption{Add second caption if you like}
    % \end{subfigure}
    % \begin{subfigure}
    %     \centering
    %     \includegraphics[width=1.0\columnwidth]{Bad_Fit.png}
    %     %\caption{Add third caption if you like}
    % \end{subfigure}
    \caption{Example SEDs with their best-fit model. Image cutouts (5\arcsec~$\times$~5\arcsec) of the LAEs in different bands (\bw, \ia, $R$, $I$, $F110W$, $F139M$, and $F160W$ from left to right) are shown in the top panel for each LAE. The green circles on the images are the apertures used for performing photometry. The top, middle, and bottom panels show examples of good, acceptable, and uncertain SED fits, respectively. The blue circles in each plot are flux densities in \bw(corrected for \lya~flux), $R$, $I$, $F110W$, and $F160W$ bands (from left to right in each panel), that are used for the SED fitting. The minimum $\chi^2$ best-fit SED is overplotted, with the $\chi^2$ value shown on the plot. The red points show the \ia~and $F139M$ flux densities, which probe the \lya~and \oii~emission lines, respectively. The pink shaded regions mark the wavelength range of their filter passbands.}
    \label{fig:sed_examples}
\end{figure}

For one particular LAE, `FIELD2\_149', the \lya~flux is high compared to the \bw~flux due to a non-detection in the continuum (S/N in \bw~$\sim$ 0.6). As a result, correcting the \bw~results in a negative value. Given this nonphysical flux value, we use the $B_{W}$ magnitude as it is, but with 2$\sigma$ error bars when fitting the SED. This ensures a larger range for the $B_{W}$ fluxes to vary, while decreasing its weight relative to other bands.

\subsection{SED Fitting} \label{subsec:sed_fitting}

\begin{figure*}[t!]
	\centering
 	\includegraphics[width = 1.0\textwidth]{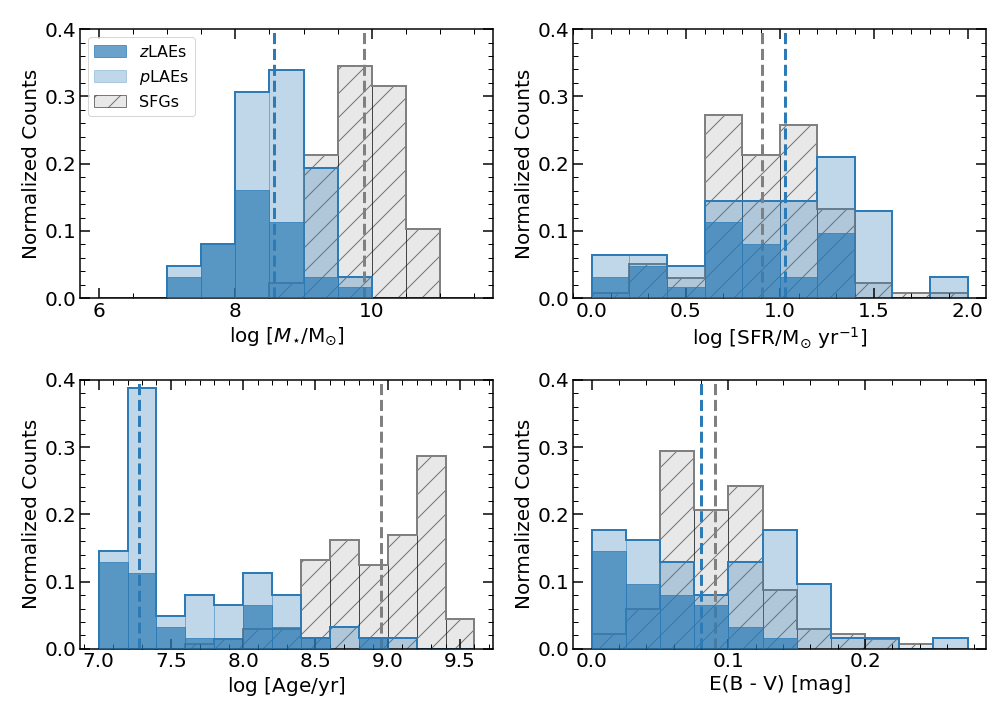}
    \caption{Distribution of stellar mass ({\it top left}), star-formation-rate ({\it top right}), age ({\it bottom left}), and dust reddening ({\it bottom right}) of the LAEs computed from the SED fitting. In each of the plots, the dark shaded blue are the LAEs with confirmed spectroscopic redshift ($z$LAEs), while the light shaded blue are the photometric candidates ($p$LAEs). The distributions of typical SFGs are overplotted in hatched gray. The median values of each parameter for the LAEs and SFGs are marked by dotted blue and gray vertical lines, respectively. The histograms in all the panels are normalized by the total number of sources.}
    \label{fig:SED_Props}
\end{figure*}

\begin{table*}[t!]
    \centering
    \caption{Median values of parameters for different sets of galaxies. The numbers in square brackets denote the 25th and 75th percentile for each parameter. }
    \begin{tabular}{|c|c|c|c|c|c|}
    \hline
    {\bf Galaxy Sample}  & {\bf Number of Sources}   &   {\bf $\log~M_{\star}~[\msun]$} &   {\bf SFR~$[\msun~{\rm yr^{-1}}]$}  &   {\bf Age~[Myr]}   &   {\bf E(B$-$V)~[mag]} \\ \hhline{|=|=|=|=|=|=|}
    {\bf $z$LAEs}        &  27   &   8.3 [8.0, 8.7]                    &   5.4 [3.6, 13.0]                  &    19 [15, 101]      & 0.04 [0.01, 0.075]      \\
    {\bf $p$LAEs}      &   35   &    8.7 [8.5, 9.1]                    &   15.5 [8.1, 25.8]                &    25 [19, 90]     & 0.13 [0.075, 0.155]      \\
    {\bf KS Test p-value}&   $-$   & 0.009                             &   0.006                            &    0.179            &  0.280              \\ \hhline{|=|=|=|=|=|=|}
    {\bf All LAEs}       &    62   &    8.6 [8.3, 8.9]                    &   10.6 [4.7, 20.3]                 &    19 [19, 101]     & 0.08 [0.04, 0.13]      \\
    {\bf MOSDEF}         &  136    &   9.9 [9.6, 10.2]                   &   8.0 [6.0, 14.0]                  &     910 [404, 1609]  &  0.09 [0.06, 0.11]\\ 
    {\bf KS Test p-value}& $-$   & 1.1 $\times~10^{-33}$             &   0.019                            &    2.4 $\times~10^{-34}$ &  0.005            \\ \hline 
    \end{tabular}
    \label{tab:params}
\end{table*}

We fit the spectral energy distribution (SED) of the LAEs using the \citet{BC03} stellar population models. We correct the broad \bw~filter for \lya~contribution (Section~\ref{subsubsec:bw_corr}) and do not use \ia~or $F139M$ photometry for the SED fitting. We assume a Salpeter Initial Mass Function (IMF) \citep{Salpeter_IMF} and a constant star formation history. We consider a stellar metallicity of, ${\rm Z_{\star}} = 0.28~{\rm Z}_{\odot}$, as sub-solar metallicity models provide a better fit to the photospheric line blanketing, that is observed in the rest-frame UV spectra of typical star-forming galaxies at z $\gtrsim$ 2 \citep{Steidel+2016, Cullen+2020, Topping+2020a, Topping+2020b, Kashino+2022, Reddy+2022}. We also assume an SMC extinction curve \citep{SMC_Extinction_Curve}, as motivated in other studies of high-redshift star-forming galaxies \citep[][]{Reddy+2018, Shivaei+2020}. The reddening values are allowed to vary from 0 $\le$ E(B~$-$~V) $\le$ 0.6. We note that changing the star formation history to exponentially increasing or varying the stellar metallicity does not significantly affect our results. We allow the ages to vary from 10 Myr to the age of the Universe at the redshift of each galaxy. When fitting for stellar populations, we use the spectroscopic redshifts for the $z$LAEs, while we set z = 2.65 for the $p$LAEs. The best fit is selected as the one with the minimum $\chi^{2}$ with respect to the photometry. We refit the best-fit model multiple times by perturbing the photometry within the uncertainties, and recomputing the parameter values each time. In most cases where the parameters do not reach the edge of the SED grid, the estimated parameters from perturbed data are normally distributed. The uncertainties in these parameters are then taken as the standard deviation of these different measurements. More details about the fitting procedure are described by \citet{Reddy+2015}.

From the SED fitting, we obtain estimates of stellar masses ($M_{\star}$), ages, star-formation-rates (SFRs), and dust reddening of the individual LAEs. We also consider the properties of SFGs (Section~\ref{subsec:mosdef}) using the same SED fitting procedure \citep{Reddy+2015}. Age is the least constrained parameter and should be considered with caution.

Figure~\ref{fig:sed_examples} shows three examples of SED fits, along with 5\arcsec $\times$ 5\arcsec~image cutouts of the LAEs in different bands. The flux densities shown as red circles probe the \lya~and \oii~emission lines at the LAE redshift and are not used for the SED fitting. The top panel shows an example of an LAE, which has a noticeable emission in all seven bands, including faint emission in $R$ and $I$. In this case, the SED model provides a reasonable fit to the photometry. In the middle panel, we show a case where the LAE is not visible in $R$ and $I$. The resulting fit depends on \bw, $F110W$, and $F160W$ photometry. Most of the sources in our sample have such comparable data. Due to the faintness of the galaxy in the bottom panel, it is only observable in the \ia~band that probes the \lya~emission line. The faint photometry of this LAE in all the other filters results in an uncertain SED fit. Nine LAEs in our sample ($\approx$ 14\%) have such uncertain fits. Photometric uncertainties are taken into account in the fitting procedure and in our physical interpretation of the results.

\subsubsection{Stellar Populations of LAEs} \label{subsubsec:stellar_props}

\begin{figure}[t]
    \centering
    \includegraphics[width = 1.0\columnwidth]{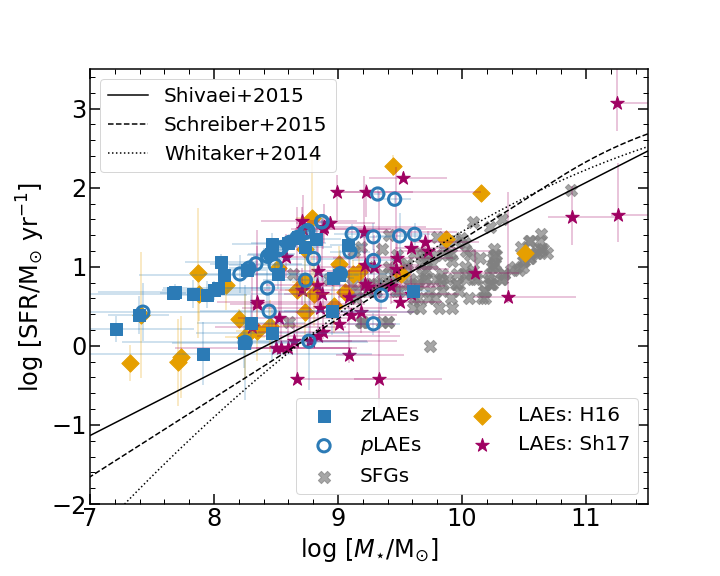}
    \caption{Position of LAEs relative to the Star Forming Main Sequence (SFMS) on the $\log~M_{\star}$ vs $\log~{\rm SFR}$ space. The sample of $z$LAEs and $p$LAEs are plotted as solid blue squares and open blue circles, respectively. The MOSDEF comparison SFGs are shown as gray crosses. Previous results of LAEs from \citet{Hagen+2016} (shown as H16) and \citet{Shimakawa+2017} (shown as Sh17) are displayed as orange diamonds and magenta stars, respectively. The main sequence parametrization computed by \citet{Whitaker+2014}, \citet{Schreiber+2015} and \citet{Shivaei+2015} are extended to lower masses and overplotted as dotted, dashed, and solid lines, respectively.}
    \label{fig:sfms}
\end{figure}

\begin{figure*}
    \centering
    \includegraphics[width = 1.0\linewidth]{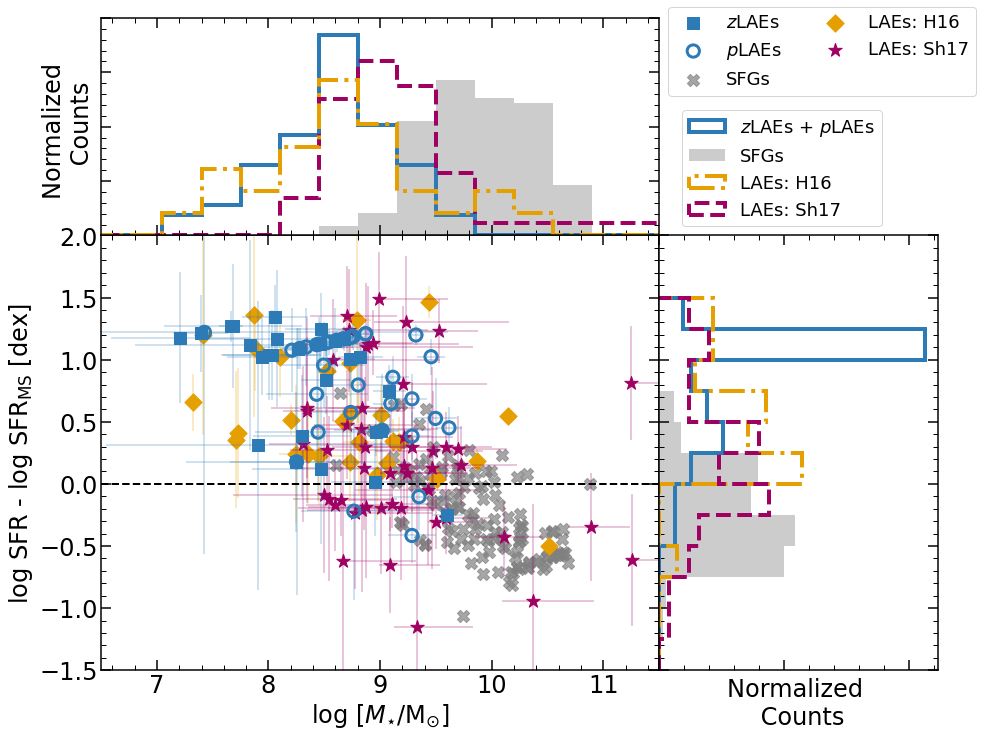}
    \caption{{\it Main:} Separation ($d_{SFMS}$) of LAEs and SFGs from the Star Forming Main Sequence (SFMS) parametrization by \citet{Shivaei+2015} as a function of stellar mass. The colors and markers are the same as Figure~\ref{fig:sfms}. {\it Top:} Histogram of Stellar masses for the different samples of galaxies. {\it Right:} Histogram of $d_{\rm SFMS}$ for the different samples of galaxies.}
    \label{fig:sfms_dist}
\end{figure*}

Figure~\ref{fig:SED_Props} shows the distribution of properties of LAEs and SFGs, derived from SED fitting. The median values for the individual parameters for both the samples are shown as dashed vertical lines. In Table~\ref{tab:params}, we show the quartile values of these parameters for $z$LAEs, $p$LAEs, all LAEs together, and SFGs.

We find that the sample of LAEs (both $z$LAEs and $p$LAEs) contain young populations. The minimum age of the sample is restricted to be 10 Myr via SED fitting. However, it is possible that the galaxies with this borderline age from SED fitting might have ages less than 10 Myr. These are most probably compact galaxies with short dynamical timescales (see Section~\ref{subsec:sizes}). All LAEs are found to be younger than 1 Gyr, with a median of $\approx$ 20~Myr. Given the typical uncertainties ($\gtrsim$ 40~Myr) in this parameter, any interpretations regarding the age should be carefully considered. The LAEs also have low dust reddening values, 0 $\le$ E(B$-$V) $\le$ 0.26 mag, with a median of 0.08 mag. The $p$LAEs are on average older and have more dust content than the $z$LAEs (Table~\ref{tab:params}).

The sample of LAEs have stellar masses ranging from 7.2 $\le \log (M_{\star}/\msun) \le$ 9.6, with a median of $\log(M_{\star}/\msun) \approx$ 8.6. The SFRs measured for the LAEs range from $\approx$ 0.8 $\msun~{\rm yr^{-1}}$ to $\approx$ 100 $\msun~{\rm yr^{-1}}$, with a median of $\approx$10 $\msun~{\rm yr^{-1}}$. These are in agreement with the measurements of stellar masses and SFRs in typical LAEs detected using narrow-band techniques at z $\sim$ 2 $-$ 3 \citep{Nilsson+2011, Vargas+2014, Sandberg+2015, Shimakawa+2017, Hao+2018, Kusakabe+2018}. We also find that the $p$LAEs are on average more massive and have higher SFRs compared to $z$LAEs. 

Table~\ref{tab:params} shows the KS test results between the $z$LAEs and $p$LAEs for the different parameters. Even though both the samples are selected using the same criterion, the p-values for stellar mass and SFR, in particular, suggest that the two distributions may not share the same parent sample. LAEs with fainter (or less luminous) \lya~emission tend to be redder and more massive, similar to what we have observed \citep[e.g., ][]{Pentericci+2009, Finkelstein+2011b, Hathi+2016, Du+2018, Du+2021, Reddy+2022}. Additionally, the $p$LAEs might have a possible low-redshift contamination, and the lack of spectroscopic redshifts introduces higher uncertainties in the SED fitting results.

On comparing the LAEs with the SFGs, we find that the LAEs are younger and have slightly lower dust content than SFGs. In fact, the median age of the SFGs is $\approx$ 1 Gyr, which is closer to the upper limit of the age of the LAEs. Additionally, the LAEs are on average less massive and have similar SFRs compared to SFGs. The KS test for all the parameters clearly suggests that the two samples have different underlying distributions (Table~\ref{tab:params}). We explore the location of LAEs and SFGs on the star-forming main sequence diagram below. 

\subsubsection{Star-Forming Main Sequence} \label{subsubsec:sfms}

The position of typical star-forming galaxies in $M_{\star}$ - SFR space follows a power law, often called the star-forming main sequence \citep[SFMS;][]{Brinchmann+2004, Daddi+2007, Elbaz+2007, Noeske+2007}. Some studies have found LAEs to lie above the SFMS \citep{Hagen+2014, Vargas+2014, Hagen+2016, Hao+2018}, while others found LAEs on the SFMS \citep{Shimakawa+2017, Kusakabe+2018}. The SFMS evolves with redshift, and these comparisons should be made with galaxies at similar redshifts. The location of a galaxy on the SFMS can help us understand the ``mode'' of star formation happening within it. Galaxies that lie on the SFMS are considered to be ``normal'' star-forming galaxies, while galaxies undergoing a current burst of star formation tend to reside above this sequence. Understanding the location of LAEs relative to the main sequence can provide clues to the star formation history of the LAEs.

Figure~\ref{fig:sfms} shows the $\log~[{\rm SFR/M_{\odot}~yr^{-1}}]$ versus $\log~[M_{\star}/{\rm M_{\odot}}]$ plot for the LAEs and SFGs. Two other LAE samples from \citet{Hagen+2016} at z $\sim$ 2 (referred to as H16) and \citet{Shimakawa+2017} at z $\sim$ 2.5 (referred to as Sh17) are also plotted. The stellar masses and SFRs from \citet{Shimakawa+2017} were computed from the SED fitting technique assuming a Chabrier IMF \citep{Chabrier_IMF}. We multiply them by 1.897, in order to scale them to Salpeter IMF, for consistency with the other data shown in Figure~\ref{fig:sfms}. Parametrizations for the SFMS computed by \citet{Whitaker+2014}, \citet{Schreiber+2015}, and \citet{Shivaei+2015} are overplotted and extended to lower masses (${\rm M_{\star}~\lesssim}~10^{8.5}~{\rm M_{\odot}}$) in dotted, dashed, and solid lines respectively. The artificial tight relation between the stellar mass and SFR, as demonstrated by blue points in the figure, is due to the young ages of the LAEs. Both SFR and $M_{\star}$ depend on the normalization of the best-fit SED to the photometry. As a result, these two parameters are tightly correlated, especially for a constant star formation history model. 

We find that most of the LAEs lie above the SFMS relation, indicating an elevated SFR for their stellar mass. This suggests that most LAEs may be undergoing a bursty mode of star formation. In order to explore how much the LAEs deviate from the SFMS, we plot the separation of galaxies from the SFMS ($d_{SFMS}$), parametrized by \citet{Shivaei+2015}, as a function of stellar mass in Figure~\ref{fig:sfms_dist}. The horizontal dashed line denotes the expected values when a galaxy lies on the SFMS for a given $M_{\star}$. We are assuming that the SFMS parametrization extends to lower masses as well. Seventeen out of 62 ($\approx$ 28 \%) LAEs lie within 2$\sigma$ of the SFMS, but the majority of sources ($\gtrsim$ 70\%) are well above the relation. Their star forming modes range from ``bursty'' to ``normal'' star formation. For a given stellar mass, LAEs typically have a more star-bursty nature compared to SFGs. However, the position of galaxies on the SFMS will also depend on their ages. The larger separations from SFMS indicates that for a given stellar mass, LAEs tend to be younger than SFGs. More massive LAEs lie on the main sequence compared to their less massive counterparts \citep[similar to ][]{Santos+2020}. However, it is possible that we are missing galaxies that have low masses as well as low SFRs. Continuum-based searches are magnitude-limited and tend to select higher-mass galaxies, while \lya~emission based searches select galaxies with higher \lya~emission, which is mostly a result of higher SFRs.

\begin{figure*}[t!]
	\centering
 	\includegraphics[width = 1.0\linewidth]{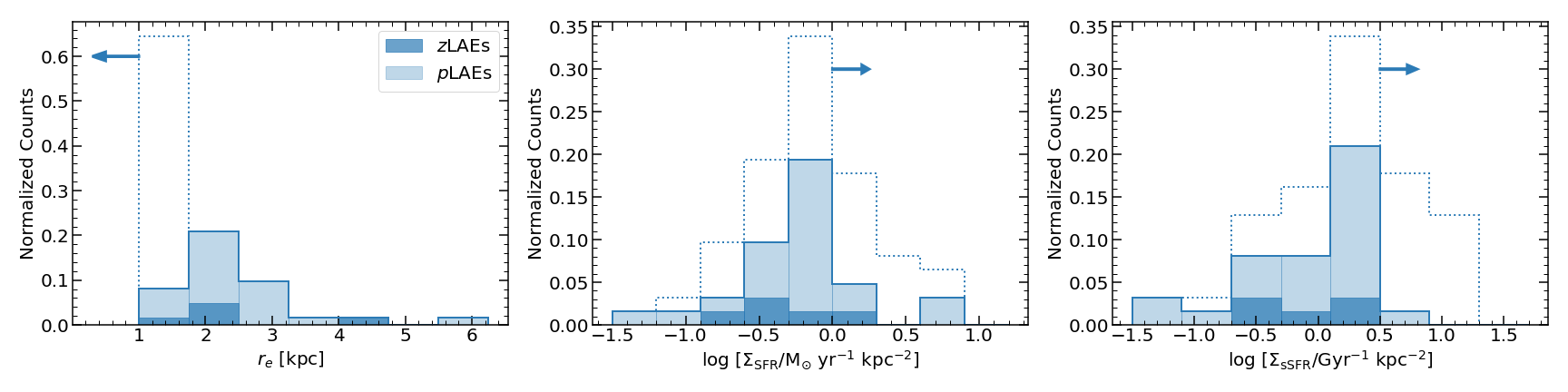}
    \caption{Distribution of sizes ($r_{e}$; {\it left}), star-formation-rate surface densities ($\Sigma_{\rm SFR}$; {\it middle}), and specific star-formation-rate surface densities ($\Sigma_{\rm sSFR}$; {\it right}) of the LAEs. The colors are the same as Figure~\ref{fig:SED_Props}. The upper limits on the sizes of unresolved galaxies and corresponding lower limits on the surface densities are shown as dotted histograms.}
    \label{fig:sizes_hist}
\end{figure*}

The top and right panels of Figure~\ref{fig:sfms_dist} show normalized histograms of stellar masses and $d_{SFMS}$ of the different galaxy populations, respectively. The observed peak in $d_{SFMS}$ is from the LAEs that border the artificial relation in Figure~\ref{fig:sfms}. Combining our results with those of LAEs in the literature, we observe that LAEs, on average, lie above the SFMS compared to typical SFGs. On the other hand, LAEs are, on average $\approx$ 1.5 dex less massive than the SFGs. This indicates that galaxies selected via \lya~emission line almost always probe the low-mass end of the galaxy mass function and have higher specific SFRs (sSFRs), on average, compared to typical SFGs. 

\subsection{Morphology of LAEs} \label{subsec:sizes}

Previous studies have shown that LAEs are compact in the rest-frame UV and optical continuum \citep{Bond+2009, Gronwall+2011, Bond+2012, Malhotra+2012, Paulino-Afonso+2018, Shibuya+2019}. We compute the sizes of our sample of LAEs using the {\it HST} $F110W$ images with the aim of estimating their SFR surface densities as well as for comparing them to typical SFGs. The sizes and SFR surface densities of MOSDEF SFGs are presented in \citet{Reddy+2022}, and we will refer to that study for comparison. 

\subsubsection{Rest-frame Optical Size Measurements}

To measure the sizes from {\it HST} images, we first compute the PSF of these images using GALFIT \citep{GALFIT}. We find stars by running Source Extractor on all the $F110W$ images based on the CLASS\_STAR parameter. We then run GALFIT on these stars with a S\'ersic2D profile \citep[][]{Sersic_Profile} by fixing the S\'ersic index, $n$ = 0.5, which is identical to a 2D-Gaussian profile. The resulting effective radius ($r_e$) is taken as the effective radius of the PSF. GALFIT fails to converge when trying to run with an input PSF, suggesting that the LAEs are barely resolved. To circumvent this, we estimate the LAE sizes by running GALFIT without an input PSF profile to the fitting tool. This results in sizes from the images that are not yet corrected for the PSF. We fix $n$ = 1 to compute the effective radius of galaxies using a S\'ersic2D profile. Given the limiting PSF resolution, there is no obvious difference in the results when we vary the value of the S\'ersic index. We correct the resulting radius for PSF by subtracting $r_e$ of the PSF from it in quadrature. In order to check the validity of our sizes, we compute $r_e$ of all the LAEs using the $F160W$ continuum images using the same method as above. The sizes derived from $F110W$ and $F160W$ images are consistent within the uncertainties.

We consider all the sources that have $r_e$ within 2$\sigma$ of the PSF as unresolved sources, and we set $r_e$ of the PSF as the upper limit to their sizes. The left panel of Figure~\ref{fig:sizes_hist} shows the distribution of these sizes. We find that 37 out of 64 LAE candidates ($\approx~60\%$) are unresolved ($r_{e}~\lesssim$ 1 kpc). If we consider the fraction of unresolved galaxies as a degree of compactness for a particular population of galaxies, then most studies of LAEs find a higher fraction of unresolved galaxies than LBGs or typical SFGs at a given stellar mass. SFGs with stellar masses $\gtrsim~10^{9}~M_{\odot}$ at z $\sim$ 2$-$3, are more spatially extended, with sizes in the range of $r_{e}~\sim~0.7~-~3~{\rm kpc}$ \citep{Law+2012, Shibuya+2015}. Additionally, studies have found that LAEs are smaller than typical SFGs across all redshifts \citep[][]{Paulino-Afonso+2018}.

LAEs have sizes that are independent of redshift \citep{Malhotra+2012, Kim+2021}. This is in contrast with the size evolution observed in SFGs and LBGs \citep{Shibuya+2015}. This means that galaxies selected based on their strong \lya~emission are mostly compact in nature, suggesting that \lya~escape might be related to the sizes of galaxies (Section~\ref{subsec:fesc_re}).

\subsubsection{Star-Formation-Rate Surface Densities}

The distribution of star formation in galaxies may influence the escape of \lya~photons from them (Section~\ref{sec:intro}). To probe this connection, we measure the star-formation-rate surface density, $\Sigma_{\rm SFR}$: \\

\begin{center}
    \begin{equation}
        \Sigma_{\rm SFR} [\msun~{\rm yr^{-1}~kpc^{-2}}] = \frac{{\rm SFR}}{2\pi r_{e}^{2}}
    \end{equation}
\end{center}

In the middle panel of Figure~\ref{fig:sizes_hist}, we show the distribution of $\Sigma_{\rm SFR}$ of our sample of LAEs. We find lower limits for the unresolved sources, which are shown as a dotted histogram in the figure. From resolved sources, we find that the LAEs have $\Sigma_{\rm SFR} \gtrsim 1 - 100~\msun~{\rm yr^{-1}~kpc^{-2}}$. This is higher than the values for typical SFGs at $2 \lesssim z \lesssim 3$, that have $\Sigma_{\rm SFR}~\approx~10^{-2} - 10~\msun~{\rm yr^{-1}~kpc^{-2}}$ \citep{Shibuya+2015, Reddy+2022}. 

Previous studies suggest that the gravitational potential may also play a role in \lya~escape \citep{Kim+2021, Reddy+2022}. We consider stellar mass as a proxy for this potential \citep[e.g., see][for more details]{Reddy+2022} and compute specific star-formation-rate surface density, $\Sigma_{\rm sSFR}$, to quantify this effect.  

\begin{center}
    \begin{equation}
        \Sigma_{\rm sSFR} [{\rm Gyr^{-1}~kpc^{-2}}] = \frac{\Sigma_{\rm SFR}}{M_{\star}}
    \end{equation}
\end{center}

The distribution of $\Sigma_{\rm sSFR}$ is shown in the right panel of Figure~\ref{fig:sizes_hist}, which includes lower limits for unresolved sources. The LAEs have $\Sigma_{\rm sSFR} \gtrsim 0.03 - 10~{\rm Gyr^{-1}~kpc^{-2}}$, while SFGs typically have $\Sigma_{\rm sSFR}$ $\approx$ $10^{-3} - 10~{\rm Gyr^{-1}~kpc^{-2}}$ \citep[][]{Reddy+2022}.

\subsection{Proxies for $Ly\alpha$ Escape} \label{subsec:escape_lya}

\begin{figure*}[t!]
	\centering
 	\includegraphics[width = 1.0\textwidth]{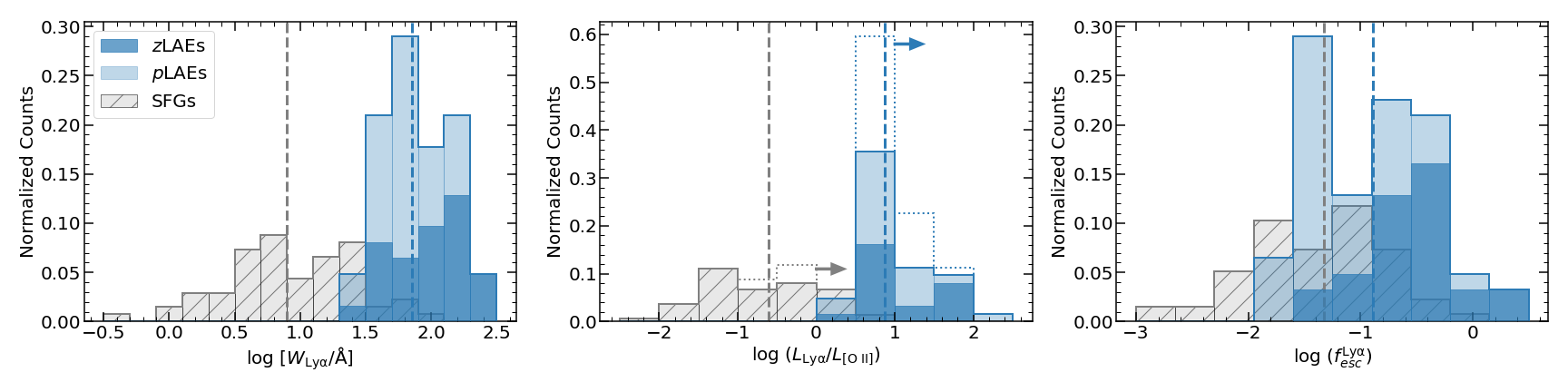}
    \caption{Three different measurements for studying the escape of \lya~in galaxies: {\it Left:}  \lya~equivalent width, $W_{\rm Ly\alpha}$; {\it Middle:} Ratio of \lya~to~\oii~luminosity; {\it Right:} Escape fraction, \fesc. The colors are the same as Figure~\ref{fig:SED_Props}. The median values for LAEs and SFGs are indicated by dashed blue and gray vertical lines, respectively.}
    \label{fig:fesc_hist}
\end{figure*}

\begin{figure}[b!]
    \centering
    \includegraphics[width = 1.0\columnwidth]{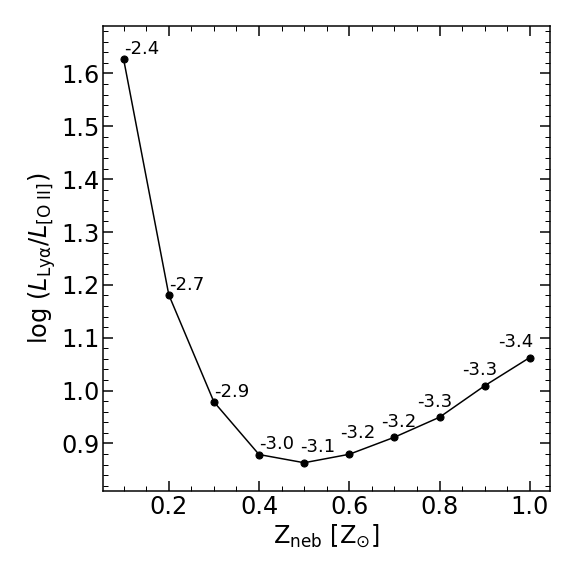}
    \caption{Predicted relationship between \lya/\oii~and nebular metallicity, ${\rm Z_{neb}}$, based on CLOUDY \citep{CLOUDY_Models} photoionization modeling. This assumes the anti-correlation between ionization parameter, $\log {\rm(U)}$, and ${\rm Z_{neb}}$ from \citet{PerezMontero2014}, with values of the former indicated at each point. See text for further details.}
    \label{fig:lya_oii_zneb}
\end{figure}

We quantify the escape of \lya~photons using three different measurements: the \lya~equivalent width ($W_{\rm Ly\alpha}$); the ratio of \lya~to \oii~luminosity (\lya/\oii); and the escape fraction (\fesc). It is important to note that we are measuring the \lya~escape only along the line-of-sight, and are not accounting for photons that are resonantly scattered into the diffuse halo. Even though these proxies are sensitive to the gas covering fraction and the dust content in galaxies, they are different measures of the observed \lya~emission. 

The equivalent width of the \lya~emission ($W_{\rm Ly\alpha}$) is the least model-dependent proxy for the \lya~escape \citep{Sobral+2019}. The calculation of $W_{\rm Ly\alpha}$ for LAEs and SFGs is described in Section~\ref{subsec:em_lines}. In the left panel of Figure~\ref{fig:fesc_hist}, we show the distribution of $W_{\rm Ly\alpha}$ for LAEs and SFGs. The LAEs have higher $W_{\rm Ly\alpha}$ ranging from 20$-$250 \AA~with a median of $\approx~70$~\AA. In contrast, the SFGs with observable \lya~emission have a median $W_{\rm Ly\alpha}$ of $\approx 7.8$~\AA. This difference is mostly by selection, since galaxies with high $W_{\rm Ly\alpha}$ are likely to be selected via narrow-band imaging and are more likely to be spectroscopically confirmed.

The ratio of \lya~to \oii~luminosity is another possible proxy for the escape of \lya, which depends on the distribution of gas and dust in the galaxy. Unlike \lya~emission, \oii~photons are not resonant in nature, and therefore \lya/\oii~can be used as an independent measure of \lya~escape. The \lya~and \oii~flux measurements are presented in Section~\ref{subsec:em_lines}. The middle panel of Figure~\ref{fig:fesc_hist} shows the distribution of \lya/\oii~ratio of LAEs and SFGs. In cases where there is no significant \oii~emission, we have lower limits on the \lya/\oii~ratio, as shown by the dotted histograms in the plot. We find that $\log$ (\lya/\oii) ranges from 0.14$-$2.06, with a median $\ge$ 0.91. The distribution also shows that LAEs have higher \lya/\oii, on average, compared to SFGs, due to the higher \lya~escape in LAEs. However, \oii~emission depends on the ionization parameter and metallicity of galaxies, as discussed below.

Figure~\ref{fig:lya_oii_zneb} shows how the intrinsic \lya/\oii~is predicted to vary with nebular metallicity, ${\rm Z_{neb}}$, based on photoionization modeling. To compute this relationship, we use the photoionization modeling code CLOUDY \citep{CLOUDY_Models} with an intrinsic ionizing spectrum set by the BPASS \citep{BPASS} constant star formation models, with a stellar metallicity of 0.2 ${\rm Z_{\odot}}$ and an age of ${\rm 10^{7.5}~yrs}$. We then use these models to compute the expected \lya/\oii~ratio as a function of ${\rm Z_{neb}}$, where the ionization parameter, $\log {\rm (U)}$, is fixed to the value predicted by the anti-correlation between $\log {\rm (U)}$ and ${\rm Z_{neb}}$ found in local H~{\sc ii} regions \citep{PerezMontero2014}. As observed in the middle panel of Figure~\ref{fig:fesc_hist}, eight LAEs ($\approx$13\%) have $\log~$(\lya/\oii)~$\gtrsim$ 1.5, on the high end of what is expected. Because these \lya/\oii~do not account for the potentially large fraction of \lya~photons that are resonantly scattered out of the photometric aperture, these ratios are effectively lower limits. This suggests that these galaxies are extremes in the parameter space shown in Figure~\ref{fig:lya_oii_zneb}. To produce such high \lya/\oii~ratios, they should either have very low nebular metallicities, ${\rm Z_{neb}} \lesssim 0.2$, and/or very high ionization parameters, $\log~{\rm(U)} \gtrsim -2.4$, similar to the values measured for other LAE samples at high redshift \citep[][]{Finkelstein+2011b, Nakajima+2012, Nakajima+2013}.

Finally, \lya~escape fraction (\fesc), which is defined as the ratio of observed \lya~luminosity to the intrinsic \lya~luminosity produced in a galaxy, is the most commonly used parameter to study the escape of \lya~in galaxies. The estimate of escape fraction depends on several assumptions such as metallicities, star formation histories, and dust attenuation curves. We calculate the intrinsic \lya~luminosity using the SFR computed from the SED fitting and the CLOUDY models \citep{CLOUDY_Models}. The ionizing photon luminosity per unit SFR for this model is ${\rm Q(H^{0})} =$ 9.259 $\times~{\rm 10^{52}~s^{-1}}$. Assuming Case-B recombination, the intrinsic H$\alpha$ luminosity is, ${\rm L_{H\alpha}~[ergs~s^{-1}]~=~1.36~\times~10^{-12}~Q(H^{0})}$ \citep{Leitherer+1995}, and the SFR [$\msun~{\rm yr^{-1}}$] $\approx~2.09~\times~{\rm 10^{-42}~L(H\alpha)~[ergs~s^{-1}}]$  \citep{Reddy+2022}. We use this relation to compute the intrinsic H$\alpha$ luminosity from the SED-based SFR. This assumes that the ionizing photons neither escape nor get absorbed by dust prior to photoionizing hydrogen. Assuming a stellar metallicity of 0.2 ${\rm Z_{\odot}}$, an age of ${\rm 10^{7.5}~yr}$, a nebular metallicity of $0.3~{\rm Z_{\odot}}$, and an ionization parameter, $\log({\rm U})$ of -2.0, yields \lya/H$\alpha$ $\approx~8.9$ from these models. We use this ratio to compute the intrinsic \lya~luminosity from H$\alpha$ luminosity, which, along with the observed \lya~luminosities (Section~\ref{subsec:em_lines}), is used to calculate \fesc~for the LAEs and SFGs. 

\begin{figure*}[t!]
	\centering
 	\includegraphics[width = 1.0\textwidth]{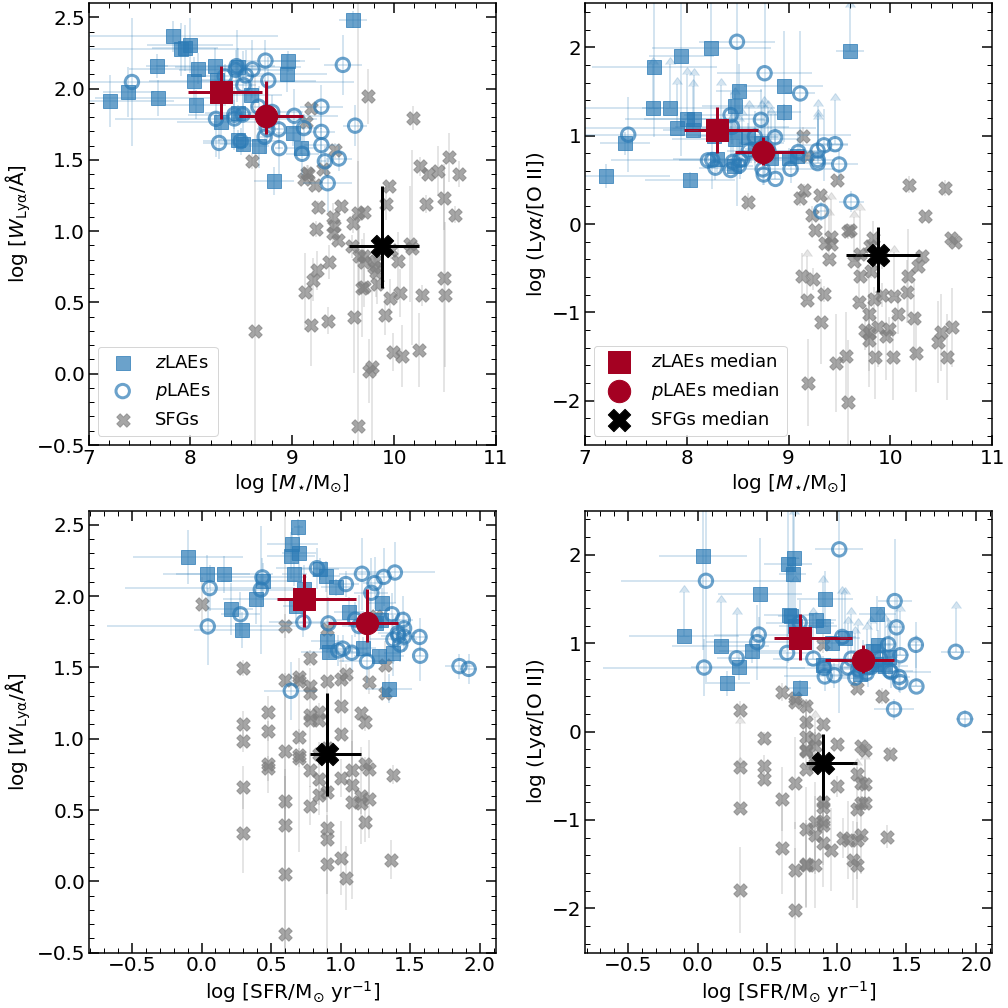}
        \caption{Dependence of $W_{\rm Ly\alpha}$ ({\it left}) and \lya/\oii~luminosity ratio ({\it right}) on the stellar mass ({\it top}) and SFR ({\it bottom}). The $z$LAEs, $p$LAEs, and the continuum-selected SFGs are shown as filled blue squares, open blue circles, and gray crosses, respectively. The red square, red circle, and black cross along with their corresponding error bars are the median and interquartile range for the subsets.}
    \label{fig:fesc_props}
\end{figure*}

\begin{table*}[t!]
    \centering
    \caption{Spearman $\rho$ parameter for different correlations. The p-values for each of the parameters are shown in parentheses.}
    \label{tab:correlations}
    \begin{tabular}{|c|c|c|c|c|}
    \hline
    \multirow{2}{*}{\bf Proxy for \lya~escape}   &    \multicolumn{2}{c|}{$\log~(M_{\star}~[\msun])$}   &   \multicolumn{2}{c|}{$\log~({\rm SFR~[\msun~yr^{-1}]})$} \\
    \cline{2-5} & {\bf Only LAEs} & {\bf LAEs+MOSDEF} & {\bf Only LAEs}  &   {\bf LAEs+MOSDEF} \\ 
    \hhline{|=|=|=|=|=|}
    $W_{\rm Ly\alpha}$   &    -0.47 (0.0001)    &    -0.72 (1.4 $\times~10^{-21}$)     &     -0.50 (2.9 $\times~10^{-5}$)      &    -0.02 (0.80)   \\ 
    \lya/\oii  &    -0.24 (0.06)      &    -0.73 (7.4 $\times~10^{-19}$)     &     -0.39 (0.002)                     &    -0.05 (0.59)    \\ \hline
    \end{tabular}
    
\end{table*}

\begin{figure*}[t!]
    \centering
    \includegraphics[width = 1.0\textwidth]{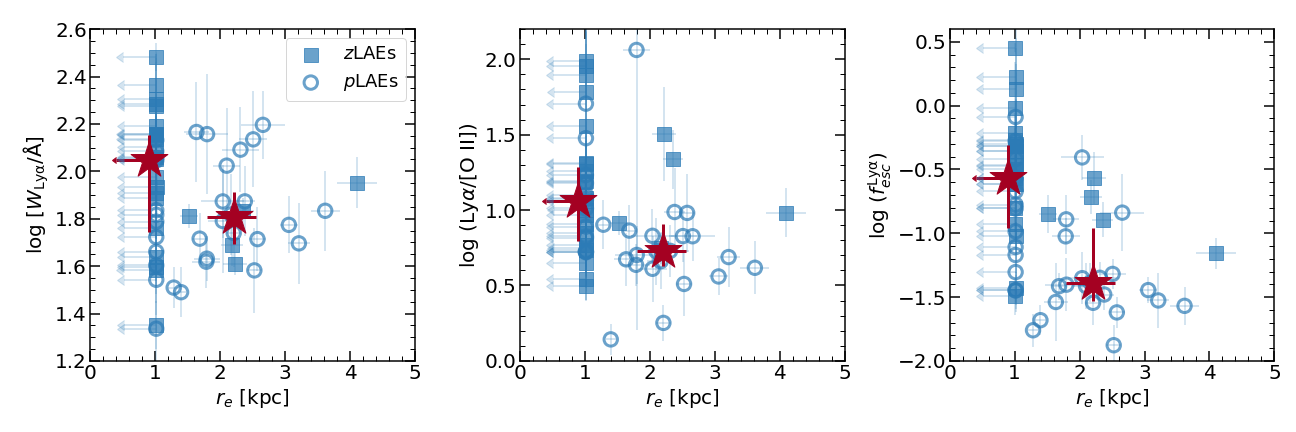}
    \caption{Equivalent width, $W_{\rm Ly\alpha}$ ({\it left}), \lya/\oii~luminosity ratio ({\it middle}), and escape fraction, \fesc~({\it right}) as a function of effective radius $r_{e}$. The symbols and colors are the same as Figure~\ref{fig:fesc_props}. In each panel, the left and right stars are the median values for the unresolved and resolved sources, respectively. The error bars denote the interquartile ranges for the individual parameters.} 
    \label{fig:fesc_re}
\end{figure*}

Three of the LAEs in our sample have computed escape fractions greater than 1. Two of these sources (DENS1\_86 and DENS1\_356) have extremely faint continua, resulting in poor SED fits. The third source (FIELD4\_124) has high \lya~luminosity ($\approx~{\rm 10^{43}~ergs~s^{-1}}$) and is possibly an active galaxy (AGN) \citep{Sobral+2018, Zhang+2021}.

The right panel of Figure~\ref{fig:fesc_hist} shows the distribution of escape fraction of LAEs in comparison with the SFGs. Given the observed $W_{\rm Ly\alpha}$, these distributions are typical for galaxies at similar and lower redshifts \citep{Matthee+2016, Yang+2017, Weiss+2021, Reddy+2022}. The median escape fraction of the SFGs ($\approx$ 4.8\%) is similar to the escape fraction expected at z $\approx$ 2$-$3 \citep{Hayes+2010, Sobral+2017}. The LAEs however have a median escape fraction of $\approx$ 13.5\%, which is significantly higher and is closer to the escape fraction observed at z $\gtrsim$ 6 (see Figures 1 and 4 in \citet{Hayes+2011}). It has been observed that \lya~escape is correlated with the escape of ionizing photons \citep{Marchi+2017, Marchi+2018, Steidel+2018, Pahl+2021}, and galaxies with high ionizing escape fractions are often bright LAEs at similar redshifts \citep{Naidu+2022}. The comparable escape fractions of the LAEs to the escape fraction observed at z $\gtrsim$ 6 suggests that these are probably low-redshift analogs of galaxies that contributed to reionization. 

\section{\lya~escape and galaxy properties} \label{sec:props_dependence}
\subsection{Dependence on Stellar Mass and SFR} \label{subsec:fesc_mass_sfr}

To understand the physical driver of high \lya~escape in LAEs compared to typical SFGs, we search for any significant trends with galaxy properties. Figure~\ref{fig:fesc_props} shows how $W_{\rm Ly\alpha}$ and the measured \lya/\oii~luminosity ratio depend on the stellar masses and SFRs of LAEs and SFGs. In each panel, the filled blue squares and open blue circles represent $z$LAEs and $p$LAEs, respectively, and gray crosses denote SFGs. Furthermore, the red square, red circle, and the black cross and their corresponding error bars mark the median and interquartile ranges of the individual samples. We study the strength of correlations using the Spearman correlation test (Table~\ref{tab:correlations}). Note that we are not considering \fesc~in this analysis, as the calculation of \fesc~is dependent on the SFRs derived from the SED fitting. As mentioned in the previous section, we consider $W_{\rm Ly\alpha}$ and \lya/\oii~as proxies for \lya~escape.

The top panels show $W_{\rm Ly\alpha}$ and \lya/\oii~ as a function of stellar mass. As expected, LAEs have higher $W_{\rm Ly\alpha}$ and \lya/\oii~ratios. From Table~\ref{tab:correlations}, we see that there is a possible anti-correlation between $W_{\rm Ly\alpha}$ vs $M_{\star}$, as well as between \lya/\oii~vs $M_{\star}$, when we consider both the LAE sample and MOSDEF SFGs together. When we consider the LAE sample alone, there is a moderate anti-correlation between \lya~escape proxies and stellar mass. Focusing on the median points in these panels, we see that \lya~escape has a clear dependence on the stellar mass of star-forming galaxies. This is similar to what has been observed in many previous studies \citep{Matthee+2016, Oyarzun+2016, Oyarzun+2017, Santos+2020, Weiss+2021}. However, the sample may be incomplete in the lower-left regions of these two panels due to the sensitivity limits of both narrowband and continuum-based surveys. 

The bottom panels of Figure~\ref{fig:fesc_props} show the dependence of $W_{\rm Ly\alpha}$ and \lya/\oii~on the SFR of galaxies. When all LAEs and SFGs are considered, the Spearman $\rho$ coefficient for $W_{\rm Ly\alpha}$ vs SFR is -0.02, while the coefficient for \lya/\oii~vs SFR is -0.05. This suggests that there is no obvious correlation between \lya~escape and SFR for these galaxies. This non-dependence of \lya~escape on SFR is in contrast with several previous studies on LAEs and SFGs \citep[][]{Matthee+2016, Oyarzun+2017, Weiss+2021}. This is not surprising because the detection of correlation depends on the dynamic range in \lya~escape and SFR probed by the different samples. As the SFR increases, the intrinsic \lya~emission will increase along with the gas and dust column densities. This increase in gas and dust densities, in turn, will influence the emergent \lya~emission. The observed non-dependence of \lya~escape on SFR, along with the diverse range of SFRs in LAEs (Section~\ref{subsubsec:stellar_props}) suggest that star formation alone is not the primary driver for the escape of \lya~photons. The following subsections discuss the dependence of \lya~escape on sizes and star-formation-rate surface densities.

\subsection{Dependence on the Sizes} \label{subsec:fesc_re}
Figure~\ref{fig:fesc_re} shows the dependence of $W_{\rm Ly\alpha}$, \lya/\oii~luminosity ratio, and \fesc~on the compactness of galaxies. All the points plotted near $r_{e} \approx$ 1 kpc are unresolved galaxies that have sizes $\lesssim$ 1 kpc. The \lya~escape in LAEs spans a broad range of values, and is higher for unresolved galaxies on average, compared to resolved galaxies. The left and right stars in each of the panels show the median values of the \lya~escape proxies for unresolved and resolved sources, respectively. The error bars denote the interquartile range of the different distributions. These median values suggest a possible 2$\sigma$ anti-correlation between \fesc~and size of the galaxy. 

From Figure~\ref{fig:fesc_re}, we also find that the fraction of resolved $p$LAEs ($\approx$ 63\%) is higher than the fraction of resolved $z$LAEs ($\approx$ 18\%). The $z$LAEs are spectroscopically confirmed due to the stronger \lya~emission from them, compared to $p$LAEs. This also suggests a dependence of \fesc~on galaxy sizes.

\citet{Bond+2012} studied the rest-frame UV sizes of z $\simeq$ 2.1 and z $\simeq$ 3.1 LAEs and found a systematic trend that higher $W_{\rm Ly\alpha}$ LAEs have smaller median sizes compared to the lower $W_{\rm Ly\alpha}$ samples. This dependence of \lya~escape on galaxy sizes is also seen in typical SFGs.  \citet{Law+2012} studied $\approx$200 star-forming galaxies and uncovered a higher fraction of \lya~emission in galaxies with compact morphologies. More recently, \citet{Weiss+2021} found that \fesc~is anti-correlated with $r_{e}$ for [OIII]-emitting galaxies \citep[also see][]{Reddy+2022}. \citet{Kim+2021} studied nearby Green Pea galaxies which emit \lya~emission and found that \fesc~is dependent on their sizes. \citet{Paulino-Afonso+2018} also found an anti-correlation between $W_{\rm Ly\alpha}$ and the UV sizes of high-redshift LAEs from 2 $\lesssim~z~\lesssim$ 6. These studies suggest that this dependence is independent of redshift. Therefore, galaxy sizes clearly play an important role in the escape of \lya~photons.

\subsection{Dependence on Star-Formation-Rate Surface Densities} \label{subsec:fesc_sigma}

\begin{figure*}[t!]
    \centering
    \includegraphics[width = 1.0\textwidth]{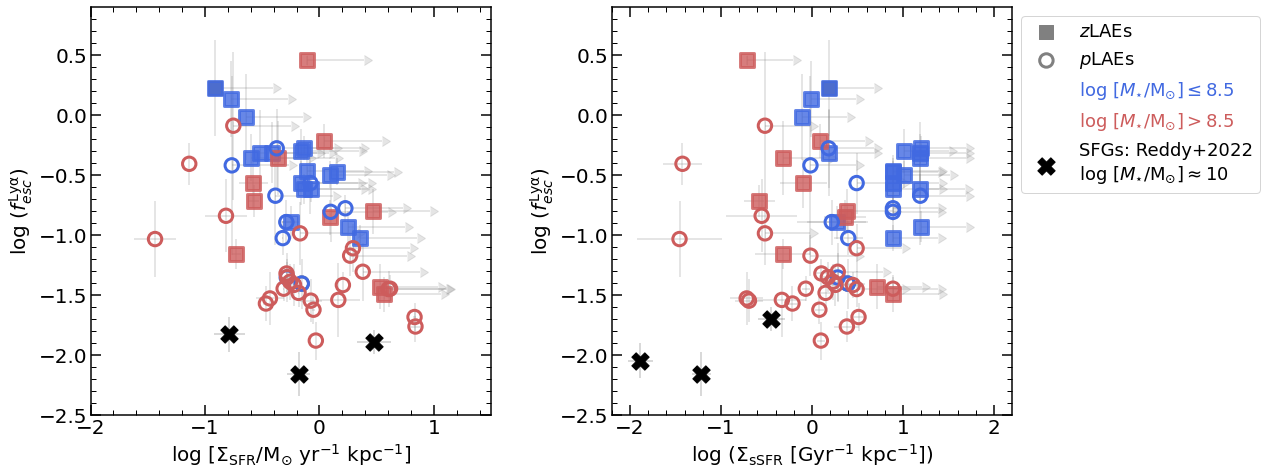}
    \caption{Escape Fraction as a function of star-formation-rate surface density, $\Sigma_{\rm SFR}$ ({\it left}) and specific star-formation-rate surface density, $\Sigma_{\rm sSFR}$ ({\it right}). The closed squares and open circles are $z$LAEs and $p$LAEs, respectively. Low-mass LAEs ($\log~[M_{\star}/{\rm M_{\odot}}] \le 8.5$) are shown in blue, while high-mass LAEs ($\log~[M_{\star}/{\rm M_{\odot}}] > 8.5$) are shown in red. The black crosses are from \citet{Reddy+2022}, which are computed using the composite spectra of the SFGs.}
    \label{fig:fesc_sig}
\end{figure*}

Observational studies suggest a possible correlation between \lya~escape fraction and $\Sigma_{\rm SFR}$ \citep[][]{Heckman+2011, Verhamme+2017, Marchi+2019, Reddy+2022}. Furthermore, the gravitational potential of the galaxy, encoded in specific star-formation-rate surface density, $\Sigma_{\rm sSFR}$ can also influence the escape of \lya~photons. In Figure~\ref{fig:fesc_sig}, we compare our sample of LAEs with those obtained for MOSDEF SFGs from \citet{Reddy+2022} (similar to their Figure 21). The $z$LAEs and $p$LAEs are shown as closed squares and open circles, respectively. We further divide the LAEs into low-mass (${\rm \log~[M_{\star}/M_{\odot}]} \le$ 8.5) and high-mass (${\rm \log~[M_{\star}/M_{\odot}]} >$ 8.5) sources, shown by blue and red markers, respectively. The SFGs have a median stellar mass, ${\rm \log~[M_{\star}/M_{\odot}]} \approx$ 10.     

In Figure~\ref{fig:fesc_sig}, the LAEs that have only lower limits to their $\Sigma_{\rm SFR}$ and $\Sigma_{\rm sSFR}$ are the sources that are unresolved, with sizes $\lesssim$ 1~kpc. From the left panel, we see that the sources that are resolved have similar $\Sigma_{\rm SFR}$, but higher \fesc, compared to SFGs. This suggests that there is an additional factor beyond $\Sigma_{\rm SFR}$ that modulate \fesc. For galaxies with a fixed $\Sigma_{\rm SFR}$, we see that the sources with lower masses have higher \fesc. This can be also seen from the right panel of the figure, where LAEs have higher $\Sigma_{\rm sSFR}$, on average, compared to SFGs. This highlights the effect of stellar mass (and therefore the gravitational potential) on the \lya~escape (Section~\ref{subsec:fesc_mass_sfr} and Figure~\ref{fig:fesc_props}).

These results reinforce the argument that the distribution of star formation is a key ingredient for \lya~escape. The feedback associated with compact star formation will lead to outflows, that will in turn create low-density channels in the ISM. These low-density columns provide pathways for the \lya~photons to propagate and escape. The gravitational potential (or stellar mass) also plays a crucial role in this scenario. The shallow potential well of low-mass sources makes it harder for them to retain the gas that is pushed out via winds and outflows. As we are measuring the \lya~photons along the line-of-sight, this physical interpretation is highly dependent on the orientation of galaxies. Large-scale deep observations of galaxies across the multi-wavelength regime will give further clues about the dependence of \lya~escape on gravitational potential and sizes of galaxies. 

\subsection{Dependence on the Environment} \label{subsec:environment}

\begin{figure*}[t!]
	\centering
 	\includegraphics[width = 1.0\textwidth]{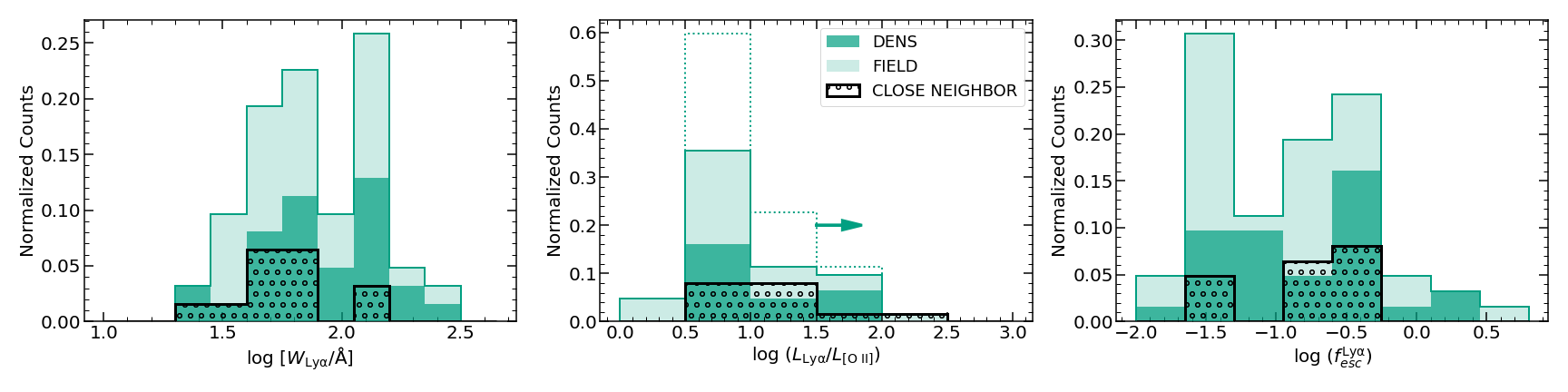}
        \caption{Distribution of $W_{\rm Ly\alpha}$ ({\it left}), \lya/\oii~ratio ({\it middle}), and \fesc~({\it right}) for LAEs in {\it DENS} (shown as dark green bins) and {\it FIELD} (shown as light green bins) regions. The hatched black bins are 14 LAEs that have close neighbors or extended features as seen visually in the {\it HST} images.}
    \label{fig:fesc_env}
\end{figure*}

\begin{table*}[t!]
    \centering
    \caption{p-values from KS-test between sources in {\it DENS} and {\it FIELD} regions}
    \label{tab:env_check}
    \begin{tabular}{|c|c|c|c|c|c|}
    \hline
    LAEs Group     &    {\bf $\log~[M_{\star}/\msun]$} &   {\bf SFR~$[\msun~{\rm yr^{-1}}]$} &  $\log~(f_{esc})$  &   $\log~(W_{\rm Ly\alpha})$   &   $\log$ (\lya/\oii) \\
    \hline
    $z$LAEs        &     0.35 & 0.45   & 0.27              &          0.98             &           0.61       \\
    All LAEs       &    0.18 & 0.39 &     0.17              &          0.51             &           0.89       \\
    \hline
    \end{tabular}
\end{table*}

In this subsection, we investigate whether \lya~escape has any dependence on the environment of LAEs. In Section~\ref{subsec:hstbased}, we mentioned that we have {\it HST} imaging of LAEs in seven fields, three of which are in dense regions, as measured by \citet{Prescott+2008}. Thus, we divide our LAEs into high-density ({\it DENS}) and low-density ({\it FIELD}) sources, based on their proximity to the \lya~blob. Some LAE candidates have either a close companion or exhibit an extended component visible in the {\it HST} images. These companions and/or components are typically not resolved in the ground-based images, and may indicate a close physical companion. Out of the 62 LAEs in our sample, we identify 14 sources which exhibit such components within 1.5\arcsec. We consider the \lya~escape of this subset in comparison with the rest of the LAEs.

Figure~\ref{fig:fesc_env} shows the distribution of different \lya~escape proxies for LAEs in low and high density environments. The dark and light green histograms are LAEs in the {\it DENS} and {\it FIELD} regions, respectively. We perform KS tests for the different proxies between all LAEs in these two groups. We also repeat the tests by considering just the $z$LAEs, since the spectroscopic redshifts yield more accurate environmental information (Table~\ref{tab:env_check}). From the p-values, we cannot reject the null hypothesis that the \lya~escape for {\it DENS} and {\it FIELD} regions are similar. Additionally, the black histogram with circles in the figure are the LAEs with close neighbors and/or extended components. The p-values from the KS tests between these 14 LAEs and the rest of the LAEs for $W_{\rm Ly\alpha}$, \lya/\oii~ratio, and \fesc~are 0.26, 0.35, and 0.42, respectively. This suggests that given our rough definition of environment, \lya~escape does not depend on the local environment of the LAEs in this sample.

Table~\ref{tab:env_check} also shows the p-values from the KS tests for stellar masses and SFRs between LAEs in {\it DENS} and {\it FIELD} regions. For both $z$LAEs and all LAEs together, we find that the stellar masses and SFRs are statistically similar for LAEs in these different environments. We cannot make any comparisons between the sizes of these galaxies because of the high fraction of unresolved sources and small numbers of resolved LAEs in the individual groups. However, given that most of the LAEs in our sample are compact, the $\Sigma_{\rm sSFR}$ of the LAEs in {\it DENS} and {\it FIELD} environments are likely similar. This argument supports our earlier result that the star formation in a low-mass, compact galaxy is key to the \lya~escape in galaxies, irrespective of the local environment.

\section{Conclusions} \label{sec:conclusions} 

In an effort to understand the mechanisms that lead to \lya~escape in LAEs, we study 62 LAEs from seven {\it HST} fields using multi-wavelength photometry. Out of these, 27 sources have confirmed spectroscopic redshifts between 2.55 $\le$ z $\le$ 2.75. In addition to broad-band photometry in \bw, $R$, $I$, $F110W$, and $F160W$ filters, we also have medium-band photometry in \ia~and $F139M$ bands that probe the \lya~and \oii~emission lines, respectively in this redshift range. We considered a comparison sample of 136 typical SFGs that are part of the MOSDEF/LRIS sample (Section~\ref{subsec:mosdef}). 

We obtained the stellar masses, SFRs, ages, and dust content of all the LAEs using the SED fitting technique (Section~\ref{subsec:sed_fitting}). We also computed LAE sizes by running GALFIT on the {\it HST} $F110W$ images (Section~\ref{subsec:sizes}). Finally, we considered three proxies for the \lya~escape: the \lya~equivalent width ($W_{\rm Ly\alpha}$); the \lya/\oii~luminosity ratio; and the \lya~escape fraction (\fesc) (Section~\ref{subsec:escape_lya}). Using these proxies and studying their correlations with the LAE properties, we conclude:

\begin{itemize}
    \item LAEs typically probe low-mass (7.2 $\le$ $\log(M_{\star}/\msun)$ $\le$ 9.6), young (age $\le$ 1 Gyr), star-forming galaxies (0.8 $\msun~{\rm yr^{-1}~\le~SFR~\le~100~\msun~yr^{-1}}$) with low dust content (E(B$-$V) $\le$ 0.26 mag). $p$LAEs are more massive, star-forming, older, and redder compared to $z$LAEs. Furthermore, LAEs and SFGs have different underlying distributions. LAEs have lower ages, lower masses, similar SFRs, and less dust content compared to SFGs at similar redshifts (Section~\ref{subsubsec:stellar_props}). 
    
    \item LAEs have a wide range of SFRs, and have higher sSFRs compared to SFGs. On average, less massive LAEs lie above the SFMS, compared to their massive counterparts (Section~\ref{subsubsec:sfms})

    \item Almost 60\% of the LAEs are unresolved, even with {\it HST} resolution, indicating that these galaxies are compact ($r_{e}~\le$ 1 kpc). LAEs also have comparable $\Sigma_{\rm SFR}$ and higher $\Sigma_{\rm sSFR}$ compared to SFGs (Section~\ref{subsec:sizes}).
    
    \item LAEs have higher $W_{\rm Ly\alpha}$, higher \lya/\oii~luminosity ratios, and higher \fesc~compared to SFGs. By comparing the \lya/\oii~ratios with expected model values, we found that some LAEs are extreme galaxies with low nebular metallicities (${\rm Z_{neb}  \lesssim 0.2 Z_{\odot}}$) and/or high ionization parameters ( $\log {\rm(U)} \gtrsim -2.4$). Additionally, the \fesc~of LAEs is similar to the escape fraction observed at z $\gtrsim$ 6, suggesting that these LAEs may be low-redshift analogs of galaxies that contributed to reionization (Section~\ref{subsec:escape_lya}).
    
    \item The escape of \lya~in galaxies is anti-correlated with stellar mass, but shows no obvious dependence on SFR (Section~\ref{subsec:fesc_mass_sfr}).
    
    \item The \lya~escape has a wide range of values for unresolved LAEs, and is on average higher for unresolved LAEs compared to their resolved counterparts (Section~\ref{subsec:fesc_re}).
    
    \item For a given $\Sigma_{\rm SFR}$, the lower-mass LAEs have higher \fesc~than their more massive counterparts. This is consistent with a scenario where compact star formation in a low gravitational potential facilitates the escape of \lya, by creating low-column-density channels in the ISM via feedback. (Section~\ref{subsec:fesc_sigma}).

    \item Based on the local density of the {\it HST} fields, we do not observe any obvious dependence of nearby environment on the \lya~escape from galaxies (Section~\ref{subsec:environment}).
    \end{itemize}

Upcoming surveys such as the One-hundred square-degree DECam Imaging in Narrowbands (ODIN) survey, the Legacy Survey of Space and Time with the Vera Rubin Telescope \citep{LSST}, and upcoming deep JWST NIRSPEC observations \citep{NIRSPEC_JWST}, will provide us with large scale observations of LAEs in different environments across different epochs. These new datasets will help us understand the physics of \lya~escape in more detail.

\facilities{HST (WFC3), Mayall, Subaru (SuprimeCam)}

\software{AstroDrizzle, Astropy, GALFIT, IRAF, Matplotlib, NumPy, Source Extractor}

\acknowledgements{We thank the anonymous referee for thoroughly reading the manuscript and helping us improve this work. The authors thank the MOSDEF team for providing \oii~measurements and SED parameters for the comparison SFG sample. We also thank Ryan Sanders, Alice Shapley, and Michael Topping for providing data from the followup LRIS observations of MOSDEF galaxies. RP also thanks Gurtina Besla, Ryan Endsley, Xiaohui Fan, and Rob Kennicutt for their helpful comments on this project.

This research is primarily the result of observations from the Hubble Space Telescope and was supported by HST-GO-13000 (PI: Sungryong Hong). RP was supported by the University of Arizona, and in part by NOIRLab, which is managed by the Association of Universities for Research in Astronomy (AURA). AD's and SJ's research is supported by NSF's NOIRLab, which is operated by the Association of Universities for Research in Astronomy (AURA) under a cooperative agreement with National Science Foundation. AD's research is also supported in part by a Fellowship from the John Simon Guggenheim Memorial Foundation and by the Institute of Theory and Computation at the Harvard-Smithsonian Center for Astrophysics. 

The research also made use of observations from the NOAO Deep Wide-Field Survey (NDWFS; NOAO Prop. ID no. 1999B-0027; B.Jannuzi and A.Dey, Co-PIs), which used observations made with the Mosaic-3 camera at the Mayall 4 m telescope at Kitt Peak National Observatory, National Optical Astronomy Observatory, which is operated by the Association of Universities of Research in Astronomy (AURA) under cooperative agreement with the National Science Foundation. The authors are honored to be permitted to conduct astronomical research on Iolkam Du'ag (Kitt Peak), a mountain with particular significance to the Tohono O'odham. 

This research is based in part on data collected at the Subaru Telescope, which is operated by the National Astronomical Observatory of Japan. We are honored and grateful for the opportunity of observing the Universe from Maunakea, which has cultural, historical, and national significance in Hawaii. 

Some of the data presented in this paper were obtained from the Mikulski Archive for Space Telescopes (MAST) at the Space Telescope Science Institute. These observations can be accessed via \dataset[10.17909/48vk-jr50]{http://dx.doi.org/10.17909/48vk-jr50}.}

% {\small
% }

\bibliography{LAE.bib}{}

\appendix

\begin{figure*}[b!]
    \centering
    \includegraphics[width=0.9\textwidth]{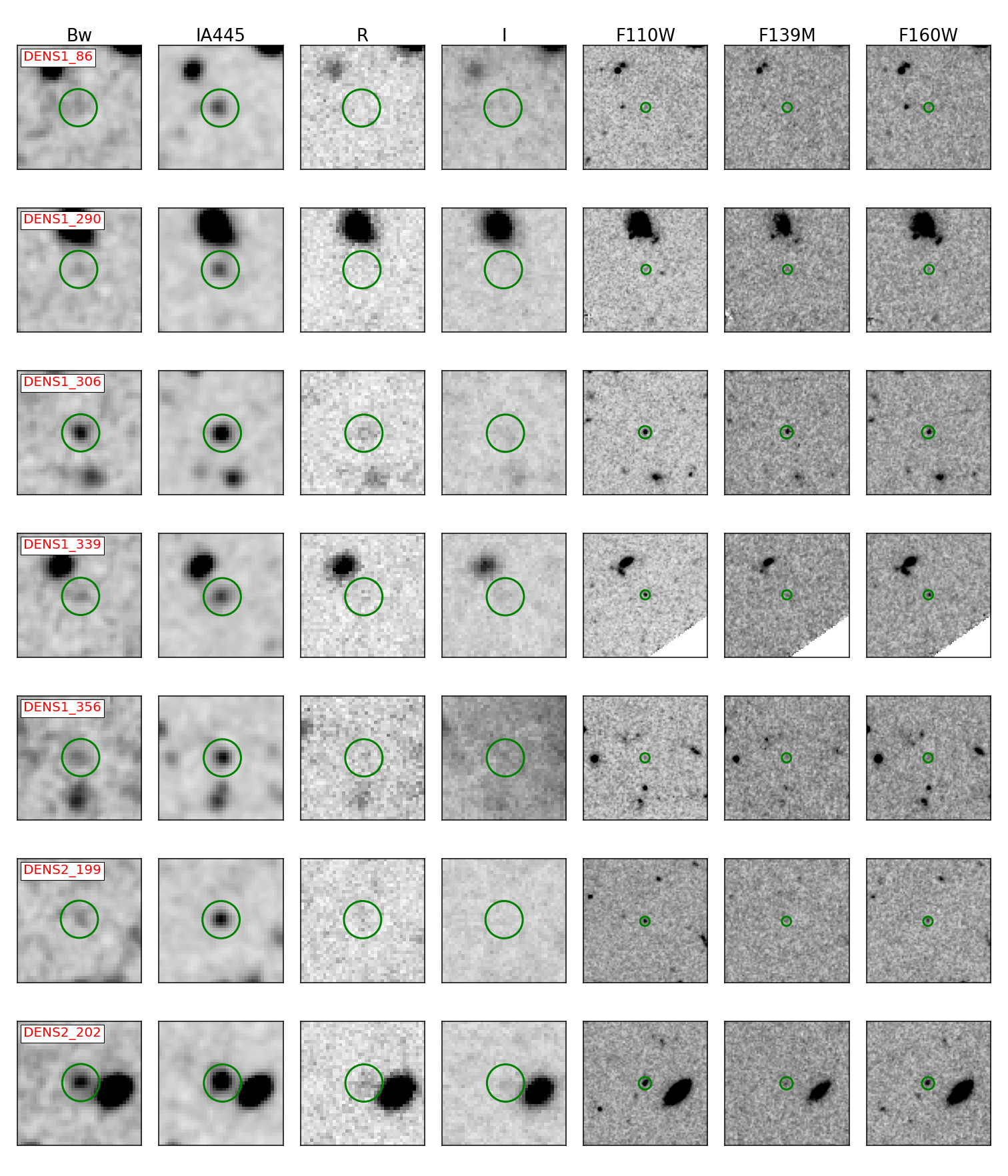}
    \caption{\small Image-Cutouts (10\arcsec $\times$ 10\arcsec) of $z$LAEs in \bw, \ia, $R$, $I$, $F110W$, $F139M$, and $F160W$ from left to right. The green circles denote the apertures used for performing photometry.}
    \label{fig:zLAEs1}
\end{figure*}

\renewcommand{\thefigure}{\arabic{figure} (Cont.)}
\addtocounter{figure}{-1}

\begin{figure*}[t!]
    \centering
    \includegraphics[width=0.9\textwidth]{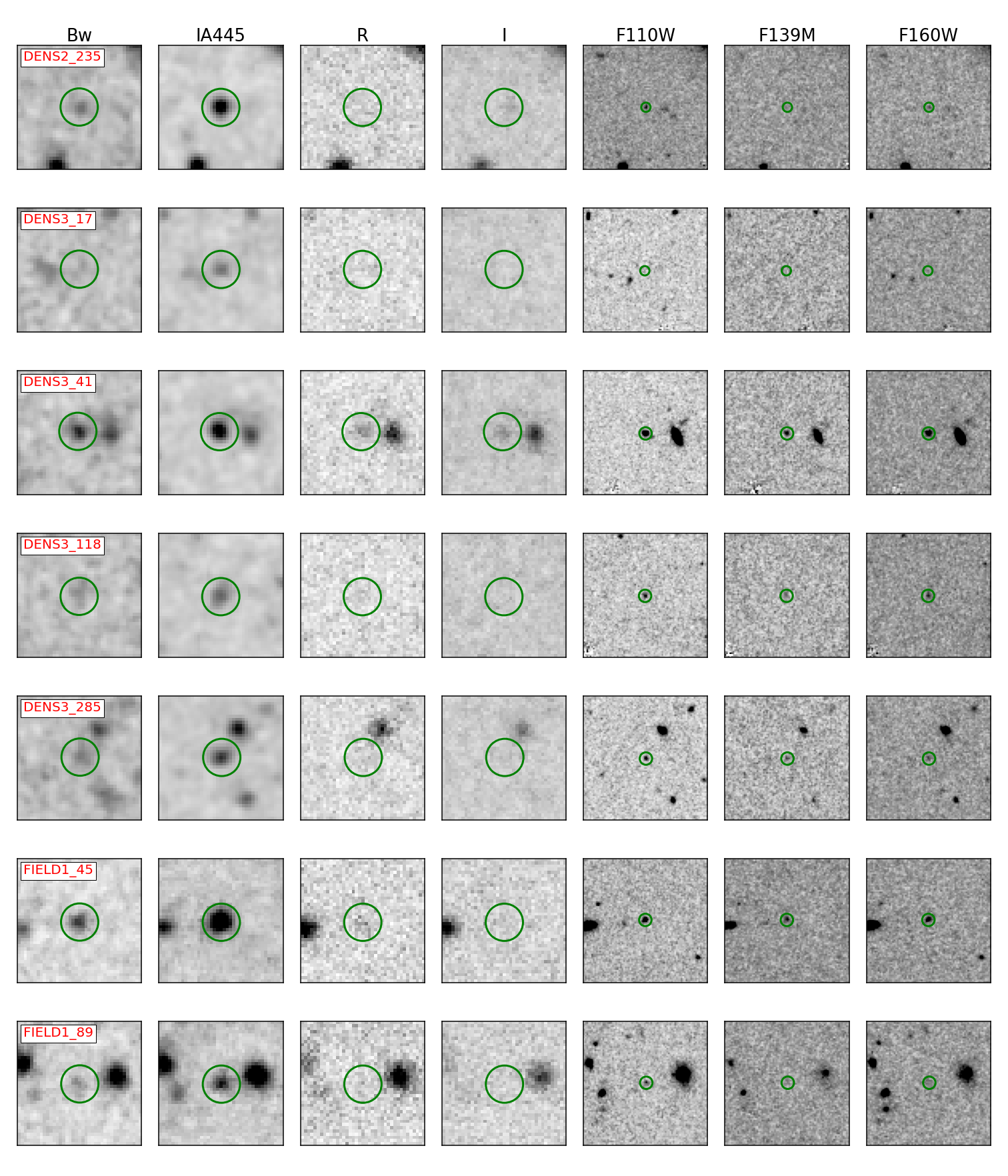}
    \caption{\small Image-Cutouts (10\arcsec $\times$ 10\arcsec) of $z$LAEs}
    \label{fig:zLAEs2}
\end{figure*}

\addtocounter{figure}{-1}

\begin{figure*}[t!]
    \centering
    \includegraphics[width=0.9\textwidth]{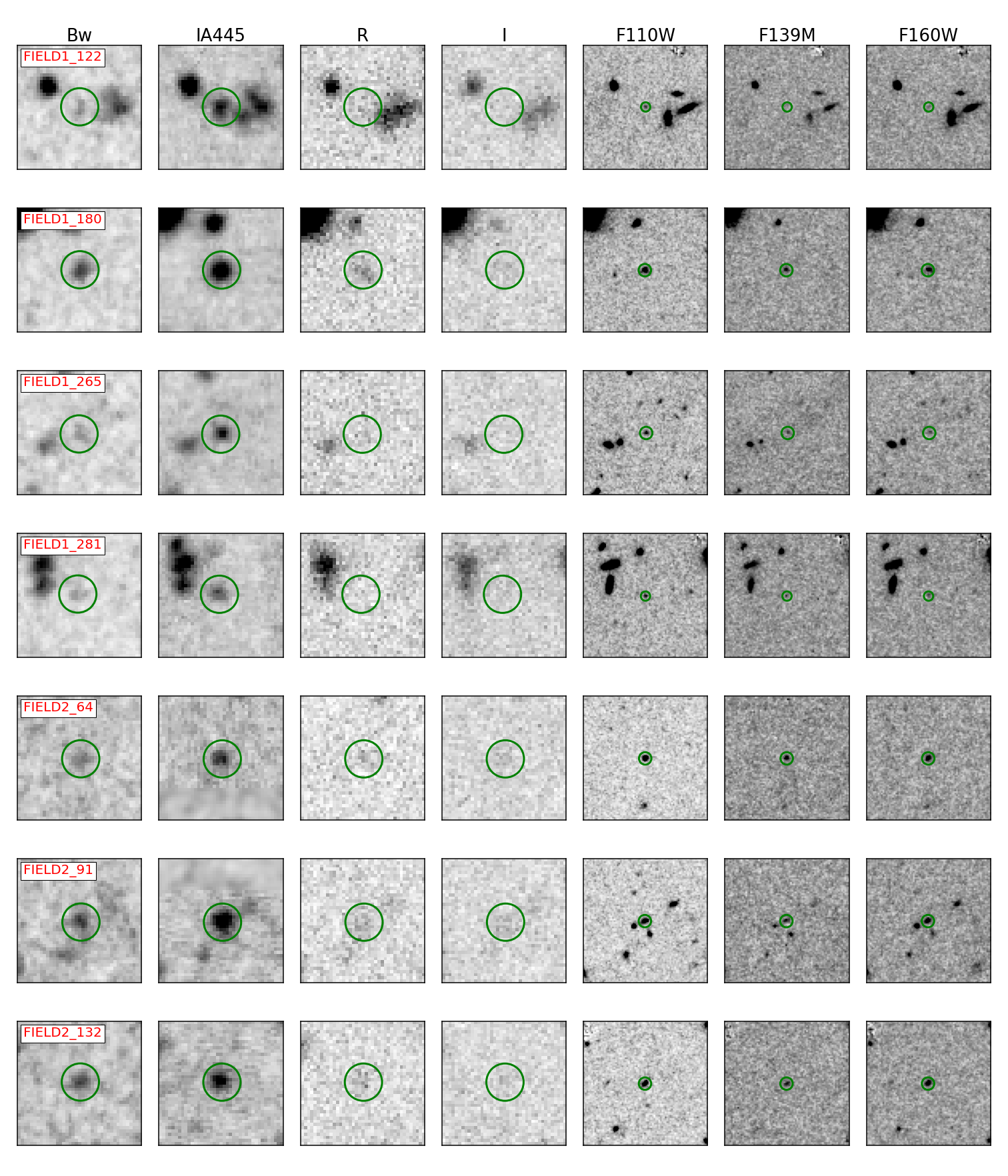}
    \caption{\small Image-Cutouts (10\arcsec $\times$ 10\arcsec) of $z$LAEs}
    \label{fig:zLAEs3}
\end{figure*}

\addtocounter{figure}{-1}

\begin{figure*}[t!]
    \centering
    \includegraphics[width=0.9\textwidth]{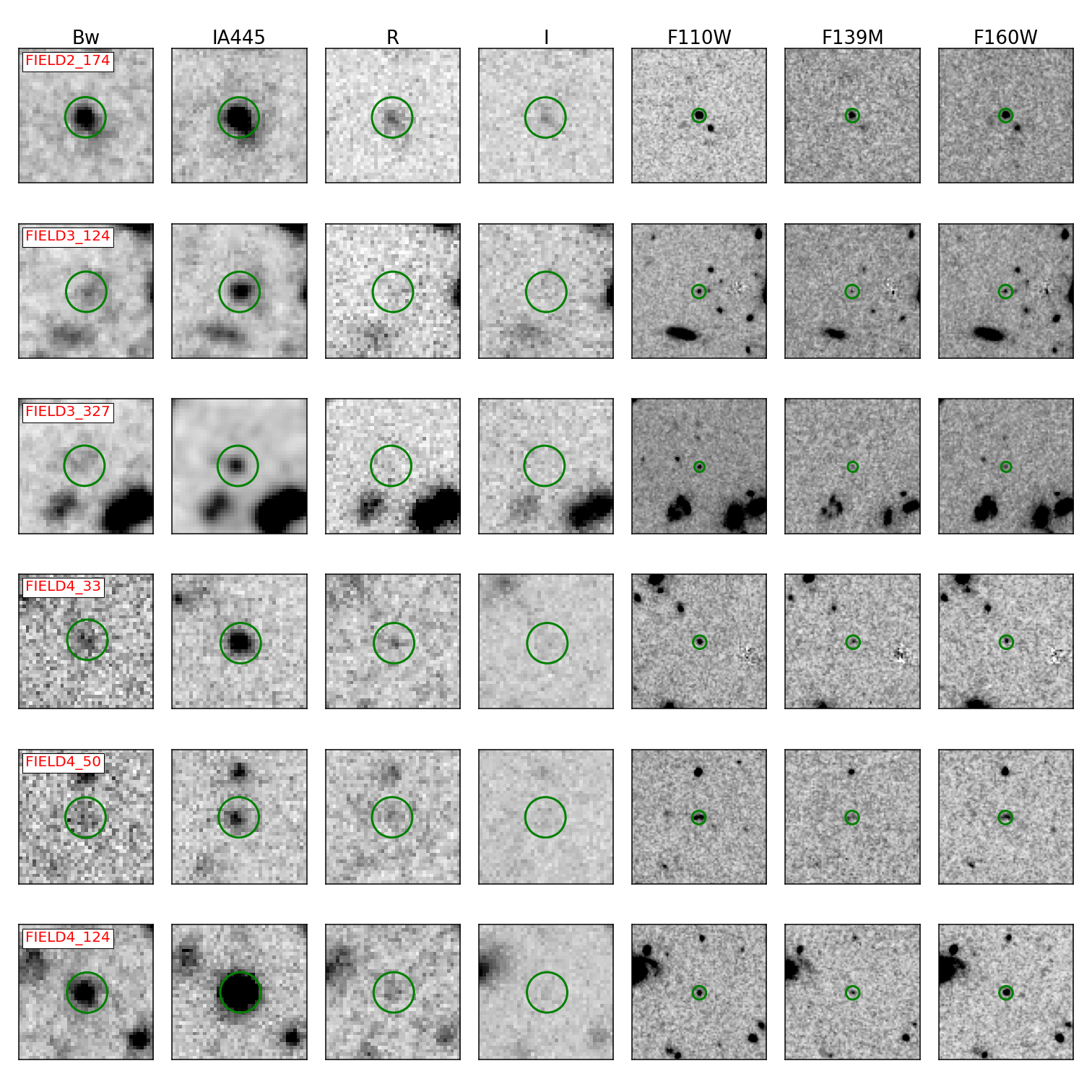}
    \caption{\small Image-Cutouts (10\arcsec $\times$ 10\arcsec) of $z$LAEs}
    \label{fig:zLAEs4}
\end{figure*}

\renewcommand{\thefigure}{\arabic{figure}}

\begin{figure*}[t!]
    \centering
    \includegraphics[width=0.9\textwidth]{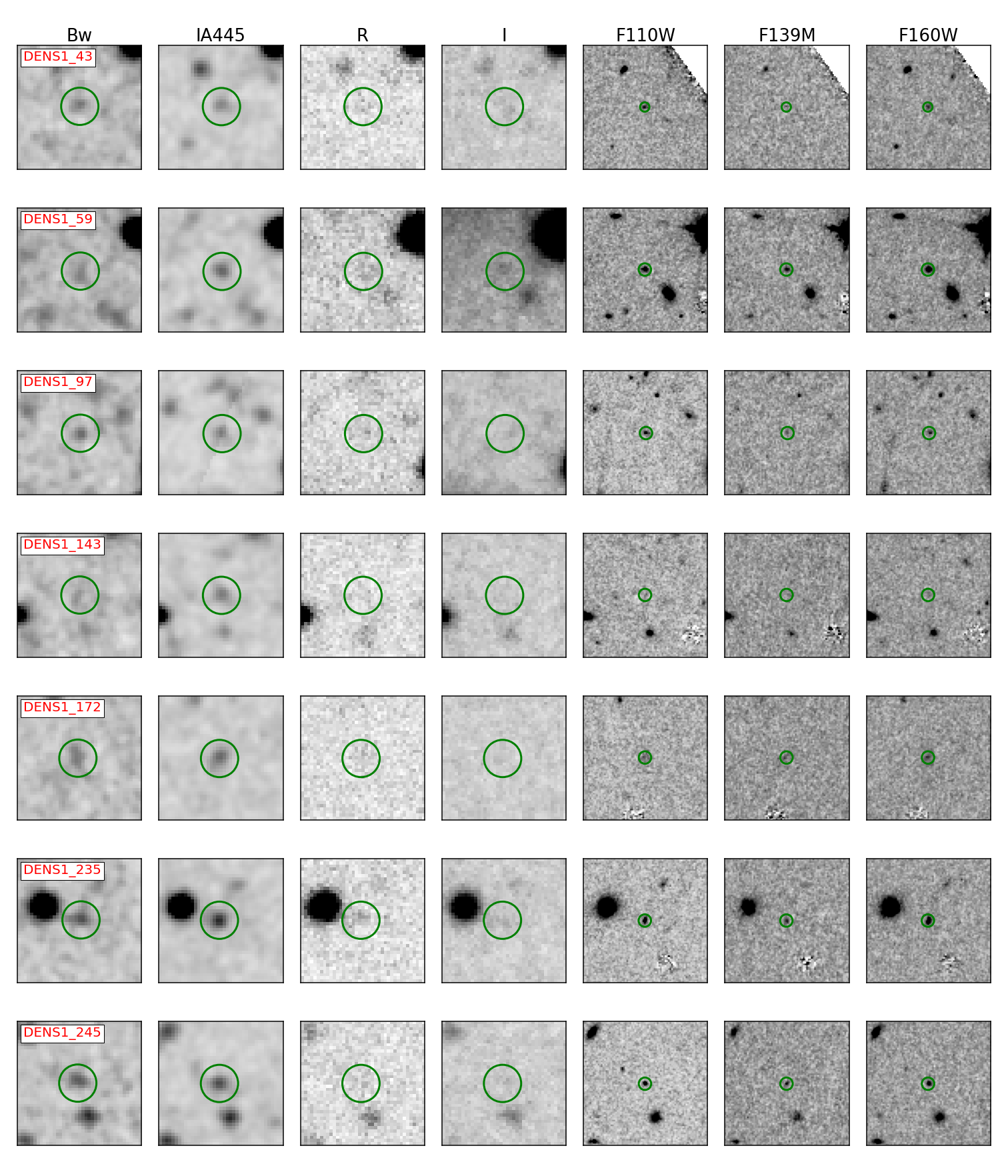}
    \caption{\small Image-Cutouts (10\arcsec $\times$ 10\arcsec) of $p$LAEs in \bw, \ia, $R$, $I$, $F110W$, $F139M$, and $F160W$ from left to right. The green circles denote the apertures used for performing photometry.}
    \label{fig:pLAEs1}
\end{figure*}

\renewcommand{\thefigure}{\arabic{figure} (Cont.)}
\addtocounter{figure}{-1}

\begin{figure*}[t!]
    \centering
    \includegraphics[width=0.9\textwidth]{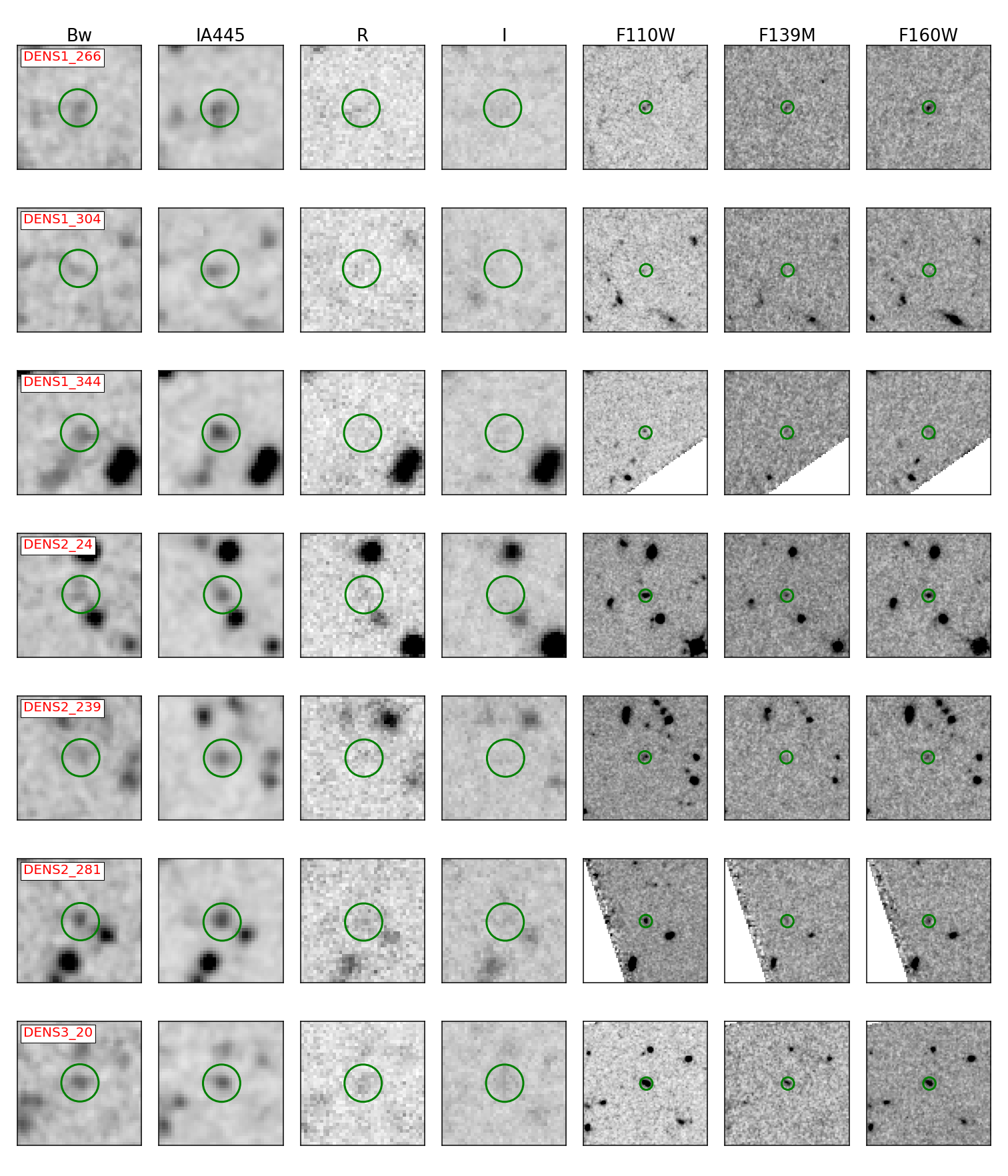}
    \caption{\small Image-Cutouts (10\arcsec $\times$ 10\arcsec) of $p$LAEs}
    \label{fig:pLAEs2}
\end{figure*}

\addtocounter{figure}{-1}

\begin{figure*}[t!]
    \centering
    \includegraphics[width=0.9\textwidth]{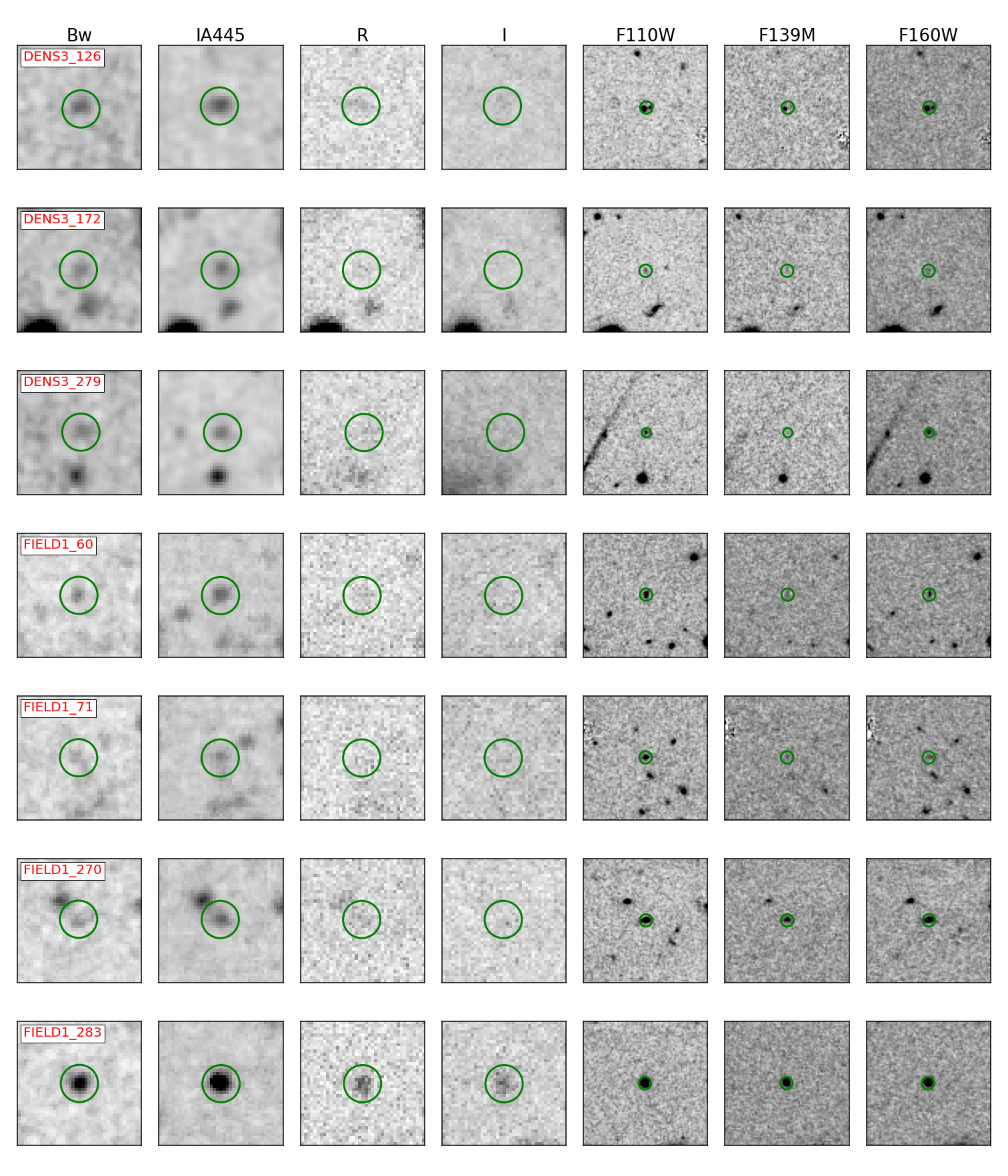}
    \caption{\small Image-Cutouts (10\arcsec $\times$ 10\arcsec) of $p$LAEs}
    \label{fig:pLAEs3}
\end{figure*}

\addtocounter{figure}{-1}

\begin{figure*}[t!]
    \centering
    \includegraphics[width=0.9\textwidth]{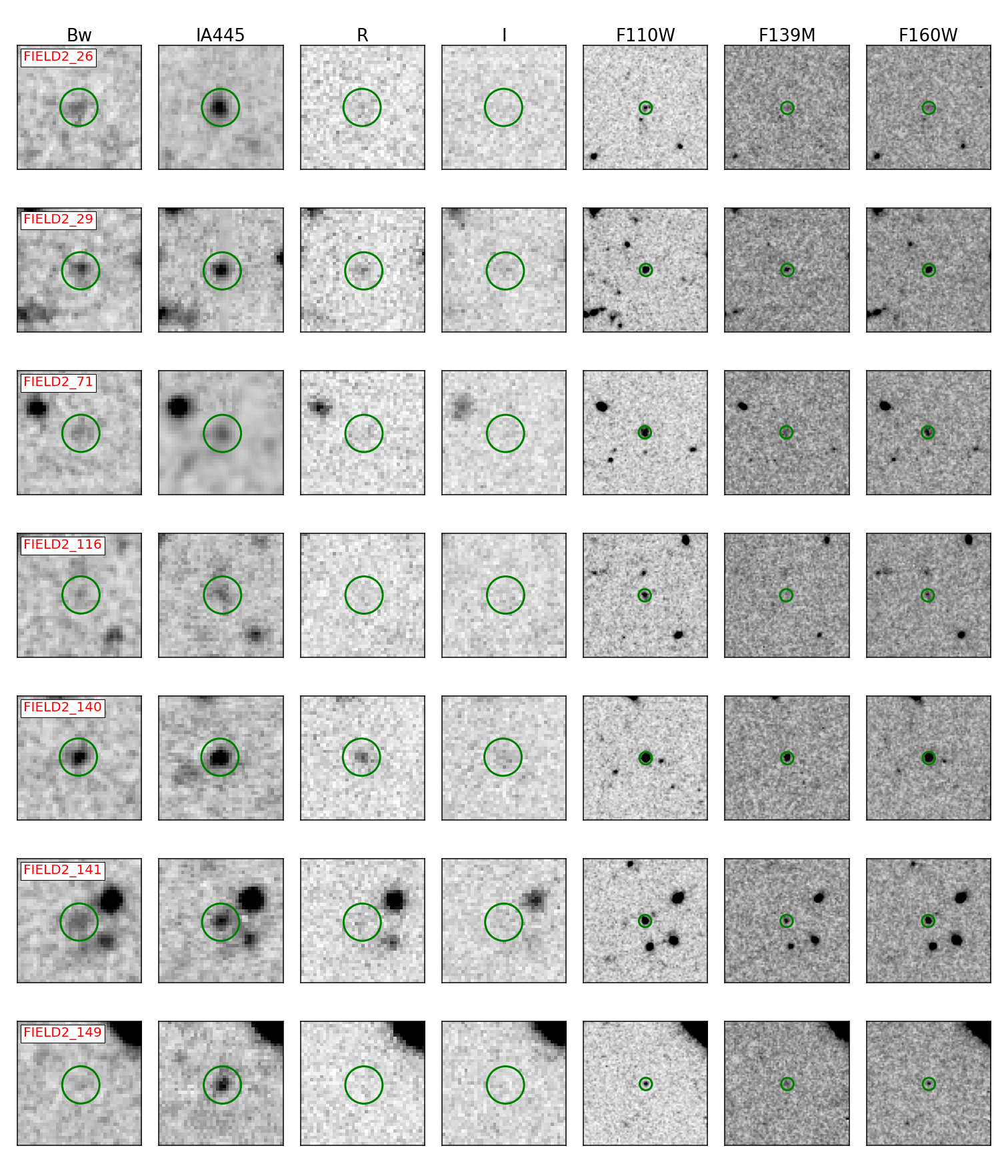}
    \caption{\small Image-Cutouts (10\arcsec $\times$ 10\arcsec) of $p$LAEs}
    \label{fig:pLAEs4}
\end{figure*}

\addtocounter{figure}{-1}

\begin{figure*}[t!]
    \centering
    \includegraphics[width=0.9\textwidth]{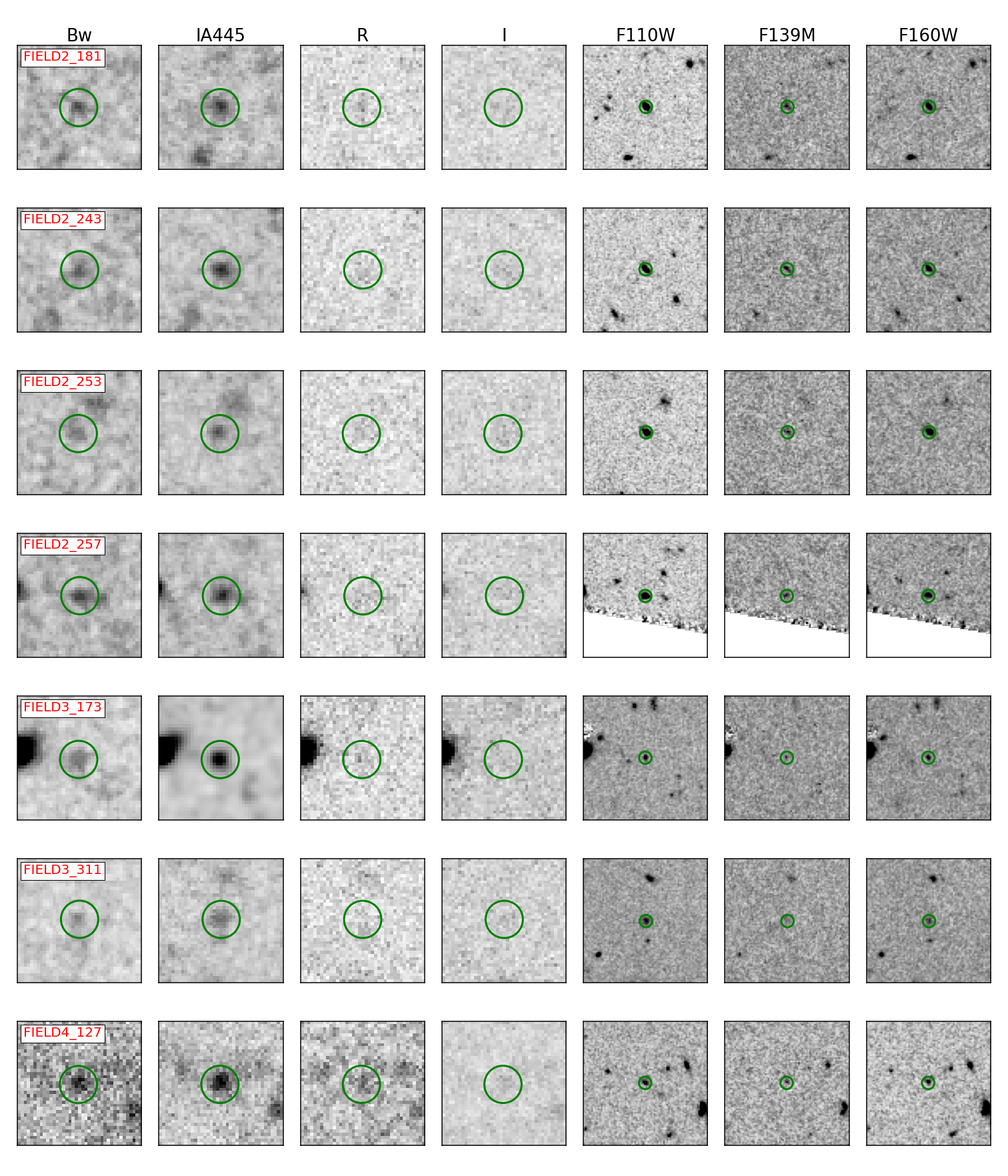}
    \caption{\small Image-Cutouts (10\arcsec $\times$ 10\arcsec) of $p$LAEs}
    \label{fig:pLAEs5}
\end{figure*}

\newpage

\begin{table*}[h!]
    \centering
    \caption{Position and Photometry of the LAEs}
    \label{tab:phot}
    \resizebox{0.85\textwidth}{!}{%
    \begin{tabular}{*{11}{|c}|}  %{|l|c|c|c|c|c|c|c|c|c|c|}
    \hline
    {\bf ID}   &   {\bf R.A.}   &    {\bf Dec}   &   {\bf $z_{spec}$}   &   {\bf \bw}   &   {\bf \ia}   &   {\bf $R$}   &   {\bf $I$}   &   {\bf   $F110W$}   &    {\bf $F139M$}   &   {\bf $F160W$} \\ \hline
    \multicolumn{11}{|c|}{$z$LAEs}                         \\ \hline
    
DENS1\_86  &  218.54699  &  33.33212   &   2.6599   &   26.89 $\pm$ 0.32   &   25.59 $\pm$ 0.09   &   28.13 $\pm$ 1.48   &   28.16 $\pm$ 1.87   &   27.87 $\pm$ 0.46   &   27.76 $\pm$ 1.00   &    27.52 $\pm$ 0.40 \\ 
DENS1\_290  &  218.54867  &  33.31537   &   2.6890   &   27.10 $\pm$ 0.37   &   25.52 $\pm$ 0.08   &   26.63 $\pm$ 0.62   &   25.93 $\pm$ 0.51   &   28.32 $\pm$ 0.64   &   26.82 $\pm$ 0.55   &    28.43 $\pm$ 0.75 \\ 
DENS1\_306  &  218.54201  &  33.31323   &   2.6593   &   25.58 $\pm$ 0.10   &   24.94 $\pm$ 0.05   &   25.52 $\pm$ 0.26   &   25.43 $\pm$ 0.36   &   25.37 $\pm$ 0.08   &   25.01 $\pm$ 0.17   &    25.45 $\pm$ 0.10 \\ 
DENS1\_339  &  218.53821  &  33.30828   &   2.6905   &   26.47 $\pm$ 0.22   &   25.30 $\pm$ 0.07   &   26.15 $\pm$ 0.43   &   25.35 $\pm$ 0.33   &   26.24 $\pm$ 0.12   &   26.97 $\pm$ 0.61   &    26.12 $\pm$ 0.13 \\ 
DENS1\_356  &  218.54893  &  33.33707   &   2.6662   &   26.54 $\pm$ 0.24   &   25.41 $\pm$ 0.08   &   27.41 $\pm$ 0.75   &   28.49 $\pm$ 2.17   &   27.78 $\pm$ 0.43   &   27.13 $\pm$ 0.67   &    26.94 $\pm$ 0.25 \\ 
DENS2\_199  &  218.54365  &  33.26862   &   2.6650   &   26.61 $\pm$ 0.25   &   25.34 $\pm$ 0.08   &   28.90 $\pm$ 2.12   &   26.84 $\pm$ 0.75   &   26.65 $\pm$ 0.09   &   27.09 $\pm$ 0.65   &    26.79 $\pm$ 0.22 \\ 
DENS2\_202  &  218.53539  &  33.26821   &   2.6638   &   25.15 $\pm$ 0.07   &   24.40 $\pm$ 0.03   &   24.97 $\pm$ 0.17   &   25.03 $\pm$ 0.25   &   25.62 $\pm$ 0.05   &   25.39 $\pm$ 0.22   &    25.44 $\pm$ 0.09 \\ 
DENS2\_235  &  218.55726  &  33.26553   &   2.6630   &   26.80 $\pm$ 0.29   &   25.07 $\pm$ 0.06   &   27.14 $\pm$ 0.88   &   26.52 $\pm$ 0.79   &   27.02 $\pm$ 0.13   &   27.21 $\pm$ 0.70   &    26.96 $\pm$ 0.26 \\ 
DENS3\_17  &  218.49395  &  33.31581   &   2.6615   &   27.17 $\pm$ 0.39   &   25.98 $\pm$ 0.13   &   26.68 $\pm$ 0.65   &   27.51 $\pm$ 1.39   &   27.85 $\pm$ 0.27   &   26.56 $\pm$ 0.45   &    29.27 $\pm$ 1.30 \\ 
DENS3\_41  &  218.50143  &  33.31221   &   2.6172   &   25.49 $\pm$ 0.10   &   25.00 $\pm$ 0.06   &   25.17 $\pm$ 0.20   &   24.70 $\pm$ 0.20   &   25.00 $\pm$ 0.03   &   25.11 $\pm$ 0.18   &    24.79 $\pm$ 0.05 \\ 
DENS3\_118  &  218.51525  &  33.30501   &   2.6443   &   26.72 $\pm$ 0.28   &   25.55 $\pm$ 0.09   &   26.66 $\pm$ 0.64   &   28.48 $\pm$ 2.20   &   26.44 $\pm$ 0.11   &   26.01 $\pm$ 0.37   &    25.91 $\pm$ 0.14 \\ 
DENS3\_285  &  218.50141  &  33.28827   &   2.6509   &   26.34 $\pm$ 0.20   &   25.50 $\pm$ 0.09   &   26.14 $\pm$ 0.43   &   26.86 $\pm$ 0.97   &   26.83 $\pm$ 0.16   &   26.00 $\pm$ 0.37   &    26.54 $\pm$ 0.23 \\ 
FIELD1\_45  &  218.19319  &  33.25234   &   2.6700   &   25.41 $\pm$ 0.10   &   24.26 $\pm$ 0.03   &   25.83 $\pm$ 0.35   &   27.00 $\pm$ 1.30   &   25.55 $\pm$ 0.06   &   25.05 $\pm$ 0.17   &    25.25 $\pm$ 0.08 \\ 
FIELD1\_89  &  218.18495  &  33.24749   &   2.6863   &   26.59 $\pm$ 0.28   &   25.48 $\pm$ 0.09   &   27.45 $\pm$ 0.75   &   26.85 $\pm$ 1.16   &   26.98 $\pm$ 0.22   &   27.37 $\pm$ 0.75   &    26.68 $\pm$ 0.27 \\ 
FIELD1\_122  &  218.19026  &  33.24321   &   2.7191   &   26.50 $\pm$ 0.26   &   24.90 $\pm$ 0.05   &   25.95 $\pm$ 0.39   &   26.77 $\pm$ 0.75   &   26.90 $\pm$ 0.14   &   30.90 $\pm$ 3.66   &    28.12 $\pm$ 0.64 \\ 
FIELD1\_180  &  218.19246  &  33.23632   &   2.6678   &   25.34 $\pm$ 0.10   &   24.51 $\pm$ 0.04   &   25.18 $\pm$ 0.20   &   25.59 $\pm$ 0.53   &   25.14 $\pm$ 0.04   &   24.74 $\pm$ 0.13   &    24.97 $\pm$ 0.06 \\ 
FIELD1\_265  &  218.19829  &  33.22367   &   2.6958   &   26.23 $\pm$ 0.21   &   25.50 $\pm$ 0.09   &   27.41 $\pm$ 1.08   &   27.22 $\pm$ 1.44   &   26.38 $\pm$ 0.13   &   25.49 $\pm$ 0.24   &    26.71 $\pm$ 0.27 \\ 
FIELD1\_281  &  218.18990  &  33.25455   &   2.6859   &   27.30 $\pm$ 0.49   &   25.74 $\pm$ 0.11   &   27.45 $\pm$ 0.75   &   26.77 $\pm$ 0.75   &   26.93 $\pm$ 0.15   &   26.97 $\pm$ 0.62   &    27.16 $\pm$ 0.31 \\ 
FIELD2\_64  &  217.76887  &  33.18873   &   2.6766   &   26.18 $\pm$ 0.21   &   25.55 $\pm$ 0.12   &   25.96 $\pm$ 0.38   &   25.30 $\pm$ 0.44   &   25.26 $\pm$ 0.04   &   25.04 $\pm$ 0.17   &    25.12 $\pm$ 0.07 \\ 
FIELD2\_91  &  217.78369  &  33.18538   &   2.6773   &   25.66 $\pm$ 0.13   &   24.56 $\pm$ 0.05   &   25.20 $\pm$ 0.20   &   24.77 $\pm$ 0.29   &   25.42 $\pm$ 0.04   &   25.16 $\pm$ 0.19   &    25.25 $\pm$ 0.08 \\ 
FIELD2\_132  &  217.78603  &  33.18136   &   2.7051   &   25.64 $\pm$ 0.13   &   24.95 $\pm$ 0.07   &   25.55 $\pm$ 0.29   &   25.70 $\pm$ 0.58   &   25.47 $\pm$ 0.05   &   24.95 $\pm$ 0.16   &    25.14 $\pm$ 0.07 \\ 
FIELD2\_174  &  217.80338  &  33.17495   &   2.7036   &   24.76 $\pm$ 0.06   &   24.08 $\pm$ 0.03   &   24.55 $\pm$ 0.12   &   24.42 $\pm$ 0.21   &   24.71 $\pm$ 0.02   &   24.27 $\pm$ 0.09   &    24.51 $\pm$ 0.04 \\ 
FIELD3\_124  &  218.08984  &  33.34466   &   2.7400   &   25.89 $\pm$ 0.16   &   24.90 $\pm$ 0.06   &   26.68 $\pm$ 0.69   &   27.76 $\pm$ 1.88   &   26.17 $\pm$ 0.10   &   26.20 $\pm$ 0.44   &    26.19 $\pm$ 0.18 \\ 
FIELD3\_327  &  218.10093  &  33.31795   &   2.5950   &   26.15 $\pm$ 0.19   &   25.41 $\pm$ 0.09   &   26.92 $\pm$ 0.77   &   26.78 $\pm$ 0.75   &   26.72 $\pm$ 0.12   &   26.70 $\pm$ 0.50   &    26.87 $\pm$ 0.24 \\ 
FIELD4\_33  &  218.24553  &  33.68381   &   2.6186   &   25.61 $\pm$ 0.17   &   24.48 $\pm$ 0.04   &   25.90 $\pm$ 0.44   &   25.78 $\pm$ 0.56   &   25.83 $\pm$ 0.09   &   25.29 $\pm$ 0.21   &    25.93 $\pm$ 0.14 \\ 
FIELD4\_50  &  218.24910  &  33.68082   &   2.6272   &   26.15 $\pm$ 0.26   &   25.27 $\pm$ 0.09   &   26.30 $\pm$ 0.59   &   26.10 $\pm$ 0.72   &   25.48 $\pm$ 0.06   &   25.15 $\pm$ 0.18   &    25.27 $\pm$ 0.08 \\ 
FIELD4\_124  &  218.25557  &  33.66906   &   2.7122   &   24.64 $\pm$ 0.06   &   23.18 $\pm$ 0.01   &   26.05 $\pm$ 0.50   &   26.78 $\pm$ 1.10   &   25.61 $\pm$ 0.07   &   25.70 $\pm$ 0.29   &    24.74 $\pm$ 0.05 \\  \hline
    \multicolumn{11}{|c|}{$p$LAEs}                         \\ \hline
DENS1\_43  &  218.52519  &  33.33647   &   $-$   &   27.74 $\pm$ 0.60   &   26.44 $\pm$ 0.19   &   27.41 $\pm$ 0.75   &   27.12 $\pm$ 1.10   &   26.18 $\pm$ 0.12   &   26.70 $\pm$ 0.50   &    25.96 $\pm$ 0.11 \\ 
DENS1\_59  &  218.53597  &  33.33473   &   $-$   &   27.49 $\pm$ 0.50   &   26.20 $\pm$ 0.15   &   26.99 $\pm$ 0.77   &   26.76 $\pm$ 0.75   &   25.36 $\pm$ 0.08   &   24.81 $\pm$ 0.14   &    24.73 $\pm$ 0.05 \\ 
DENS1\_97  &  218.56296  &  33.33126   &   $-$   &   27.26 $\pm$ 0.42   &   26.03 $\pm$ 0.13   &   26.71 $\pm$ 0.64   &   26.76 $\pm$ 0.75   &   25.58 $\pm$ 0.10   &   25.96 $\pm$ 0.35   &    25.87 $\pm$ 0.13 \\ 
DENS1\_143  &  218.53259  &  33.32746   &   $-$   &   27.26 $\pm$ 0.42   &   26.41 $\pm$ 0.18   &   27.41 $\pm$ 0.75   &   26.70 $\pm$ 0.88   &   27.54 $\pm$ 0.49   &   26.70 $\pm$ 0.62   &    27.03 $\pm$ 0.36 \\ 
DENS1\_172  &  218.52629  &  33.32519   &   $-$   &   26.47 $\pm$ 0.23   &   25.80 $\pm$ 0.11   &   26.32 $\pm$ 0.49   &   30.13 $\pm$ 3.67   &   25.72 $\pm$ 0.11   &   25.18 $\pm$ 0.19   &    25.29 $\pm$ 0.08 \\ 
DENS1\_235  &  218.54419  &  33.32148   &   $-$   &   26.10 $\pm$ 0.17   &   25.46 $\pm$ 0.08   &   25.93 $\pm$ 0.37   &   25.10 $\pm$ 0.27   &   25.26 $\pm$ 0.07   &   25.27 $\pm$ 0.21   &    24.78 $\pm$ 0.05 \\ 
DENS1\_245  &  218.55910  &  33.32026   &   $-$   &   26.24 $\pm$ 0.19   &   25.54 $\pm$ 0.09   &   26.65 $\pm$ 0.64   &   25.68 $\pm$ 0.43   &   25.56 $\pm$ 0.10   &   25.54 $\pm$ 0.26   &    25.28 $\pm$ 0.08 \\ 
DENS1\_266  &  218.55763  &  33.31803   &   $-$   &   26.24 $\pm$ 0.19   &   25.66 $\pm$ 0.10   &   25.66 $\pm$ 0.29   &   25.44 $\pm$ 0.36   &   25.82 $\pm$ 0.12   &   25.32 $\pm$ 0.21   &    25.17 $\pm$ 0.07 \\ 
DENS1\_304  &  218.55536  &  33.31371   &   $-$   &   27.96 $\pm$ 0.70   &   26.35 $\pm$ 0.17   &   27.41 $\pm$ 0.75   &   26.76 $\pm$ 0.75   &   27.88 $\pm$ 0.63   &   27.27 $\pm$ 0.75   &    27.93 $\pm$ 0.68 \\ 
DENS1\_344  &  218.54156  &  33.30592   &   $-$   &   26.34 $\pm$ 0.20   &   25.40 $\pm$ 0.08   &   27.25 $\pm$ 0.92   &   27.19 $\pm$ 1.15   &   26.53 $\pm$ 0.22   &   26.02 $\pm$ 0.38   &    26.33 $\pm$ 0.20 \\ 
DENS2\_24  &  218.54337  &  33.28530   &   $-$   &   26.57 $\pm$ 0.24   &   25.85 $\pm$ 0.12   &   26.86 $\pm$ 0.74   &   25.89 $\pm$ 0.50   &   25.40 $\pm$ 0.04   &   25.21 $\pm$ 0.20   &    25.27 $\pm$ 0.08 \\ 
DENS2\_239  &  218.56262  &  33.26529   &   $-$   &   27.16 $\pm$ 0.39   &   25.86 $\pm$ 0.13   &   27.31 $\pm$ 0.75   &   26.84 $\pm$ 0.75   &   26.32 $\pm$ 0.10   &   26.50 $\pm$ 0.54   &    25.90 $\pm$ 0.14 \\ 
DENS2\_281  &  218.57589  &  33.26184   &   $-$   &   26.01 $\pm$ 0.15   &   25.36 $\pm$ 0.08   &   25.85 $\pm$ 0.36   &   25.89 $\pm$ 0.50   &   25.87 $\pm$ 0.07   &   25.52 $\pm$ 0.26   &    26.03 $\pm$ 0.16 \\ 
DENS3\_20  &  218.49754  &  33.31536   &   $-$   &   26.64 $\pm$ 0.26   &   25.97 $\pm$ 0.13   &   26.01 $\pm$ 0.39   &   25.18 $\pm$ 0.29   &   24.84 $\pm$ 0.03   &   24.87 $\pm$ 0.15   &    24.88 $\pm$ 0.06 \\ 
DENS3\_126  &  218.51845  &  33.30441   &   $-$   &   26.13 $\pm$ 0.17   &   25.48 $\pm$ 0.09   &   25.95 $\pm$ 0.38   &   25.74 $\pm$ 0.44   &   24.99 $\pm$ 0.03   &   24.76 $\pm$ 0.13   &    24.66 $\pm$ 0.05 \\ 
DENS3\_172  &  218.50684  &  33.29998   &   $-$   &   26.89 $\pm$ 0.31   &   25.82 $\pm$ 0.12   &   27.38 $\pm$ 1.02   &   26.82 $\pm$ 0.75   &   27.08 $\pm$ 0.20   &   26.62 $\pm$ 0.59   &    26.50 $\pm$ 0.23 \\ 
DENS3\_279  &  218.51691  &  33.28891   &   $-$   &   26.82 $\pm$ 0.30   &   25.89 $\pm$ 0.12   &   26.46 $\pm$ 0.55   &   27.16 $\pm$ 1.15   &   26.52 $\pm$ 0.09   &   27.32 $\pm$ 0.76   &    25.81 $\pm$ 0.10 \\ 
FIELD1\_60  &  218.21814  &  33.25018   &   $-$   &   26.74 $\pm$ 0.32   &   25.94 $\pm$ 0.13   &   26.60 $\pm$ 0.63   &   26.01 $\pm$ 0.72   &   25.63 $\pm$ 0.07   &   25.79 $\pm$ 0.31   &    25.72 $\pm$ 0.12 \\ 
FIELD1\_71  &  218.21986  &  33.24939   &   $-$   &   27.12 $\pm$ 0.43   &   26.20 $\pm$ 0.16   &   27.73 $\pm$ 1.27   &   25.89 $\pm$ 0.65   &   25.59 $\pm$ 0.06   &   25.69 $\pm$ 0.29   &    26.19 $\pm$ 0.18 \\ 
FIELD1\_270  &  218.20566  &  33.22301   &   $-$   &   26.46 $\pm$ 0.25   &   25.75 $\pm$ 0.11   &   26.07 $\pm$ 0.42   &   25.73 $\pm$ 0.60   &   24.83 $\pm$ 0.03   &   24.23 $\pm$ 0.08   &    24.27 $\pm$ 0.03 \\ 
FIELD1\_283  &  218.20151  &  33.25264   &   $-$   &   24.97 $\pm$ 0.07   &   24.46 $\pm$ 0.09   &   24.69 $\pm$ 0.14   &   24.10 $\pm$ 0.16   &   23.81 $\pm$ 0.01   &   23.38 $\pm$ 0.04   &    23.70 $\pm$ 0.02 \\ 
FIELD2\_26  &  217.78132  &  33.19250   &   $-$   &   26.02 $\pm$ 0.19   &   25.09 $\pm$ 0.08   &   25.67 $\pm$ 0.32   &   26.34 $\pm$ 0.90   &   26.13 $\pm$ 0.08   &   25.79 $\pm$ 0.32   &    25.93 $\pm$ 0.14 \\ 
FIELD2\_29  &  217.79498  &  33.19222   &   $-$   &   26.04 $\pm$ 0.19   &   25.43 $\pm$ 0.10   &   25.64 $\pm$ 0.31   &   25.40 $\pm$ 0.47   &   25.20 $\pm$ 0.04   &   24.87 $\pm$ 0.15   &    24.86 $\pm$ 0.06 \\ 
FIELD2\_71  &  217.77752  &  33.18753   &   $-$   &   26.70 $\pm$ 0.32   &   25.83 $\pm$ 0.15   &   26.18 $\pm$ 0.45   &   26.13 $\pm$ 0.77   &   25.23 $\pm$ 0.04   &   24.96 $\pm$ 0.16   &    25.13 $\pm$ 0.07 \\ 
FIELD2\_116  &  217.79680  &  33.18165   &   $-$   &   27.00 $\pm$ 0.39   &   26.03 $\pm$ 0.18   &   26.41 $\pm$ 0.55   &   25.77 $\pm$ 0.60   &   25.87 $\pm$ 0.07   &   26.00 $\pm$ 0.37   &    25.92 $\pm$ 0.14 \\ 
FIELD2\_140  &  217.78061  &  33.17919   &   $-$   &   25.31 $\pm$ 0.10   &   24.85 $\pm$ 0.06   &   24.99 $\pm$ 0.18   &   24.81 $\pm$ 0.30   &   23.97 $\pm$ 0.01   &   24.01 $\pm$ 0.07   &    23.71 $\pm$ 0.02 \\ 
FIELD2\_141  &  217.78641  &  33.17908   &   $-$   &   25.82 $\pm$ 0.15   &   25.28 $\pm$ 0.09   &   25.99 $\pm$ 0.40   &   25.79 $\pm$ 0.63   &   25.01 $\pm$ 0.03   &   25.07 $\pm$ 0.17   &    24.89 $\pm$ 0.06 \\ 
FIELD2\_149  &  217.76941  &  33.17862   &   $-$   &   28.94 $\pm$ 1.09   &   25.84 $\pm$ 0.08   &   29.43 $\pm$ 2.30   &   27.38 $\pm$ 0.75   &   26.51 $\pm$ 0.12   &   25.78 $\pm$ 0.31   &    26.55 $\pm$ 0.24 \\
FIELD2\_181  &  217.77873  &  33.17458   &   $-$   &   26.26 $\pm$ 0.22   &   25.52 $\pm$ 0.11   &   25.83 $\pm$ 0.35   &   25.47 $\pm$ 0.50   &   24.93 $\pm$ 0.03   &   24.75 $\pm$ 0.13   &    24.78 $\pm$ 0.05 \\ 
FIELD2\_243  &  217.78797  &  33.16508   &   $-$   &   26.18 $\pm$ 0.20   &   25.43 $\pm$ 0.10   &   26.17 $\pm$ 0.46   &   25.12 $\pm$ 0.38   &   25.01 $\pm$ 0.03   &   24.74 $\pm$ 0.13   &    25.02 $\pm$ 0.06 \\ 
FIELD2\_253  &  217.78553  &  33.16276   &   $-$   &   26.55 $\pm$ 0.27   &   25.73 $\pm$ 0.14   &   25.99 $\pm$ 0.40   &   25.10 $\pm$ 0.38   &   25.06 $\pm$ 0.03   &   24.96 $\pm$ 0.15   &    24.61 $\pm$ 0.04 \\ 
FIELD2\_257  &  217.78823  &  33.16198   &   $-$   &   25.97 $\pm$ 0.17   &   25.36 $\pm$ 0.10   &   25.85 $\pm$ 0.36   &   25.42 $\pm$ 0.47   &   25.05 $\pm$ 0.03   &   24.95 $\pm$ 0.15   &    25.00 $\pm$ 0.06 \\ 
FIELD3\_173  &  218.09895  &  33.33332   &   $-$   &   26.05 $\pm$ 0.18   &   25.11 $\pm$ 0.07   &   26.62 $\pm$ 0.62   &   26.78 $\pm$ 0.75   &   25.84 $\pm$ 0.08   &   25.92 $\pm$ 0.36   &    25.84 $\pm$ 0.14 \\ 
FIELD3\_311  &  218.07962  &  33.32042   &   $-$   &   26.80 $\pm$ 0.33   &   25.83 $\pm$ 0.13   &   27.81 $\pm$ 1.32   &   26.00 $\pm$ 0.70   &   25.45 $\pm$ 0.05   &   25.53 $\pm$ 0.25   &    25.52 $\pm$ 0.10 \\ 
FIELD4\_127  &  218.23677  &  33.66879   &   $-$   &   25.62 $\pm$ 0.17   &   24.96 $\pm$ 0.06   &   25.54 $\pm$ 0.37   &   25.52 $\pm$ 0.45   &   25.63 $\pm$ 0.07   &   25.51 $\pm$ 0.25   &    25.47 $\pm$ 0.10 \\ 
     \hline
    \end{tabular}%
}
\end{table*}

%%%%%%%%%%%%%%%%%%%%%%%%%%%%%%%%%%%%%%%%%%%%%%%%%%%%%%%%%%%%%%%%%%%%%%%%%%%%%%%%%%%%%%%%%%%%%%%%%%%%%%%%%%%%%%%%%%%%%%%%%%%%%%%
%%%%%%%%%%%%%%%%%%%%%%%%%%%%%%%%%%%%%%%%%%%%%%%%%%%%%%%%%%%%%%%%%%%%%%%%%%%%%%%%%%%%%%%%%%%%%%%%%%%%%%%%%%%%%%%%%%%%%%%%%%%%%%%%

\begin{table*}[t!]
    \centering
    \caption{Properties of the Lyman-Alpha Emitters}
    \label{tab:sed_props}
    \resizebox{0.9\textwidth}{!}{%
    \begin{tabular}{*{7}{|c}|}
    \hline
{\bf ID}   &   {\bf $\log~(M_{\star}~[\msun])$}   &   {\bf SFR [$\msun~{\rm yr^{-1}}$]}   &   {\bf Age [Myr]}   &   {\bf E(B-V) [mag]}   &   {\bf $L_{\rm Ly\alpha}~{\rm [10^{42}~ergs~s^{-1}]}$}   & {\bf $L_{\rm [OII]}~{\rm [10^{41}~ergs~s^{-1}]}$} \\ \hline
    \multicolumn{7}{|c|}{$z$LAEs} \\ \hline
DENS1\_86  &   7.9 $\pm$ 1.4   &   0.8 $\pm$ 0.7   &   101.0 $\pm$ 518.6   &   0.00 $\pm$ 0.02   &   2.99~$\pm$~0.44   &    $<$~2.5    \\
DENS1\_290   &   7.4 $\pm$ 0.2   &   2.5 $\pm$ 1.4   &   10.0 $\pm$ 0.5   &   0.01 $\pm$ 0.03   &   2.69~$\pm$~0.51     &   3.3~$\pm$~0.8  \\
DENS1\_306   &   8.5 $\pm$ 0.2   &   14.6 $\pm$ 3.1   &   19.0 $\pm$ 21.8   &   0.06 $\pm$ 0.01   &   3.10~$\pm$~0.40     &   6.9~$\pm$~1.1  \\
DENS1\_339  &   8.1 $\pm$ 0.3   &   11.4 $\pm$ 2.9   &   10.0 $\pm$ 24.3   &   0.08 $\pm$ 0.03   &   3.03~$\pm$~0.43   &    $<$~2.6    \\
DENS1\_356   &   8.2 $\pm$ 0.6   &   1.1 $\pm$ 0.8   &   160.0 $\pm$ 332.0   &   0.00 $\pm$ 0.01   &   3.32~$\pm$~0.42     &   0.3~$\pm$~0.8  \\
DENS2\_199  &   8.0 $\pm$ 0.4   &   5.0 $\pm$ 1.4   &   19.0 $\pm$ 61.9   &   0.07 $\pm$ 0.03   &   3.86~$\pm$~0.51   &    $<$~2.5    \\
DENS2\_202   &   8.5 $\pm$ 0.2   &   8.2 $\pm$ 3.8   &   40.0 $\pm$ 33.5   &   0.01 $\pm$ 0.01   &   4.96~$\pm$~0.43     &   1.6~$\pm$~1.1  \\
DENS2\_235  &   7.8 $\pm$ 1.0   &   4.5 $\pm$ 1.9   &   15.0 $\pm$ 209.9   &   0.09 $\pm$ 0.04   &   5.08~$\pm$~0.44   &    $<$~2.5    \\
DENS3\_17   &   7.2 $\pm$ 0.6   &   1.6 $\pm$ 0.6   &   10.0 $\pm$ 50.6   &   0.00 $\pm$ 0.03   &   1.62~$\pm$~0.42     &   4.6~$\pm$~0.9  \\
DENS3\_41  &   8.8 $\pm$ 0.2   &   22.2 $\pm$ 8.4   &   30.0 $\pm$ 28.3   &   0.09 $\pm$ 0.02   &   1.85~$\pm$~0.39   &    $<$~3.3    \\
DENS3\_118   &   9.0 $\pm$ 0.4   &   2.8 $\pm$ 1.9   &   321.0 $\pm$ 523.5   &   0.06 $\pm$ 0.04   &   2.75~$\pm$~0.41     &   0.8~$\pm$~1.1  \\
DENS3\_285   &   8.3 $\pm$ 0.4   &   2.0 $\pm$ 1.6   &   101.0 $\pm$ 143.6   &   0.00 $\pm$ 0.02   &   2.14~$\pm$~0.43     &   3.9~$\pm$~1.1  \\
FIELD1\_45   &   9.0 $\pm$ 0.2   &   7.1 $\pm$ 2.3   &   127.0 $\pm$ 89.8   &   0.04 $\pm$ 0.02   &   9.77~$\pm$~0.56     &   5.4~$\pm$~1.1  \\
FIELD1\_89  &   8.5 $\pm$ 0.6   &   1.5 $\pm$ 1.2   &   202.0 $\pm$ 433.8   &   0.01 $\pm$ 0.03   &   3.16~$\pm$~0.43   &    $<$~3.4    \\
FIELD1\_122  &   7.7 $\pm$ 0.1   &   4.7 $\pm$ 1.0   &   10.0 $\pm$ 2.3   &   0.03 $\pm$ 0.02   &   5.60~$\pm$~0.47   &    $<$~2.8    \\
FIELD1\_180   &   8.7 $\pm$ 0.2   &   18.1 $\pm$ 6.0   &   30.0 $\pm$ 30.3   &   0.07 $\pm$ 0.01   &   5.70~$\pm$~0.44     &   6.9~$\pm$~1.1  \\
FIELD1\_265   &   8.0 $\pm$ 0.4   &   5.4 $\pm$ 1.0   &   19.0 $\pm$ 49.2   &   0.04 $\pm$ 0.02   &   2.93~$\pm$~0.51     &   9.4~$\pm$~1.1  \\
FIELD1\_281   &   7.9 $\pm$ 0.3   &   4.4 $\pm$ 1.8   &   19.0 $\pm$ 29.9   &   0.10 $\pm$ 0.05   &   2.70~$\pm$~0.42     &   0.3~$\pm$~0.9  \\
FIELD2\_64   &   8.7 $\pm$ 0.1   &   23.7 $\pm$ 3.3   &   19.0 $\pm$ 7.9   &   0.12 $\pm$ 0.02   &   1.71~$\pm$~0.52     &   2.4~$\pm$~1.1  \\
FIELD2\_91   &   8.5 $\pm$ 0.3   &   19.6 $\pm$ 6.0   &   15.0 $\pm$ 30.1   &   0.08 $\pm$ 0.02   &   5.62~$\pm$~0.53     &   2.6~$\pm$~1.2  \\
FIELD2\_132   &   9.0 $\pm$ 0.2   &   8.0 $\pm$ 2.1   &   127.0 $\pm$ 78.8   &   0.04 $\pm$ 0.02   &   3.42~$\pm$~0.54     &   6.1~$\pm$~1.2  \\
FIELD2\_174   &   9.1 $\pm$ 0.1   &   19.0 $\pm$ 7.9   &   64.0 $\pm$ 34.6   &   0.04 $\pm$ 0.01   &   6.72~$\pm$~0.54     &   11.4~$\pm$~1.2  \\
FIELD3\_124  &   8.1 $\pm$ 0.7   &   7.9 $\pm$ 1.9   &   15.0 $\pm$ 80.4   &   0.04 $\pm$ 0.02   &   5.66~$\pm$~0.56   &    $<$~3.7    \\
FIELD3\_327   &   7.7 $\pm$ 0.6   &   4.8 $\pm$ 1.1   &   10.0 $\pm$ 47.3   &   0.02 $\pm$ 0.02   &   2.62~$\pm$~0.45     &   0.4~$\pm$~0.8  \\
FIELD4\_33   &   8.3 $\pm$ 0.2   &   9.2 $\pm$ 2.3   &   19.0 $\pm$ 17.0   &   0.05 $\pm$ 0.02   &   6.98~$\pm$~0.49     &   7.0~$\pm$~1.1  \\
FIELD4\_50   &   8.6 $\pm$ 0.3   &   19.9 $\pm$ 4.7   &   19.0 $\pm$ 43.0   &   0.13 $\pm$ 0.03   &   3.11~$\pm$~0.48     &   3.2~$\pm$~1.1  \\
FIELD4\_124  &   9.6 $\pm$ 0.1   &   5.0 $\pm$ 0.4   &   806.0 $\pm$ 324.8   &   0.00 $\pm$ 0.00   &   31.67~$\pm$~0.80   &    $<$~3.5    \\ \hline
\multicolumn{7}{|c|}{$p$LAEs}  \\ \hline
DENS1\_43  &   8.5 $\pm$ 0.2   &   14.2 $\pm$ 5.3   &   19.0 $\pm$ 19.7   &   0.20 $\pm$ 0.06   &   1.26~$\pm$~0.38   &    $<$~2.5    \\
DENS1\_59   &   9.5 $\pm$ 0.2   &   24.6 $\pm$ 16.3   &   127.0 $\pm$ 212.2   &   0.26 $\pm$ 0.06   &   1.59~$\pm$~0.38     &   3.4~$\pm$~1.1  \\
DENS1\_97  &   8.5 $\pm$ 0.1   &   17.5 $\pm$ 3.4   &   19.0 $\pm$ 0.0   &   0.16 $\pm$ 0.04   &   1.76~$\pm$~0.39   &    $<$~3.3    \\
DENS1\_143   &   8.3 $\pm$ 1.0   &   1.1 $\pm$ 1.9   &   160.0 $\pm$ 597.8   &   0.02 $\pm$ 0.04   &   0.95~$\pm$~0.40     &   1.8~$\pm$~1.1  \\
DENS1\_172   &   9.0 $\pm$ 0.4   &   8.2 $\pm$ 5.6   &   127.0 $\pm$ 344.1   &   0.10 $\pm$ 0.04   &   1.71~$\pm$~0.45     &   4.1~$\pm$~1.1  \\
DENS1\_235  &   9.3 $\pm$ 0.2   &   12.0 $\pm$ 3.8   &   160.0 $\pm$ 134.5   &   0.10 $\pm$ 0.02   &   1.82~$\pm$~0.40   &    $<$~3.4    \\
DENS1\_245  &   8.8 $\pm$ 0.3   &   12.7 $\pm$ 5.4   &   50.0 $\pm$ 81.0   &   0.11 $\pm$ 0.02   &   2.22~$\pm$~0.40   &    $<$~3.3    \\
DENS1\_266   &   9.3 $\pm$ 0.2   &   4.4 $\pm$ 1.9   &   508.0 $\pm$ 470.6   &   0.04 $\pm$ 0.03   &   1.02~$\pm$~0.41     &   1.3~$\pm$~1.1  \\
DENS1\_304   &   7.4 $\pm$ 2.2   &   2.7 $\pm$ 2.2   &   10.0 $\pm$ 165.8   &   0.07 $\pm$ 0.08   &   1.28~$\pm$~0.54     &   1.2~$\pm$~1.1  \\
DENS1\_344   &   8.4 $\pm$ 0.6   &   2.7 $\pm$ 2.0   &   101.0 $\pm$ 260.3   &   0.03 $\pm$ 0.03   &   3.25~$\pm$~0.44     &   2.6~$\pm$~1.1  \\
DENS2\_24   &   8.7 $\pm$ 0.1   &   23.4 $\pm$ 3.1   &   19.0 $\pm$ 6.1   &   0.15 $\pm$ 0.02   &   1.74~$\pm$~0.42     &   1.8~$\pm$~1.1  \\
DENS2\_239  &   8.7 $\pm$ 0.3   &   6.8 $\pm$ 4.6   &   80.0 $\pm$ 193.5   &   0.15 $\pm$ 0.05   &   2.20~$\pm$~0.40   &    $<$~3.3    \\
DENS2\_281   &   8.3 $\pm$ 0.2   &   9.6 $\pm$ 2.4   &   19.0 $\pm$ 15.4   &   0.06 $\pm$ 0.02   &   2.04~$\pm$~0.42     &   4.7~$\pm$~1.1  \\
DENS3\_20  &   8.9 $\pm$ 0.0   &   37.3 $\pm$ 3.6   &   19.0 $\pm$ 0.0   &   0.16 $\pm$ 0.02   &   1.11~$\pm$~0.39   &    $<$~3.4    \\
DENS3\_126   &   9.1 $\pm$ 0.1   &   26.0 $\pm$ 7.4   &   50.0 $\pm$ 34.6   &   0.15 $\pm$ 0.02   &   2.08~$\pm$~0.39     &   0.7~$\pm$~1.1  \\
DENS3\_172   &   8.8 $\pm$ 0.5   &   1.1 $\pm$ 1.6   &   508.0 $\pm$ 696.7   &   0.01 $\pm$ 0.03   &   2.09~$\pm$~0.42     &   0.4~$\pm$~1.1  \\
DENS3\_279  &   9.3 $\pm$ 0.2   &   1.9 $\pm$ 0.6   &   1015.0 $\pm$ 661.0   &   0.03 $\pm$ 0.03   &   1.68~$\pm$~0.40   &    $<$~2.5    \\
FIELD1\_60  &   8.5 $\pm$ 0.1   &   16.4 $\pm$ 3.4   &   19.0 $\pm$ 2.5   &   0.13 $\pm$ 0.03   &   1.51~$\pm$~0.41   &    $<$~3.3    \\
FIELD1\_71   &   8.5 $\pm$ 0.1   &   16.6 $\pm$ 4.6   &   19.0 $\pm$ 1.1   &   0.13 $\pm$ 0.05   &   1.44~$\pm$~0.43     &   2.7~$\pm$~1.1  \\
FIELD1\_270   &   9.6 $\pm$ 0.1   &   25.8 $\pm$ 8.4   &   160.0 $\pm$ 86.3   &   0.18 $\pm$ 0.03   &   1.65~$\pm$~0.41     &   9.2~$\pm$~1.1  \\
FIELD1\_283   &   9.3 $\pm$ 0.1   &   83.3 $\pm$ 12.1   &   25.0 $\pm$ 7.1   &   0.14 $\pm$ 0.01   &   3.89~$\pm$~0.91     &   28.0~$\pm$~1.2  \\
FIELD2\_26   &   8.4 $\pm$ 0.5   &   5.4 $\pm$ 3.0   &   50.0 $\pm$ 113.7   &   0.04 $\pm$ 0.02   &   3.29~$\pm$~0.54     &   1.9~$\pm$~1.1  \\
FIELD2\_29   &   9.1 $\pm$ 0.2   &   15.5 $\pm$ 7.3   &   80.0 $\pm$ 51.2   &   0.11 $\pm$ 0.02   &   1.73~$\pm$~0.51     &   2.6~$\pm$~1.1  \\
FIELD2\_71   &   8.7 $\pm$ 0.1   &   28.0 $\pm$ 4.9   &   19.0 $\pm$ 1.7   &   0.16 $\pm$ 0.03   &   1.70~$\pm$~0.50     &   4.1~$\pm$~1.1  \\
FIELD2\_116  &   8.4 $\pm$ 0.1   &   13.6 $\pm$ 3.8   &   19.0 $\pm$ 11.4   &   0.13 $\pm$ 0.04   &   1.35~$\pm$~0.51   &    $<$~3.3    \\
FIELD2\_140  &   9.5 $\pm$ 0.1   &   71.6 $\pm$ 17.0   &   40.0 $\pm$ 23.7   &   0.16 $\pm$ 0.01   &   2.79~$\pm$~0.50   &    $<$~3.5    \\
FIELD2\_141  &   8.8 $\pm$ 0.1   &   28.6 $\pm$ 4.0   &   19.0 $\pm$ 11.0   &   0.12 $\pm$ 0.01   &   2.47~$\pm$~0.50   &    $<$~3.4    \\
FIELD2\_149   &   8.2 $\pm$ 0.3   &   8.1 $\pm$ 4.6   &   19.0 $\pm$ 15.4   &   0.15 $\pm$ 0.10   &   2.85~$\pm$~0.26     &   5.4~$\pm$~1.1  \\
FIELD2\_181   &   8.9 $\pm$ 0.0   &   37.0 $\pm$ 4.7   &   19.0 $\pm$ 3.7   &   0.16 $\pm$ 0.02   &   1.99~$\pm$~0.50     &   2.1~$\pm$~1.1  \\
FIELD2\_243   &   8.8 $\pm$ 0.0   &   28.6 $\pm$ 2.9   &   19.0 $\pm$ 0.9   &   0.13 $\pm$ 0.02   &   2.30~$\pm$~0.50     &   6.3~$\pm$~1.1  \\
FIELD2\_253  &   9.3 $\pm$ 0.1   &   24.0 $\pm$ 9.3   &   80.0 $\pm$ 52.1   &   0.17 $\pm$ 0.03   &   1.60~$\pm$~0.51   &    $<$~3.3    \\
FIELD2\_257   &   8.7 $\pm$ 0.1   &   26.9 $\pm$ 3.5   &   19.0 $\pm$ 3.4   &   0.12 $\pm$ 0.01   &   2.16~$\pm$~0.50     &   1.4~$\pm$~1.1  \\
FIELD3\_173  &   8.3 $\pm$ 0.4   &   10.9 $\pm$ 2.4   &   19.0 $\pm$ 46.7   &   0.08 $\pm$ 0.02   &   4.10~$\pm$~0.44   &    $<$~3.5    \\
FIELD3\_311  &   8.6 $\pm$ 0.1   &   20.4 $\pm$ 3.5   &   19.0 $\pm$ 2.0   &   0.14 $\pm$ 0.03   &   2.20~$\pm$~0.47   &    $<$~3.3    \\
FIELD4\_127   &   8.5 $\pm$ 0.4   &   10.4 $\pm$ 4.1   &   30.0 $\pm$ 64.0   &   0.05 $\pm$ 0.02   &   3.00~$\pm$~0.51     &   0.3~$\pm$~1.1  \\

\hline
    
    \end{tabular}%
    }
\end{table*}

%%%%%%%%%%%%%%%%%%%%%%%%%%%%%%%%%%%%%%%%%%%%%%%%%%%%%%%%%%%%%%%%%%%%%%%%%%%%%%%%%%%%%%%%%%%%%%%%%%%%%%%%%%%%%%%%%%%%%%%%%%%%%%%
%%%%%%%%%%%%%%%%%%%%%%%%%%%%%%%%%%%%%%%%%%%%%%%%%%%%%%%%%%%%%%%%%%%%%%%%%%%%%%%%%%%%%%%%%%%%%%%%%%%%%%%%%%%%%%%%%%%%%%%%%%%%%%%%

\begin{table*}[h!]
    \centering
    \caption{Properties of the Lyman-Alpha Emitters}
    \label{tab:size_props}
    \resizebox{0.5\textwidth}{!}{%
    \begin{tabular}{*{4}{|c}|}
    \hline
{\bf ID}   &   {\bf $r_{e}~[kpc])$}   &   {\bf $\Sigma_{\rm SFR}$ [$\msun~{\rm yr^{-1}~kpc^{-2}}$]}   &   {\bf $\Sigma_{\rm sSFR}$ [${\rm Gyr^{-1}~kpc^{-2}}$]} \\ \hline
    \multicolumn{4}{|c|}{$z$LAEs} \\ \hline
DENS1\_86  &    $<$ 1.0   &   $>$ 0.12   &   $>$1.53    \\
DENS1\_290  &    $<$ 1.0   &   $>$ 0.38   &   $>$15.57    \\
DENS1\_306  &    $<$ 1.0   &   $>$ 2.26   &   $>$7.76    \\
DENS1\_339  &    $<$ 1.0   &   $>$ 1.78   &   $>$15.57    \\
DENS1\_356  &    $<$ 1.0   &   $>$ 0.17   &   $>$0.96    \\
DENS2\_199  &    $<$ 1.0   &   $>$ 0.78   &   $>$7.77    \\
DENS2\_202  &   2.2 $\pm$ 0.2   &   0.26 $\pm$ 0.08    &    0.81 $\pm$ 0.48  \\
DENS2\_235  &    $<$ 1.0   &   $>$ 0.70   &   $>$10.23    \\
DENS3\_17  &    $<$ 1.0   &   $>$ 0.25   &   $>$15.49    \\
DENS3\_41  &    $<$ 1.0   &   $>$ 3.42   &   $>$5.12    \\
DENS3\_118  &    $<$ 1.0   &   $>$ 0.43   &   $>$0.48    \\
DENS3\_285  &    $<$ 1.0   &   $>$ 0.30   &   $>$1.52    \\
FIELD1\_45  &    $<$ 1.0   &   $>$ 1.11   &   $>$1.21    \\
FIELD1\_89  &    $<$ 1.0   &   $>$ 0.23   &   $>$0.77    \\
FIELD1\_122  &    $<$ 1.0   &   $>$ 0.73   &   $>$15.65    \\
FIELD1\_180  &   1.5 $\pm$ 0.1   &   1.24 $\pm$ 0.19    &    2.30 $\pm$ 1.22  \\
FIELD1\_265  &    $<$ 1.0   &   $>$ 0.84   &   $>$7.81    \\
FIELD1\_281  &    $<$ 1.0   &   $>$ 0.69   &   $>$7.80    \\
FIELD2\_64  &    $<$ 1.0   &   $>$ 3.68   &   $>$7.79    \\
FIELD2\_91  &   2.4 $\pm$ 0.2   &   0.56 $\pm$ 0.09    &    1.89 $\pm$ 1.45  \\
FIELD2\_132  &   2.2 $\pm$ 0.2   &   0.27 $\pm$ 0.03    &    0.26 $\pm$ 0.10  \\
FIELD2\_174  &    $<$ 1.0   &   $>$ 2.96   &   $>$2.44    \\
FIELD3\_124  &    $<$ 1.0   &   $>$ 1.24   &   $>$10.38    \\
FIELD3\_327  &    $<$ 1.0   &   $>$ 0.74   &   $>$15.31    \\
FIELD4\_33  &    $<$ 1.0   &   $>$ 1.41   &   $>$7.71    \\
FIELD4\_50  &   4.1 $\pm$ 0.3   &   0.19 $\pm$ 0.02    &    0.47 $\pm$ 0.33  \\
FIELD4\_124  &    $<$ 1.0   &   $>$ 0.77   &   $>$0.19    \\
    \hline
    \multicolumn{4}{|c|}{$p$LAEs}  \\ \hline
DENS1\_43  &   1.8 $\pm$ 0.3   &   0.70 $\pm$ 0.00    &    2.47 $\pm$ 1.08  \\
DENS1\_59  &   1.6 $\pm$ 0.2   &   1.46 $\pm$ 0.64    &    0.47 $\pm$ 0.29  \\
DENS1\_97  &   2.3 $\pm$ 0.3   &   0.52 $\pm$ 0.03    &    1.49 $\pm$ 0.30  \\
DENS1\_143  &    $<$ 1.0   &   $>$ 0.17   &   $>$0.96    \\
DENS1\_172  &   6.0 $\pm$ 0.8   &   0.04 $\pm$ 0.02    &    0.03 $\pm$ 0.04  \\
DENS1\_235  &    $<$ 1.0   &   $>$ 1.86   &   $>$0.96    \\
DENS1\_245  &    $<$ 1.0   &   $>$ 1.97   &   $>$3.09    \\
DENS1\_266  &    $<$ 1.0   &   $>$ 0.68   &   $>$0.30    \\
DENS1\_304  &    $<$ 1.0   &   $>$ 0.41   &   $>$15.46    \\
DENS1\_344  &    $<$ 1.0   &   $>$ 0.42   &   $>$1.52    \\
DENS2\_24  &   2.4 $\pm$ 0.2   &   0.66 $\pm$ 0.00    &    1.41 $\pm$ 0.25  \\
DENS2\_239  &   2.7 $\pm$ 0.3   &   0.15 $\pm$ 0.06    &    0.28 $\pm$ 0.25  \\
DENS2\_281  &   1.8 $\pm$ 0.2   &   0.48 $\pm$ 0.00    &    2.50 $\pm$ 0.91  \\
DENS3\_20  &   2.5 $\pm$ 0.1   &   0.93 $\pm$ 0.00    &    1.25 $\pm$ 0.12  \\
DENS3\_126  &    $<$ 1.0   &   $>$ 4.02   &   $>$3.09    \\
DENS3\_172  &    $<$ 1.0   &   $>$ 0.18   &   $>$0.30    \\
DENS3\_279  &   2.0 $\pm$ 0.3   &   0.07 $\pm$ 0.00    &    0.04 $\pm$ 0.02  \\
FIELD1\_60  &   2.2 $\pm$ 0.2   &   0.56 $\pm$ 0.01    &    1.70 $\pm$ 0.28  \\
FIELD1\_71  &   2.1 $\pm$ 0.2   &   0.60 $\pm$ 0.05    &    1.81 $\pm$ 0.54  \\
FIELD1\_270  &   2.2 $\pm$ 0.1   &   0.84 $\pm$ 0.18    &    0.20 $\pm$ 0.07  \\
FIELD1\_283  &   1.4 $\pm$ 0.1   &   6.79 $\pm$ 0.22    &    3.25 $\pm$ 0.53  \\
FIELD2\_26  &    $<$ 1.0   &   $>$ 0.83   &   $>$3.09    \\
FIELD2\_29  &    $<$ 1.0   &   $>$ 2.39   &   $>$1.92    \\
FIELD2\_71  &   3.6 $\pm$ 0.2   &   0.34 $\pm$ 0.02    &    0.61 $\pm$ 0.09  \\
FIELD2\_116  &   2.0 $\pm$ 0.2   &   0.52 $\pm$ 0.03    &    1.92 $\pm$ 0.54  \\
FIELD2\_140  &   1.3 $\pm$ 0.1   &   6.92 $\pm$ 0.77    &    2.42 $\pm$ 0.76  \\
FIELD2\_141  &   1.7 $\pm$ 0.1   &   1.61 $\pm$ 0.00    &    2.82 $\pm$ 0.68  \\
FIELD2\_149  &    $<$ 1.0   &   $>$ 1.25   &   $>$7.75    \\ 
FIELD2\_181  &   2.6 $\pm$ 0.1   &   0.89 $\pm$ 0.03    &    1.20 $\pm$ 0.14  \\
FIELD2\_243  &   3.1 $\pm$ 0.1   &   0.49 $\pm$ 0.00    &    0.85 $\pm$ 0.09  \\
FIELD2\_253  &   3.2 $\pm$ 0.2   &   0.37 $\pm$ 0.10    &    0.19 $\pm$ 0.08  \\
FIELD2\_257  &    $<$ 1.0   &   $>$ 4.15   &   $>$7.75    \\
FIELD3\_173  &    $<$ 1.0   &   $>$ 1.69   &   $>$7.75    \\
FIELD3\_311  &   2.5 $\pm$ 0.2   &   0.52 $\pm$ 0.00    &    1.27 $\pm$ 0.20  \\
FIELD4\_127  &   1.8 $\pm$ 0.2   &   0.51 $\pm$ 0.09    &    1.65 $\pm$ 1.44  \\
    \hline
    \end{tabular}%
    }
\end{table*}

%%%%%%%%%%%%%%%%%%%%%%%%%%%%%%%%%%%%%%%%%%%%%%%%%%%%%%%%%%%%%%%%%%%%%%%%%%%%%%%%%%%%%%%%%%%%%%%%%%%%%%%%%%%%%%%%%%%%%%%%%%%%%%%
%%%%%%%%%%%%%%%%%%%%%%%%%%%%%%%%%%%%%%%%%%%%%%%%%%%%%%%%%%%%%%%%%%%%%%%%%%%%%%%%%%%%%%%%%%%%%%%%%%%%%%%%%%%%%%%%%%%%%%%%%%%%%%%%

\begin{table*}[t!]
    \centering
    \caption{\lya~escape proxies for the Lyman-Alpha Emitters}
    \label{tab:esc_props}
    \resizebox{0.45\textwidth}{!}{%
    \begin{tabular}{*{4}{|c}|}
    \hline
{\bf ID}   &   {\bf $W_{\rm Ly\alpha}$ [\AA]}   &   {\bf \lya/\oii}    &     {\bf \fesc}  \\ \hline
    \multicolumn{4}{|c|}{$z$LAEs} \\ \hline
DENS1\_86  &   188.2 $\pm$ 81.36    &    $>$ 12.11    &    1.67 $\pm$ 1.54   \\
DENS1\_290  &   95.0 $\pm$ 38.25    &    8.15 $\pm$ 2.54    &    0.49 $\pm$ 0.28  \\
DENS1\_306  &   43.4 $\pm$ 7.37    &    4.48 $\pm$ 0.94    &    0.09 $\pm$ 0.02  \\
DENS1\_339  &   77.2 $\pm$ 18.31    &    $>$ 11.50    &    0.12 $\pm$ 0.03   \\
DENS1\_356  &   143.1 $\pm$ 42.82    &    98.04 $\pm$ 240.63    &    1.36 $\pm$ 1.00  \\
DENS2\_199  &   201.4 $\pm$ 89.15    &    $>$ 15.55    &    0.34 $\pm$ 0.11   \\
DENS2\_202  &   40.8 $\pm$ 4.42    &    31.98 $\pm$ 22.79    &    0.27 $\pm$ 0.13  \\
DENS2\_235  &   232.4 $\pm$ 59.34    &    $>$ 20.59    &    0.50 $\pm$ 0.21   \\
DENS3\_17  &   81.3 $\pm$ 34.00    &    3.50 $\pm$ 1.12    &    0.44 $\pm$ 0.20  \\
DENS3\_41  &   22.5 $\pm$ 5.20    &    $>$ 5.59    &    0.04 $\pm$ 0.02   \\
DENS3\_118  &   127.1 $\pm$ 38.41    &    36.00 $\pm$ 52.07    &    0.44 $\pm$ 0.31  \\
DENS3\_285  &   57.9 $\pm$ 16.52    &    5.41 $\pm$ 1.84    &    0.48 $\pm$ 0.40  \\
FIELD1\_45  &   154.9 $\pm$ 26.59    &    18.25 $\pm$ 4.01    &    0.61 $\pm$ 0.20  \\
FIELD1\_89  &   142.3 $\pm$ 42.19    &    $>$ 9.24    &    0.97 $\pm$ 0.79   \\
FIELD1\_122  &   143.1 $\pm$ 27.77    &    $>$ 20.26    &    0.54 $\pm$ 0.12   \\
FIELD1\_180  &   64.7 $\pm$ 7.48    &    8.21 $\pm$ 1.48    &    0.14 $\pm$ 0.05  \\
FIELD1\_265  &   112.5 $\pm$ 43.58    &    3.12 $\pm$ 0.66    &    0.24 $\pm$ 0.06  \\
FIELD1\_281  &   192.7 $\pm$ 72.70    &    78.96 $\pm$ 204.40    &    0.27 $\pm$ 0.12  \\
FIELD2\_64  &   39.6 $\pm$ 14.13    &    7.00 $\pm$ 3.92    &    0.03 $\pm$ 0.01  \\
FIELD2\_91  &   67.7 $\pm$ 8.97    &    21.66 $\pm$ 9.85    &    0.13 $\pm$ 0.04  \\
FIELD2\_132  &   48.7 $\pm$ 9.72    &    5.61 $\pm$ 1.39    &    0.19 $\pm$ 0.06  \\
FIELD2\_174  &   38.3 $\pm$ 3.72    &    5.88 $\pm$ 0.76    &    0.16 $\pm$ 0.07  \\
FIELD3\_124  &   137.5 $\pm$ 34.43    &    $>$ 15.46    &    0.32 $\pm$ 0.08   \\
FIELD3\_327  &   86.0 $\pm$ 24.95    &    60.42 $\pm$ 111.36    &    0.24 $\pm$ 0.07  \\
FIELD4\_33  &   115.1 $\pm$ 15.58    &    9.99 $\pm$ 1.71    &    0.34 $\pm$ 0.09  \\
FIELD4\_50  &   89.6 $\pm$ 22.19    &    9.62 $\pm$ 3.56    &    0.07 $\pm$ 0.02  \\
FIELD4\_124  &   303.1 $\pm$ 41.55    &    $>$ 90.80    &    2.85 $\pm$ 0.22   \\
\hline
\multicolumn{4}{|c|}{$p$LAEs}  \\ \hline
DENS1\_43  &   143.5 $\pm$ 83.51    &    $>$ 5.04    &    0.04 $\pm$ 0.02   \\
DENS1\_59  &   146.5 $\pm$ 71.37    &    4.72 $\pm$ 1.93    &    0.03 $\pm$ 0.02  \\
DENS1\_97  &   123.4 $\pm$ 51.16    &    $>$ 5.37    &    0.04 $\pm$ 0.01   \\
DENS1\_143  &   61.3 $\pm$ 38.22    &    5.30 $\pm$ 3.99    &    0.38 $\pm$ 0.66  \\
DENS1\_172  &   64.4 $\pm$ 27.43    &    4.21 $\pm$ 1.60    &    0.09 $\pm$ 0.07  \\
DENS1\_235  &   40.1 $\pm$ 11.28    &    $>$ 5.31    &    0.07 $\pm$ 0.03   \\
DENS1\_245  &   68.5 $\pm$ 19.03    &    $>$ 6.68    &    0.08 $\pm$ 0.04   \\
DENS1\_266  &   21.7 $\pm$ 10.19    &    7.76 $\pm$ 7.30    &    0.10 $\pm$ 0.06  \\
DENS1\_304  &   111.2 $\pm$ 114.88    &    10.25 $\pm$ 10.05    &    0.21 $\pm$ 0.20  \\
DENS1\_344  &   135.4 $\pm$ 42.92    &    12.53 $\pm$ 5.65    &    0.53 $\pm$ 0.40  \\
DENS2\_24  &   74.7 $\pm$ 25.32    &    9.71 $\pm$ 6.56    &    0.03 $\pm$ 0.01  \\
DENS2\_239  &   157.0 $\pm$ 51.91    &    $>$ 6.71    &    0.15 $\pm$ 0.10   \\
DENS2\_281  &   41.5 $\pm$ 10.50    &    4.34 $\pm$ 1.38    &    0.09 $\pm$ 0.03  \\
DENS3\_20  &   38.2 $\pm$ 15.68    &    $>$ 3.24    &    0.01 $\pm$ 0.00   \\
DENS3\_126  &   52.8 $\pm$ 12.22    &    29.97 $\pm$ 48.80    &    0.04 $\pm$ 0.01  \\
DENS3\_172  &   113.8 $\pm$ 42.65    &    50.77 $\pm$ 134.47    &    0.82 $\pm$ 1.16  \\
DENS3\_279  &   74.6 $\pm$ 26.48    &    $>$ 6.75    &    0.39 $\pm$ 0.16   \\
FIELD1\_60  &   65.7 $\pm$ 26.12    &    $>$ 4.58    &    0.04 $\pm$ 0.01   \\
FIELD1\_71  &   105.6 $\pm$ 58.71    &    5.33 $\pm$ 2.71    &    0.04 $\pm$ 0.02  \\
FIELD1\_270  &   54.9 $\pm$ 17.95    &    1.79 $\pm$ 0.49    &    0.03 $\pm$ 0.01  \\
FIELD1\_283  &   30.9 $\pm$ 7.52    &    1.39 $\pm$ 0.33    &    0.02 $\pm$ 0.01  \\
FIELD2\_26  &   65.6 $\pm$ 15.56    &    17.09 $\pm$ 10.15    &    0.27 $\pm$ 0.16  \\
FIELD2\_29  &   34.9 $\pm$ 11.93    &    6.53 $\pm$ 3.37    &    0.05 $\pm$ 0.03  \\
FIELD2\_71  &   68.1 $\pm$ 26.40    &    4.14 $\pm$ 1.67    &    0.03 $\pm$ 0.01  \\
FIELD2\_116  &   61.7 $\pm$ 30.98    &    $>$ 4.08    &    0.04 $\pm$ 0.02   \\
FIELD2\_140  &   32.3 $\pm$ 6.52    &    $>$ 8.00    &    0.02 $\pm$ 0.01   \\
FIELD2\_141  &   52.0 $\pm$ 13.08    &    $>$ 7.31    &    0.04 $\pm$ 0.01   \\
FIELD2\_149  &   858.0 $\pm$ 565.77    &    5.25 $\pm$ 1.16    &    0.16 $\pm$ 0.09  \\
FIELD2\_181  &   51.7 $\pm$ 16.06    &    9.58 $\pm$ 5.72    &    0.02 $\pm$ 0.01  \\
FIELD2\_243  &   59.4 $\pm$ 16.52    &    3.63 $\pm$ 1.00    &    0.04 $\pm$ 0.01  \\
FIELD2\_253  &   49.7 $\pm$ 19.57    &    $>$ 4.88    &    0.03 $\pm$ 0.01   \\
FIELD2\_257  &   45.8 $\pm$ 12.82    &    15.11 $\pm$ 12.01    &    0.04 $\pm$ 0.01  \\
FIELD3\_173  &   121.5 $\pm$ 24.68    &    $>$ 11.77    &    0.17 $\pm$ 0.04   \\
FIELD3\_311  &   136.4 $\pm$ 63.59    &    $>$ 6.72    &    0.05 $\pm$ 0.01   \\
FIELD4\_127  &   42.8 $\pm$ 9.40    &    115.28 $\pm$ 492.40    &    0.13 $\pm$ 0.06  \\
    \hline
    \end{tabular}%
    }
\end{table*}

%%%%%%%%%%%%%%%%%%%%%%%%%%%%%%%%%%%%%%%%%%%%%%%%%%%%%%%%%%%%%%%%%%%%%%%%%%%%%%%%%%%%%%%%%%%%%%%%%%%%%%%%%%%%%%%%%%%%%%%%%%%%%%%
%%%%%%%%%%%%%%%%%%%%%%%%%%%%%%%%%%%%%%%%%%%%%%%%%%%%%%%%%%%%%%%%%%%%%%%%%%%%%%%%%%%%%%%%%%%%%%%%%%%%%%%%%%%%%%%%%%%%%%%%%%%%%%%%

%%%%%%%%%%%%%%%%%%%%%%%%%%%%%%%%%%%%%%%%%%%%%%%%%%%%%%%%%%%%%%%%%%%%%%%%%%%%%%%%%%%%%%%%%%%%%%%%%%%%%%%%%%%%%%%%%%%%%%%%%%%%%%%
%%%%%%%%%%%%%%%%%%%%%%%%%%%%%%%%%%%%%%%%%%%%%%%%%%%%%%%%%%%%%%%%%%%%%%%%%%%%%%%%%%%%%%%%%%%%%%%%%%%%%%%%%%%%%%%%%%%%%%%%%%%%%%%%

\end{document}